\documentclass[manuscript,nonacm,screen]{acmart}
\usepackage{amsmath,amsfonts}

\PassOptionsToPackage{svgnames}{xcolor}

\usepackage{mathrsfs}
\usepackage{balance}  
\usepackage{csquotes}
\usepackage{enumitem}
\usepackage[capitalise]{cleveref}
\usepackage{amsthm}
\usepackage{graphicx}
\usepackage{subcaption}
\usepackage{booktabs} 
\usepackage{thmtools, thm-restate}
\usepackage{listings}

\usepackage{ltablex}

\usepackage{algorithm,caption}
\usepackage{algpseudocode}
\usepackage{rotating}

\theoremstyle{definition}
\newtheorem{example}{Example}
\newtheorem{definition}{Definition}

\newtheorem{proposition}{Proposition}

\newcommand{\mysubsubsection}[1]{\subsubsection{#1}\ \smallskip}

\usepackage{commands}
\graphicspath{ {figures/} }

\newcommand{\comm}[3]{
  \noindent{{\normalsize{\color{#2}{[[{\bfseries #1: }#3{\bfseries\ #1}]]}}}}}

\newcommand{\bernd}[1]{\comm{B}{orange}{#1}}

\newcommand{\eric}[1]{\comm{E}{green!90!blue!90}{#1}}
\renewcommand{\comm}[3]{}

\newcommand{\Cnameref}[1]{\Cref{#1}}

\definecolor{new}{RGB}{32, 32,0}

\newenvironment{myproof}{\begin{proof}}{\end{proof}}
\newenvironment{myexample}{\begin{example}}{\end{example}}

\def\ojoin{\setbox0=\hbox{$\bowtie$}%
  \rule[-.02ex]{.25em}{.4pt}\llap{\rule[\ht0]{.25em}{.4pt}}}
\def\leftouterjoin{\mathbin{\ojoin\mkern-5.8mu\bowtie}}
\def\rightouterjoin{\mathbin{\bowtie\mkern-5.8mu\ojoin}}
\def\fullouterjoin{\mathbin{\ojoin\mkern-5.8mu\bowtie\mkern-5.8mu\ojoin}}
\DeclareMathAlphabet{\mathpzc}{OT1}{pzc}{m}{it} 

\setcopyright{acmcopyright}
\copyrightyear{2021}
\acmYear{2021}
\acmDOI{10.1145/1122445.1122456}

\acmBooktitle{J. Data and Information Quality}
\acmPrice{15.00}
\acmISBN{978-1-4503-XXXX-X/18/06}

\ccsdesc[500]{Information systems~Data management systems}
\ccsdesc[300]{Information systems~Data provenance}
\ccsdesc[300]{Information systems~Inconsistent data}
\ccsdesc[500]{Information systems~Data warehouses}

\keywords{analytic queries, summarizability, data quality, multi-dimensional data model, interactive query sessions}

\begin{abstract}
\bernd{comments are active}
We present a comprehensive set of conditions and rules to control the correctness of aggregation queries within an interactive data analysis session. The goal is to extend self-service data preparation and BI tools to automatically detect semantically incorrect aggregate queries on analytic tables and views built by using the common analytic operations including filter, project, join, aggregate, union, difference, and pivot. We introduce  \textit{aggregable properties} to describe for any attribute of an analytic table, which aggregation functions correctly aggregate the attribute along which sets of dimension attributes. These properties can also be used to formally identify attributes which are \emph{summarizable} with respect to some aggregation function along a given set of dimension attributes. This is particularly helpful to detect incorrect aggregations of  measures obtained through the use of non-distributive aggregation functions like average and count. We extend the notion of summarizability by introducing a new \textit{generalized summarizability condition} to control the aggregation of attributes after any analytic operation. 
Finally, we define \textit{propagation rules} which transform aggregable properties of the query input tables into new aggregable properties for the result tables, preserving summarizability and generalized summarizability.
\end{abstract}

\begin{document}
\title[Controlling the Correctness of Analytic Queries]{Controlling the Correctness of Aggregation Operations During Sessions of Interactive Analytic Queries}

\author{Eric Simon}
\email{eric.simon@sap.com}
\affiliation{
  \institution{SAP France}
  \country{France}
}

\author{Bernd Amann}
\email{bernd.amann@lip6.fr}
\affiliation{
  \institution{LIP6 -- Sorbonne Universit\'e, CNRS}
  \country{France}
}
\author{Rutian Liu}
\email{rutian.liu.fr@gmail.com}
\affiliation{%
  \institution{SAP France, LIP6 -- Sorbonne Université, CNRS}
  \country{France}
}

\author{St\'ephane Gan\c{c}arski}
\email{stephane.gancarski@lip6.fr}
\affiliation{
  \institution{LIP6 -- Sorbonne Universit\'e, CNRS}
  \country{France}
}

\maketitle



\section{Introduction}
\label{sec:introduction}


\subsection{Problem statement and motivations}

Analytic datasets are ubiquitous in numerous application domains and 
their usage includes, for example, the classic reporting on business activities in transactional applications \cite{kimball_data_2013}, the monitoring of the behavior of on-line systems based on log analysis (e.g., Splunk~\cite{splunk_webpage}, Elasticsearch/Kibana~\cite{kibana_webpage}, Datadog~\cite{datadog_webpage}), trend analysis in finance or social networks, or the conduct of epidemiological studies in healthcare~\cite{hamzah2020coronatracker}. 
In a world where an overwhelming amount of raw data is collected and stored at an affordable price in cloud object stores (e.g., Amazon S3~\cite{s3_webpage}, Azure Blob Storage~\cite{azure_webpage}), properly aggregated and cleaned data is the data foundation layer on which "augmented" analytics are built with the help of machine learning pipelines. 

\bernd{remove the following sentence?} The creation and maintenance of analytic datasets for supporting Business Intelligence (BI) 
applications 
has traditionally been the entitlement of experienced data engineers in IT organizations. Today, the emergence of self-service data preparation and BI tools (e.g., \cite{trifacta_web, paxata_web, data-preparation_sap_web}, \cite{tableau_webpage, powerbi_web, qlik_web}) empowers business users and data scientists to directly create and mash up analytic datasets according to their needs. With these tools, data analysis becomes an interactive and iterative process whereby a user issues a data analysis action (translated into a query), receives a result, and possibly decides which action to perform next. Eventually, a user may decide to share the final analytic dataset thus obtained in the form of a reusable view. Interactive data analysis sessions facilitate the exploration and creation of analytic datasets, even for users lacking knowledge of SQL, MDX and any programming languages.

However, data experts who directly manipulate analytic datasets created by others expose themselves to possible disappointments, particularly when data aggregation -- the most common operation done by analysts -- is involved. 
Imagine a simple use case with the analytic datasets shown in \Cref{tab:exampletables3}, representing multidimensional \emph{facts} that hold \emph{measures} and refer to one or more hierarchical \emph{dimensions} \cite{jensen_multidimensional_2010}. 
The dimension table \dimt{Region} (\Cref{tab:region}) describes a list of cities. 
These cities are referenced by the fact table \factt{DEM}(ographics) which contains three dimension attributes \attr{City}, \attr{STATE}, \attr{Country} from dimension \dimt{Region} and one attribute \attr{Year} from another dimension table \dimt{TIME} (not displayed). Attributes \attr{POP} and \attr{UNEMP} are measure attributes that respectively represent the population and the unemployment rate in that city. 

\begin{table}[htb]
    \centering
    \caption{Fact and dimension tables for demographics}
    \label{tab:exampletables3}
\normalsize{
\begin{subtable}[b]{.7\linewidth}
\centering
\caption{Fact table \factt{DEM} (Demographics)}
\label{tab:dem}
\begin{tabular}[t]{l|l|l|l|r|r}
  \toprule
  \attr{City} & \attr{State} & \attr{Country} & \attr{Year} & \attr{Pop}& \attr{Unemp} (\%) \\
\midrule
Dublin & California & USA & 2017 & 61 & 3.1 \\ 
Palo Alto & California & USA & 2017 & 67 & 2.1 \\
Dublin & California & USA & 2018 & 63 & 3.0 \\ 
Palo Alto & California & USA & 2018 & 66 & 2.0 \\
San Jose & California & USA & 2018 & 1,028 & 2.2 \\
Dublin & Ohio & USA & 2018 & 44 & 3.7 \\
Washington D.C & - & USA & 2018 & 672 & 6.2 \\
Dublin & - & Ireland & 2018 & 1,348 & 6.71 \\

\bottomrule
\end{tabular}
\end{subtable}
\\\vspace{2mm}
\begin{subtable}[b]{.59\linewidth}
\centering
\caption{Dimension table \dimt{REGION}}
\label{tab:region}
\begin{tabular}[t]{l|l|l|l}
  \toprule
  \attr{City} & \attr{State} & \attr{Country} & \attr{Region} \\
\midrule
Dublin & California & USA &  North America \\ 
Palo Alto & California & USA & North America \\
San Jose & California & USA & North America \\
Dublin & Ohio & USA & North America \\
Washington D.C & - & USA & North America \\
Dublin & - & Ireland & Europe \\
\bottomrule
\end{tabular}
\end{subtable} 
}
\end{table}

\bernd{put this paragraph after the next paragraph: Now suppose.... ?}
Suppose that a business user wants to aggregate the measures in the \factt{DEM} fact table. A first concern is to  express aggregations that produce semantically correct results. For measure \attr{POP}, any common aggregation function can be used, but the dimension attributes along which aggregation can be done must be restricted to \attr{City}, \attr{STATE} and \attr{Country}. That is, aggregation can only be done within every partition of \factt{DEM} by \attr{YEAR}, otherwise the population will be double counted for the cities of 'Palo Alto' and 'Dublin' in California. For measure \attr{UNEMP}, only a limited set of aggregation functions can be applied ($\MIN$, $\MAX$), because the attribute represents a ratio that cannot be summed or averaged along any dimension. Expressing valid aggregation operations therefore requires a clear understanding of the semantics of measure attributes and the dimensions on which they depend. Ideally, the querying system should automatically control which  aggregation operation is valid using metadata properties that express the above restrictions on the \factt{DEM} table. 

\begin{figure}[htb]
    \centering
    \includegraphics[width=0.42\linewidth]{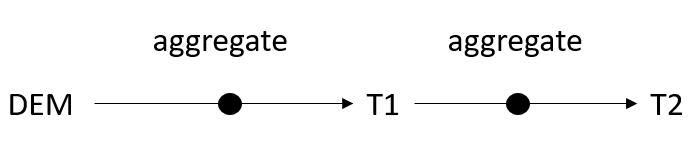}
    \caption{Interactive data analysis session 1 over  \factt{DEM}}
    \label{fig:dem-session}
\end{figure}

Now, suppose that a business user, 
in the interactive data analysis session displayed in \Cref{fig:dem-session}, first wants to count (without duplicates) the number of cities per state, and country. 
This can be achieved using a "roll-up" action which aggregates the \factt{DEM} data along attribute \attr{CITY} and attribute \attr{YEAR}. This action can be translated into a SQL aggregate query on table \factt{DEM} by doing a $\COUNTDISTINCT(\attr{CITY})$ group by \attr{state} and \attr{country} whose result table $\T_1$ is displayed in \Cref{tab:dem-count1} (the count has been renamed into \attr{NB\_CITIES} which is a measure).

\begin{table}[htbp]
\caption{Results of aggregate queries in the session of \Cref{fig:dem-session}}
\label{tab:session1}
\normalsize{
\begin{subtable}[b]{0.37\linewidth}
    \caption{Table $\T_1$}
    \label{tab:dem-count1}
    \centering
    \begin{tabular}{l|l|l}
        \toprule
         \attr{NB\_CITIES} & \attr{STATE}& \attr{COUNTRY} \\
         \midrule
         1 & Ohio & USA  \\
         3 & California & USA \\
         1 & - & USA  \\
         1 & - & Ireland  \\
         \bottomrule
    \end{tabular}
\end{subtable}
\hfill
\begin{subtable}[b]{0.27\linewidth}
    \caption{Table $T_2$}
    \label{tab:dem-sum}
    \centering
    \begin{tabular}{l|l}
        \toprule
         \attr{SUM(NB\_CITIES)} & \attr{COUNTRY} \\
         \midrule
         5 &  USA  \\
         1 & Ireland  \\
         \bottomrule
    \end{tabular}
\end{subtable}
\hfill
\begin{subtable}[b]{0.27\linewidth}
    \caption{Table $T'_2$}
    \label{tab:dem-sum2}
    \centering
    \begin{tabular}{l|l}
        \toprule
         \attr{SUM(NB\_CITIES)} & \attr{COUNTRY} \\
         \midrule
         7 &  USA  \\
         1 & Ireland  \\
         \bottomrule
    \end{tabular}
\end{subtable}
}
\end{table}

Later, suppose that the business user, in the same interactive session, 
aggregates further \attr{NB\_CITIES} by \attr{country} using function $\SUM$, yielding a new table \factt{T2}  displayed in \Cref{tab:dem-sum}. The value of \attr{SUM(NB\_CITIES)} in $\T_2$ is however hard to interpret: for country 'USA', it is neither the count with duplicates nor the count without duplicates of cities by country, if we refer to the original table \factt{DEM}. If the intention of the user was to obtain a count without duplicates of cities, the result of that interactive session is \emph{incorrect}.  On the other hand, if the first aggregate query in the session of \Cref{fig:dem-session} was counting cities \emph{with duplicates}, and the subsequent aggregate query was summing \attr{NB\_CITIES} as before, the result table $\T'_2$ of the interactive session, displayed in table \Cref{tab:dem-sum2}, would be \emph{correct}. It is easy to figure out that this problem is non-trivial for a non-expert user.

This problem is known as a \emph{summarizability} issue: we shall say that attribute \attr{CITY} is \emph{not} summarizable with respect to grouping set $\{\attr{State}, \attr{Country}\}$ and function $\COUNTDISTINCT$ using function $\SUM$. As before, a business user may expect that the querying system controls what aggregation is valid on table $\T_1$ using proper metadata associated with that table. Thus, if the user cannot compute a global count of cities \emph{without duplicates} per country using $\T_1$, she would have to backtrack within the interactive session to a result over which such a global count is expressible (in our example, backtrack to the original table \factt{DEM}). 

This summarizability issue can be generalized to an arbitrary sequence of interactive analytic queries. 
Consider the analytic datasets shown in \Cref{tab:exampletables1}.
The dimension table \dimt{SALESORG} describes a list of stores that are referenced by the fact table \factt{STORE\_SALES}  containing four dimension attributes \attr{STORE\_Id}, \attr{City}, \attr{STATE} and \attr{Country} from dimension \dimt{SalesOrg} and one attribute \attr{Year} from dimension  \dimt{TIME}. Attribute \attr{AMOUNT} is a measure attribute 
and \attr{Unit} is a detail attribute of that measure.

\begin{figure}[htb]
    \centering
    \includegraphics[width=0.9\linewidth]{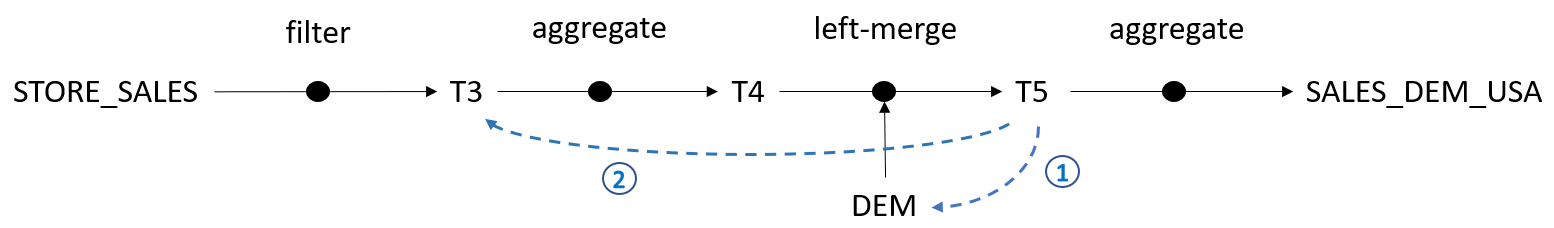}
    \caption{Interactive data analysis session 2 yielding \factt{SALES\_DEM\_USA}}
    \label{fig:sales-dem-usa}
\end{figure}

A data analyst might want to build a new analytic dataset, named \factt{SALES\_DEM\_USA}, using the interactive data analysis session shown in \Cref{fig:sales-dem-usa} (the dashed lines will be explained later). First, 
\factt{STORE\_SALES} is filtered on \attr{country} = 'USA' and \attr{year} = '2018', yielding a table named \factt{T3}. 
Then, an aggregate \attr{SUM(AMOUNT)} is computed for each partition of \attr{City}, \attr{State}, \attr{Country}, \attr{Year},
yielding a table named \factt{T4}  displayed in \Cref{tab:state-query-result1}. At this point, we can control that each tuple in \factt{T4} is \emph{correct} because it would also be a tuple in the result of the \emph{same} aggregate query computed over \factt{STORE\_SALES} (the original data).

\begin{table}[htb]
    \centering
    \caption{Fact and dimension tables for store sales}
    \label{tab:exampletables1}
    \normalsize{
\begin{subtable}[b]{0.6\linewidth}
\centering
\caption{Dimension table \dimt{SALESORG}}
\label{tab:state-salesorg1}
    \begin{tabular}[t]{l|l|l|l}
      \toprule
      \attr{Store\_Id} & \attr{City} & \attr{STATE}& \attr{Country} \\
    \midrule
    Ca\_01 & Dublin & California & USA \\
    Sa\_01 & San Jose & California & USA \\
    Oh\_01 & Dublin & Ohio & USA\\
    Wa\_01 & Washington DC & - & USA\\
    Du\_01 & Dublin & - & Ireland\\
      \bottomrule
    \end{tabular}
\end{subtable}
\\\vspace{2mm}
\begin{subtable}[b]{.85\linewidth}
    \centering
    \caption{Fact table \factt{STORE\_SALES}}
    \label{tab:state-summarizability-sales1}
    \begin{tabular}[t]{l|l|l|l|l|l|r}
      \toprule
    \attr{Store\_Id} & \attr{City} & \attr{State} & \attr{Country} & \attr{Year} & \attr{Amount} & \attr{Unit} \\
    \midrule
    Ca\_01 & Dublin & California & USA & 2018 & 5.3 & mega dollar \\
    Ca\_02 & Dublin & California & USA & 2018 & 1.4 & mega dollar \\
    Ca\_01 & Dublin & California & USA & 2017 & 3.5 & mega dollar \\
    Sa\_01 & San Jose & California & USA & 2018 & 22.8 & mega dollar \\
    Oh\_o1 & Dublin & Ohio & USA & 2018 & 1.2 & mega dollar \\
    Wa\_o1 & Washington DC & - & USA & 2018 & 16.1 & mega dollar \\
    Wa\_o2 & Washington DC & - & USA & 2018 & 27.6 & mega dollar \\
    Du\_01 & Dublin & - & Ireland & 2018 & 7.8 & mega euro \\
      \bottomrule
    \end{tabular}
\end{subtable}
}
\end{table}

Next, the schema of \factt{T4} is augmented with the measure attribute \attr{POP} of table \factt{DEM}, yielding a new table named \factt{T5} (\Cref{tab:T3}). This latter action, called a \emph{left-merge}, can be translated into a natural left outer join SQL query between \factt{T4} and \factt{DEM} on attributes \attr{City}, \attr{State}, \attr{Country} and \attr{Year}. 
Thus, in table \factt{T5}, attributes \attr{City}, \attr{State} and \attr{Country} represent attributes of both dimensions \dimt{REGION} and \dimt{SALESORG}. 
However, measure attribute \attr{Pop} depends on dimension \dimt{REGION} while attribute \attr{Sum(amount)} depends on dimension \dimt{SALESORG}.

\begin{table}[htb]
    \centering
    \caption{Result \factt{T4} in session 2 of \Cref{fig:sales-dem-usa}}
    \normalsize{
    \label{tab:state-query-result1}
    \begin{tabular}[t]{l|l|l|l|r}
      \toprule
    \attr{City} & \attr{State} & \attr{Country} & \attr{Year} & \attr{SUM(Amount)}  \\
    \midrule
    Dublin & California & USA & 2018 & 6.7  \\
    San Jose & California & USA & 2018 & 22.8  \\
    Dublin & Ohio & USA & 2018 & 1.2  \\
    Washington DC & - & USA & 2018 & 43.7  \\
   \bottomrule
    \end{tabular}
}
\end{table}

In the last step of the interactive data analysis session 2, the measure attributes \attr{SUM(AMOUNT)} and \attr{POP} of \factt{T5} are summed by \attr{State}, \attr{Country} and \attr{Year}, yielding the final result \factt{SALES\_DEM\_USA} displayed in \Cref{tab:state-sales-dem}. However, the value of \attr{SUM(POP)} in \factt{SALES\_DEM\_USA} is \emph{misleading} because it does not correspond to the population of each state as it would be obtained from the \factt{DEM} table. Indeed, the population of cities without any store, such as the city of 'Palo Alto', has not been counted. Thus, the aggregation along \attr{City} of \attr{SUM(POP)} should not be allowed on \factt{T5} (or at least, a warning must be raised that it only accounts for the population of cities in dimension \dimt{SALESORG}). 
To obtain the total population of each state, suppose that the user backtracks to the previous step of the session and expresses the summation of \attr{POP} along \attr{City} on table \factt{DEM}, yielding a new table \factt{DEM'} (in \Cref{tab:demprime}), before performing the left-merge operation. This backtracking is depicted by the dashed line labelled "1" in \Cref{fig:sales-dem-usa}.

\begin{table}[htb]
    \centering
    \caption{Results \factt{T5} and \factt{SALES\_DEM\_USA} in session of \Cref{fig:sales-dem-usa}}
    \label{tab:exampletables4}
    \normalsize{
\begin{subtable}[b]{.9\linewidth}
\centering
\caption{Result \factt{T5} in session of \Cref{fig:sales-dem-usa}}
\label{tab:T3}
\begin{tabular}[t]{l|l|l|l|l|r}
  \toprule
  \attr{City} & \attr{State} & \attr{Country} & \attr{Year} & \attr{SUM(AMOUNT)} & \attr{Pop} \\
\midrule
Dublin & California & USA & 2018 & 6.7 & 61 \\ 
San Jose & California & USA & 2018 & 22.8 & 1,028  \\
Dublin & Ohio & USA & 2018 & 1.2 & 44  \\
Washington & - & USA & 2018 & 43.7 & 672 \\ 
\bottomrule
\end{tabular}
\end{subtable}
\\\vspace{2mm}
\begin{subtable}[b]{.81\linewidth}
    \centering
    \caption{Fact table \factt{SALES\_DEM\_USA} with misleading \attr{SUM(POP)}}
    \label{tab:state-sales-dem}
    \begin{tabular}[t]{l|l|l|l|r}
      \toprule
    \attr{State} & \attr{Country} & \attr{Year} & \attr{SUM(Amount)} & \attr{SUM(POP)} \\
    \midrule
California & USA & 2018 & 29.5 & 1,089 \\ 
Ohio & USA & 2018 & 1.2 & 44  \\
 - & USA & 2018 & 43.7 & 672 \\ 
   \bottomrule
    \end{tabular}
\end{subtable}
}
\end{table}

However, after performing the left-merge of \factt{T4} with \factt{DEM'} (result is displayed in \Cref{tab:T3-bis}), the summation of \attr{SUM(POP)} should again be disallowed. Indeed, it would be \emph{incorrect} with respect to the same summation computed over table \factt{DEM'}, since population of California would be double-counted. The proper explanation is that tuples from \factt{DEM'} match multiple tuples of \factt{T4} because they don't have the same dimension granularity.

\begin{table}[htb]
\centering
\caption{Results after first backtracking in session of \Cref{fig:sales-dem-usa} }
\normalsize{
\begin{subtable}[b]{.4\linewidth}
\centering
\caption{Fact table \factt{DEM'}}
\label{tab:demprime}
\begin{tabular}[t]{l|l|l|r}
  \toprule
   \attr{State} & \attr{Country} & \attr{Year} & \attr{\SUM(Pop)} \\
\midrule
  California & USA & 2017 & 61   \\ 
  California & USA & 2017 & 128   \\
  California & USA & 2018 & 1,157   \\
  Ohio & USA & 2018 & 44   \\
  - & USA & 2018 & 672   \\
  - & Ireland & 2018 & 1,348   \\
\bottomrule
\end{tabular}
\end{subtable}
\vspace{3mm}

\begin{subtable}[b]{.8\linewidth}
\centering
\caption{Result of the left-merge of \factt{T4} with \factt{DEM'}}
\label{tab:T3-bis}
\begin{tabular}[t]{l|l|l|l|l|r}
  \toprule
  \attr{City} & \attr{State} & \attr{Country} & \attr{Year} & \attr{SUM(AMOUNT)} & \attr{SUM(Pop)} \\
\midrule
Dublin & California & USA & 2018 & 6.7 & 1,157 \\ 
San Jose & California & USA & 2018 & 22.8 & 1,157 \\
Dublin & Ohio & USA & 2018 & 1.2 & 44  \\
Washington & - & USA & 2018 & 43.7 & 672 \\ 
\bottomrule
\end{tabular}
\end{subtable}
}
\end{table}

\begin{figure}[htb]
    \centering
    \includegraphics[width=0.8\linewidth]{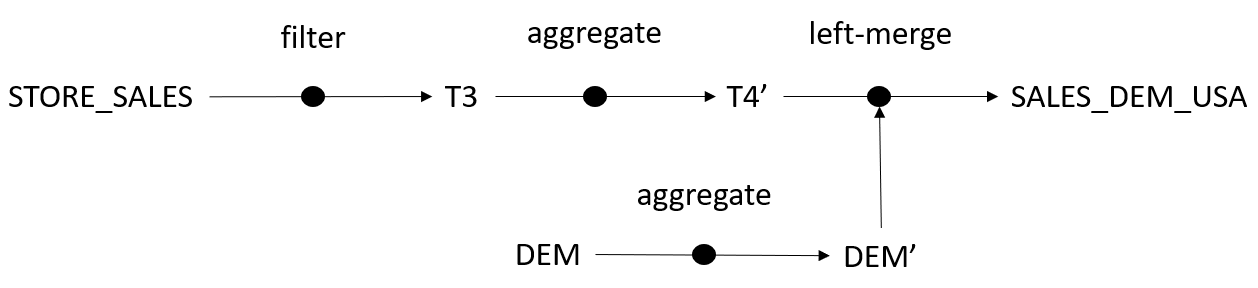}
    \caption{Final flow of interactive queries yielding a correct instance of \factt{SALES\_DEM\_USA}}
    \label{fig:sales-dem-usa_bis}
\end{figure}

Hence, the user has to backtrack to \factt{T3} (backtracking is depicted by the dashed line labelled "2" in \Cref{fig:sales-dem-usa}) and aggregate \attr{SUM(AMOUNT)} by \attr{State}, \attr{Country} and \attr{Year}, yielding a new table \factt{T4'}. After merging \factt{T4'} with \factt{DEM'}, the final table \factt{SALES\_DEM\_USA} is obtained, as displayed in \Cref{tab:state-sales-dem-bis}. 
The actual flow of interactive queries that produced the final result is displayed in \Cref{fig:sales-dem-usa_bis}.

\begin{table}[htbp]
\centering
    \caption{Result of the left-merge  of \factt{T4'} with \factt{DEM'}  with correct \uppercase{SUM(POP)}}
    \label{tab:state-sales-dem-bis}
    \normalsize{
    \begin{tabular}[t]{l|l|l|l|r}
      \toprule
    \attr{State} & \attr{Country} & \attr{Year} & \attr{SUM(Amount)} & \attr{SUM(POP)} \\
    \midrule
California & USA & 2018 & 29.5 & 1,157 \\ 
Ohio & USA & 2018 & 1.2 & 44  \\
 - & USA & 2018 & 43.7 & 672 \\ 
   \bottomrule
    \end{tabular}
}
\end{table}


The previous examples hopefully showed that it is easy for an end user such as an analyst to perform erroneous or misleading aggregation operations during an interactive data analysis session. This motivated our design of a method that automatically controls the validity of aggregation operations and provides explanations that are easy to understand for an end user.  

\subsection{Limitations of previous related work}

As mentioned before, the occurrence of an incorrect sequence of two aggregations is known as a \emph{summarizability} problem. In the original definition of the problem \cite{lenz_summarizability_1997}, an initial fact table represents "micro-data", at the finest granularity level, and a \emph{summarization query} is expressed over an attribute $\A$ of this fact table, yielding another fact table representing "macro-data". A summarization query over a fact table performs an aggregation operation $F(A)$ using a function $\AGGF$ for each partition of the table defined by a grouping set of attributes $\X$. In essence, the problem of summarizability is to determine whether, for some summarization query over attribute $\AGGF(A)$ of the macro-data using a function $\AGGG$ (possibly identical to $\AGGF$), there exists a summarization query over attribute $\A$ of the micro-data using $\AGGF$ that returns exactly the same result. If this is the case, the summarization query over $\AGGF(A)$ of the macro-data is said to be \emph{correct}. For instance, using the previous example of \Cref{fig:dem-session}, if table \factt{DEM} represents the micro-data, and table \factt{T1} represents the macro-data obtained after summarizing attribute \attr{CITY} using function $\COUNTDISTINCT$, then the query that summarizes \attr{NB\_CITIES} using function $\SUM$ is \emph{incorrect}.  

In the most general formulation of the problem, attribute $\A$ in the micro-data is defined to be summarizable with respect to a grouping set $\X$ and a function $\AGGF$ using function $\AGGG$ if for any subset of attributes $Z$ of $\X$, any aggregation $G(F(A))$ with grouping set $Z$ over the macro-data is correct (with the above meaning). Ideally, we want to determine the largest subset of attributes of $\X$ for which the above summarizability condition holds. 

To address the summarizability problem, a first group of \textit{model-based} solutions proposes to model dimension and fact tables in a restricted and controlled way so that any aggregation query over a previously aggregated fact table is always correct. 
Some solutions even propose to modify the hierarchical dimension data to enforce the restrictions that assure the summarizability of aggregate queries. \eric{changed: }See \cite{mazon_survey_2009} for a survey of these solutions and more recently \cite{CANIUPAN201217}.

A second group of \textit{constraint-based} solutions defines summarizability constraints over the schemas and the data of dimension and fact tables, which can be evaluated to determine whether an attribute of a fact table is summarizable with respect to a grouping set using an aggregation function. We focus our work on this second group of solutions which, instead of imposing constraints on the analytic data model, control the summarizability of aggregate queries to avoid incorrect results. These solutions are better suited to an environment where analytic data is created by multiple independent parties using different data modeling techniques. 

However, the detailed analysis of the best existing constraint-based solutions to the summarizability problem \cite{lenz_summarizability_1997, pedersen_foundation_2001, lehner_normal_1998, lechtenborger_multidimensional_2003, hurtado_capturing_2005}, reveals the following limitations.  Firstly, within the hierarchy of a dimension, any non-null value of an attribute must map to a single parent attribute value. This discards the use of dimension tables like \dimt{SALESORG} (in \Cref{tab:state-salesorg1}), wherein a city can have multiple states. Secondly, measure attributes in a fact table must depend on all the identifiers of the dimensions over which the fact table is defined. This discards the use of fact tables such as the result of the left-merge of \factt{T4} with \factt{DEM'} in \Cref{tab:T3-bis}, where measure attribute \attr{SUM(POP)} depends only on \attr{State}, \attr{Country} and \attr{Year}. Thirdly, it is assumed that aggregate (summarization) operations do not handle null values in their grouping set, which creates too restrictive conditions for summarizable queries in the case of SQL aggregate operations. Finally, summarizability conditions depend on the size of the dimension data, because they require either testing disjointness conditions over fact table partitions defined by some dimension attributes \cite{lenz_summarizability_1997} or reasoning on \emph{dimension constraints} whose number depends on the paths that exist in the dimension hierarchies \cite{hurtado_capturing_2005}.

Another important limitation is that the existing methods do not consider the case  when an aggregate query occurs after another type of query such as a filter or a left-merge query, as in the data analysis session of \Cref{fig:sales-dem-usa}. In real life scenarios though,  "mash-up" queries are popular because analytic data is often siloed in the context of a specific business activity (e.g., product marketing, medical care) or a particular application domain (e.g., monitoring system logs). For example, analytic data on hospitalized patients contains measures on those patients that are treated in hospitals. Other analytic data may contain measures about ambulatory patients who are treated by medical doctors in the city, or measures about patient demographics. Typical epidemiological studies therefore require mashing up this data using filter and merge operations.

Other research efforts related to the problem studied in this paper exist. Some works address the correctness issues raised by the semantics of various aggregation functions and their applicability depending on the domain of values on which they are applied \cite{lenz_olap_2001, lenz_formal_2009} or the issues caused by the SQL implementation of relational operations when they are applied over an empty set of values or over a set of values containing \emph{null} values 
\cite{guagliardo_formal_2017,console2020fragments}. In our work, we assume a standard implementation of SQL aggregate queries for \emph{null} values. 

Other works propose methods which automatically control or enforce the consistency of arithmetic and aggregation query results with respect to the scales, units and currencies associated with measure attributes in fact tables \cite{horner_taxonomy_2005, liu_rutian_semantic_2020, THANISCH2019116}. We are not addressing this problem which is complementary to the correctness issues targeted by this paper. 

Finally, recommendation-based approaches suggest queries for the interactive exploration of databases. The \emph{collaborative filtering} approach uses previously collected query logs of a dataset (SQL queries in \cite{eirinaki2013querie, khoussainova2010snipsuggest}, and OLAP queries in \cite{aligon2015collaborative, marcel2011survey}) to recommend queries on the same initial dataset. The \emph{data-driven approach} \cite{chatzopoulou2009query, BolchiniQT12, drosou2013ymaldb, JoglekarGP19, sarawagi1998discovery, singh2016dbexplorer}, recommends a single type of exploration actions whose result are expected to optimize a measure of ``interestingness'' with respect to the current analysis context of a user on a given dataset. For instance, \cite{JoglekarGP19} suggests different ''drill-down'' operations on a given table, each producing a different set of tuples. However, these works do not control the correctness of the recommended ''drill-down'' exploration actions with respect to the summarizability property we introduced before.

\subsection{Research contributions}
In this article we present a comprehensive set of conditions to control the \textit{correctness of aggregation queries} within a data analysis session consisting of a large variety of interactive queries, which includes the most common operations that are supported by self-service data preparation and BI tools, such as filter, project, inner and outer joins, aggregate, union, difference, and pivot. These operations also subsume the traditional operations used in interactive exploration sessions of OLAP cubes, such as roll-up, drill-down, dice, and slice and our correctness criteria for aggregate queries include as a special case the summarizability property addressed by previous work. 

The \emph{analytic data model} we introduce in \Cref{sec:model}  is more expressive than the other data models considered by the previous work on summarizability, because it accepts arbitrary dimension hierarchies and fact tables. In our data model, dimension tables are defined as views over non-analytic tables (that is, regular relational tables), and fact tables are initially defined as views over dimension tables and non-analytic tables. Then, new fact or dimension tables are defined as the result of interactive queries over previously defined dimension and fact tables. 

At the core of our approach is the definition of two types of metadata associated with analytic tables which help to check the correctness of analytic queries. Firstly, \emph{attribute graphs} are used to describe literal functional dependencies between the attributes of hierarchical dimensions with possible $null$  values (as in the example of dimension \dimt{SALESORG}). We showed in a previous paper \cite{liu2020discovering} how to efficiently compute attribute graphs through the analysis of dimension data samples. 
Secondly, \emph{aggregable properties} describe, for any attribute of an analytic table, which aggregation functions can be used, and along which set of dimension attributes these aggregation functions can be applied. \emph{Default rules} assist the designer of a table to define aggregable properties when the table is initially created as a view from source data (i.e.,  from non-analytic tables) and, using these properties, it is then possible to automatically control which aggregations are possible on an analytic table. 

The central technical results of this article are \emph{propagation rules}, which automatically compute the aggregable properties for a  table resulting from an interactive analytic query and thereby allow us to control the \emph{correctness} of aggregate queries at any stage of an interactive data analysis session. Our correctness criteria for aggregate queries  include the semantic properties of measure attributes (like in the first examples of aggregate queries over table \factt{DEM} that we presented before). 
Furthermore, these criteria not only subsume the sufficient conditions defined in previous work to assure that aggregate queries are expressed over summarizable attributes, but their propagation  makes it also possible to characterize the results of sessions composing two or more aggregate queries as being correct, with respect to summarizability, when previous work would view them as being incorrect.
Conversely, any aggregate query that previous works characterize as correct is also detected as correct using our aggregable properties. 
Finally, in the case of a sequence composed of any interactive query followed by an aggregate query, we introduce the novel notion of \emph{G (generalized) summarizability} to characterize correct aggregate queries. 

In summary, we make the following main research contributions:

\begin{enumerate}
    \item  In \Cref{sec:aggregable-properties-def-prop}, we extend the notion of aggregable properties introduced in \cite{liu2020discovering}, as a general means to express, for any attribute of an analytic table, which aggregation functions are correctly applicable along which sets of dimension attributes. We use aggregable properties to express the semantic properties of measures previously defined in \cite{kimball_data_2013, horner_analysis_2004, niemi_detecting_2014, stevens_theory_1946, pedersen_foundation_2001, lenz_summarizability_1997} and  provide default rules to minimize the effort of end users for defining aggregable properties on analytic tables built from source data (i.e., non-analytic data). We then provide a first set of propagation rules to automatically compute aggregable properties in the results of interactive analytic queries.    
    
    \item In \Cref{sec:distributive_functions},     we formally define summarizability conditions for attributes and in \Cref{sec:prop_summarize}, we refine our propagation rules for the case of aggregate queries to compute aggregable properties of attributes such that subsequent aggregate queries over these attributes can only be expressed if the attributes are summarizable.  In \Cref{sec:related-work}, we show that our aggregable properties subsume the summarizability conditions defined in previous work \cite{lenz_summarizability_1997, pedersen_foundation_2001, lehner_normal_1998, lechtenborger_multidimensional_2003, hurtado_capturing_2005}.
    
    \item In \Cref{sec:generalized_summarizability}, we introduce the new notion of G-summarizability that extends the summarizability property of attributes to the case of an aggregate query expressed over the result of an arbitrary analytic query. We then refine our propagation rules in \Cref{sec:control-g-sum} to compute aggregable properties such that aggregate queries over some attributes can be expressed only if these attributes are G-summarizable. 
    
    \item Finally, in \Cref{sec:related-work} we focus on previous works that propose conditions on the schema of a fact table, or on the parameters of an aggregate query expressed over that fact table, to determine if the aggregate query returns a correct result with respect to some summarizability definition. In our  analysis, we establish that our data model is more general than the data models considered by previous work.  In the case of a sequence of two aggregate queries, $Q_1$ followed by $Q_2$, our sufficient conditions to determine if $Q_2$ is correct, subsume the conditions proposed by previous work. To the best of our knowledge, no previous work addressed the case of a sequence made of an arbitrary analytic query followed by an aggregate query, which is addressed by our notion of G-summarizability.
\end{enumerate}

\section{Multi-dimensional Data Model and Analytic Queries}
\label{sec:model}
In this section, we first present our multidimensional data model, composed of dimension and fact tables, and some logical and structural constraints on dimensions expressed using an \emph{attribute graph}. We then present the types of analytic queries that can be expressed on our data model. 
We use conventional relational database notations \cite{d_ullman_database_nodate}. Each table $\XXT$ is a finite multiset of tuples over a set of domains of values $\S = \{\A_1, . . . , \A_n\}$, called \emph{attributes}, where each domain may contain a \emph{null} marker. We call $\S$ the \emph{schema} of $\XXT$.

\subsection{Dimension and fact tables}
\label{sec:dimension-fact-tables}
We consider datasets in which data is separated into \emph{non-analytic tables} and \emph{analytic tables}. Non-analytic tables correspond to  relational tables storing the  source data.  \emph{Analytic tables}, or \emph{analytic views}, are defined by queries  over non-analytic and analytic tables.

\begin{myexample}
\Cref{fig:structure-analaytic} details the definitions of two analytic tables \factt{STORE\_SALES} and \dimt{PROD}. 
The analytic table \dimt{PROD} represents a dimension that is defined by a join-project query over three non-analytic tables (represented by square rectangles). The analytic table \factt{STORE\_SALES} represents a fact table that is  defined by a join-project query over a non-analytic table \dset{ct\_SALES} and three analytic tables \dimt{TIME}, \dimt{PROD} and \dimt{STORE} representing dimensions. 
\begin{figure}[htb]
\centering
\begin{subfigure}[b]{0.45\linewidth}
\centering
\includegraphics[width=0.75\linewidth]{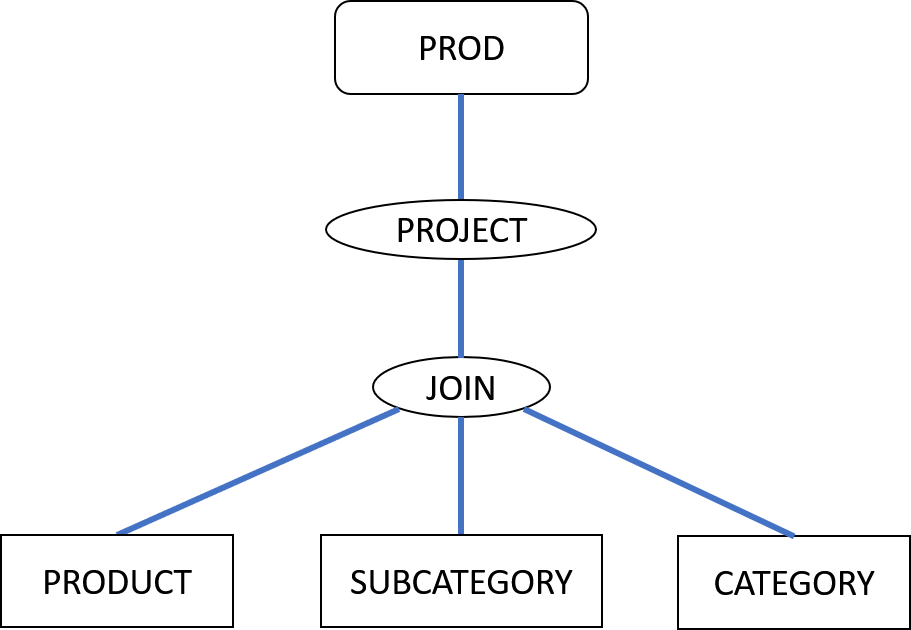}
\caption{Definition of dimension table \dimt{PROD}}
\label{fig:structure-sales}
\end{subfigure}
\hfill
\begin{subfigure}[b]{0.45\linewidth}
\centering
\includegraphics[width=0.9\linewidth]{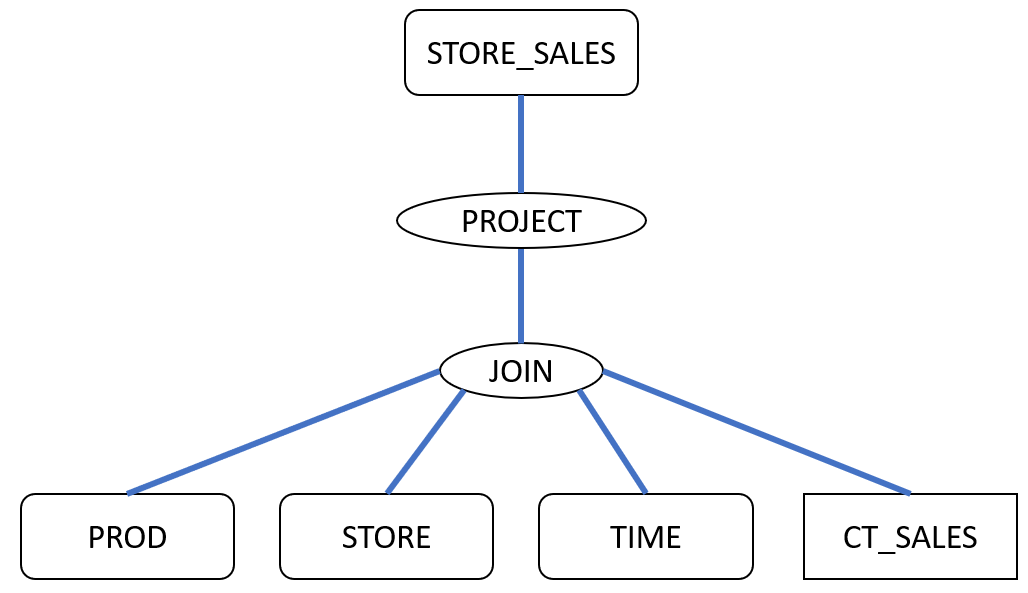}
\caption{Definition of fact table \factt{STORE\_SALES}}
\label{fig:structure-prod}
\end{subfigure}
\caption{Two analytic tables (views)}
\label{fig:structure-analaytic}
\end{figure}
\end{myexample}

Attributes in analytic tables are categorized into two types: \emph{dimension attributes} and \emph{measures}. Dimension attributes describe entities like stores, customers and dates whereas measure attributes are used to define facts about these entities. Following this distinction of attributes, analytic tables are categorized
 into two types: \emph{dimension tables} and \emph{fact tables}. An analytic table is a \emph{dimension table} if it only contains \emph{dimension attributes} 
and a \emph{fact table} if it contains at least one or more \emph{dimension attributes} from one ore more dimensions and one \emph{measure attribute}.

\begin{myexample}
\label{ex:tables_all}
A complete example of analytic tables, which will be used throughout this paper, is shown in \Cref{tab:exampletables}. Dimension table names are in italic font to distinguish them from fact tables. 
The first three tables are dimension tables identifying and describing products (dimension \dimt{PROD}), stores (dimension \dimt{SALESORG}) and dates (dimension \dimt{TIME}). 
The schema of fact tables \factt{STORE\_SALES} was already introduced in \Cref{sec:introduction}. 
Fact table \factt{PRODUCT\_LIST}, shown in \Cref{tab:prod-list2-repeat2}, describes the sold quantity (\attr{QTY}) of products (attributes \attr{PROD\_SKU}, \attr{BRAND} and \attr{COUNTRY} from dimension \dimt{PROD}), by year (attribute \attr{YEAR} from dimension \dimt{TIME}).
Attribute value "-" in these tables represents a null marker.

\begin{table}[htb]
    \centering
    \caption{Fact and dimension tables}
    
    \label{tab:exampletables}
    \normalsize{
\begin{subtable}[b]{0.7\linewidth}
    \centering
         \label{tab:dim-prod}
    \caption{Dimension table \dimt{PROD}}
   \begin{tabular}{l|l|l|l|l}
        \toprule
         \attr{PROD\_SKU} & \attr{BRAND}& \attr{COUNTRY} & \attr{SUBCATEGORY} & \attr{CATEGORY} \\
         \midrule
         coco-can-33cl & Coco Cola & USA & Soft Drinks & Drinks \\
         coco-can-25cl & Coco Cola & USA & Soft Drinks & Drinks \\
         cz-tshirt-s & Zora & Spain & Woman Tops & Clothes \\
         cz-tshirt-s & Coco Cola & USA & Woman Tops & Clothes \\
         \bottomrule
    \end{tabular}
    \end{subtable}

\vspace{2mm}

\begin{subtable}[b]{0.5\linewidth}
\centering
\caption{Dimension table \dimt{SALESORG}}
\label{tab:state-salesorg2}
    \begin{tabular}[t]{l|l|l|l}
      \toprule
      \attr{Store\_Id} & \attr{City} & \attr{STATE}& \attr{Country} \\
    \midrule
    Oh\_01 & Dublin & Ohio & USA \\
    Ca\_01 & Dublin & California & USA \\
    Ca\_02 & Palo Alto & California & USA\\
    Pa\_01 & Paris & - & France\\
    Ly\_01 & Lyon & - & France\\
    Ir\_01 & Dublin & - & Ireland \\
      \bottomrule
    \end{tabular}
\end{subtable}
\hfill
\begin{subtable}[b]{0.4\linewidth}
\centering
\caption{Dimension table \dimt{TIME}}
\label{tab:motivation-time}
    \begin{tabular}[t]{l|l|l|l}
      \toprule
      \attr{DATE} & \attr{WEEK} & \attr{MONTH} & \attr{YEAR} \\
    \midrule
    1/1/2018 & 1 & 1 & 2018 \\
    2/1/2018 & 1 & 1 & 2018 \\
    3/1/2018 & 1 & 1 & 2018 \\
    \ldots&\ldots&\ldots&\ldots\\
      \bottomrule
    \end{tabular}
\end{subtable}

\vspace{2mm}

\begin{subtable}[b]{0.7\linewidth}
    \centering
     \caption{Fact table \factt{STORE\_SALES}}
    \label{tab:state-summarizability-sales2}
    \label{tab:store_sales2}
    \begin{tabular}[t]{l|l|l|l|l|r}
      \toprule
    \attr{Store\_Id} & \attr{City} & \attr{State} & \attr{Country} & \attr{Year} & \attr{Amount} \\
    \midrule
    Oh\_01 & Dublin & Ohio & USA & 2017 & 3.2 \\
    Ca\_01 & Dublin & California & USA & 2017 & 5.3 \\
    Oh\_01 & Dublin & Ohio & USA & 2018 & 8.2 \\
    Ca\_01 & Dublin & California & USA & 2018 & 6.3 \\
    Pa\_01 & Paris & - & France & 2017 & 45.1 \\
      \bottomrule
    \end{tabular}
    \end{subtable}

\vspace{2mm}

\begin{subtable}[b]{0.7\linewidth}
    \centering
     \caption{Fact table \factt{PRODUCT\_LIST}}
    \label{tab:prod-list2-repeat2}

    \begin{tabular}{l|l|l|l|r}
        \toprule
         \attr{PROD\_SKU} & \attr{BRAND}& \attr{COUNTRY} & \attr{YEAR} & \attr{QTY} \\
         \midrule
         cz-tshirt-s & Coco Cola & USA & 2017 & 5 000 \\
         cz-tshirt-s & Coco Cola & USA & 2018 & 7 000 \\         
         cz-tshirt-s & Zora & Spain & 2017 & 5 000 \\
         cz-tshirt-s & Zora & Spain & 2018 & 7 000 \\
         coco-can-33cl & Coco Cola & USA & 2017 & 10 000 \\
         \bottomrule
    \end{tabular}
 \end{subtable}   
}
\end{table}

\end{myexample}

We consider a classical  multi-dimensional data model which organizes a set of dimension attributes $\X$ into an \emph{attribute hierarchy} noted $(X, \typeprec)$. Unlike several other models, which we shall review in \Cref{sec:state-of-art-summarizability}, we make no special assumption on the attribute hierarchy: there can be one or more bottom or top level attributes, and an attribute can have multiple parents. 

A \emph{hierarchy instance} 
of an attribute hierarchy  $\AH=(X,\typeprec)$ is a set of values $\valdom$ 
and a partial order  $\instprec$, where $\valdom$ contains for each attribute $\X_i \in X$ a non empty subset of values $\valdom_i \subseteq \valdom$ such that each order relation $v_i\instprec v_j$ preserves the ancestor/descendant relation $\typeprec^{*}$ between the corresponding attributes $\X_i$ and $\X_j$, \ie, $v_i\in \valdom_i, v_j\in \valdom_j\Rightarrow \X_i \typeprec^{*} \X_j$. We also assume that $(\valdom,\instprec)$ is \emph{transitively reduced}, \ie, there is no pair of values that is connected by an order relation ($\instprec$) and a sequence of two or more order relations.

\begin{myexample}
The left part of \Cref{fig:prod-hier-in} illustrates an \emph{attribute hierarchy} for dimension \dimt{PROD}: \attr{PROD\_SKU}$\typeprec$ \attr{BRAND} $\typeprec$ \attr{COUNTRY}, and \attr{PROD\_SKU} $\typeprec$ \attr{SUBCATEGORY} $\typeprec$ \attr{CATEGORY}, in which \attr{PROD\_SKU} is a bottom level attribute and \attr{CATEGORY} and \attr{COUNTRY} are two top level attributes. 
An instance of that attribute hierarchy is displayed on the right where attribute values are horizontally aligned with the name of each attribute. 

\begin{figure}[htb]
    \centering
    \includegraphics[width=0.75\linewidth]{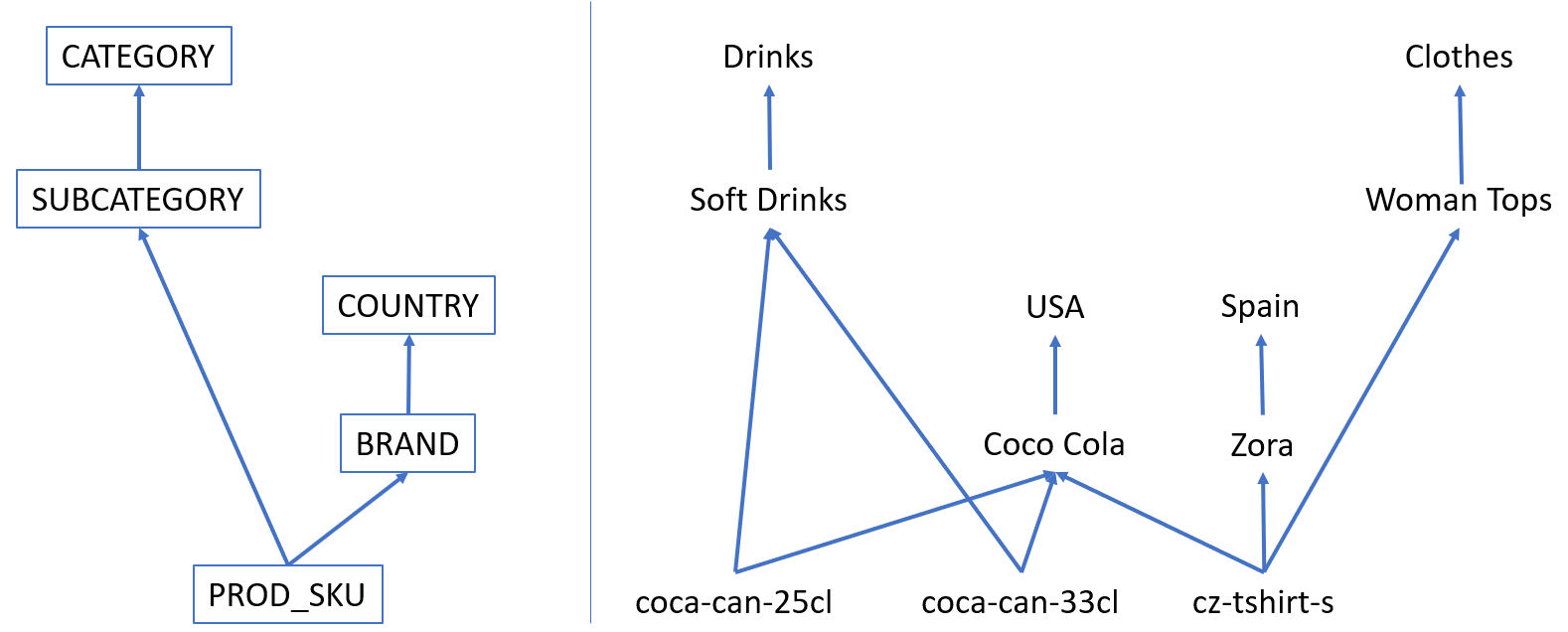}
    \caption{Attribute hierarchy and hierarchy instance defined by dimension table \dimt{PROD}}
    \label{fig:prod-hier-in}
\end{figure}

\end{myexample}

We can now formally define dimension and fact tables.

\begin{definition}[Dimension table]
A dimension table $\D$ over some attribute hierarchy $\AH=(\S,\typeprec)$ is a table $\D(\S)$ where each tuple $t$ of $\D$ corresponds to a complete path in the hierarchy instance of $\AH$. 
Attributes of $\S$ are henceforth called \emph{dimension attributes}.
\end{definition}

In practice \cite{kimball_data_2013}, dimension tables also include \emph{detail attributes} that functionally depend on one or more dimension attributes, and these dependencies are part of the metadata of the dimension table. Examples of detail attributes for the dimension table \dimt{SALESORG} could be: \attr{ZIPCODE}, \attr{COUNTRY\_CODE}, \attr{STORE\_NAME}, \attr{STORE\_SQUARE\_METERS}, etc. We shall not consider such detail attributes because they do not impact the results presented in this paper. 

\begin{definition}[Fact table]
A \emph{fact table} over a set of dimensions  $\D_1, \cdots, \D_n$  is a table $\XXT(\S)$ without any duplicate where schema $\S$ contains a non-empty subset $\X_i$ of  dimension attributes from dimension $\D_i$, and a non-empty set of attributes $\Z$ representing one or more \emph{measures}. Each tuple of values $t.X_i$ in $\XXT$ has a corresponding tuple of values in $\D_i$. 
\end{definition}

In practice \cite{kimball_data_2013}, each measure in a fact table is usually represented by one attribute having the role of \emph{Value} and a possibly empty group of attributes having the role of \emph{Detail}. The \emph{Value} attribute of a measure carries the actual value while the \emph{Detail} attributes provide optional auxiliary information on the measure such as a unit (as in example of fact table \factt{STORE\_SALES} in the introduction) or a currency. In this paper, we shall not discuss how to control the quality of aggregation queries with respect to different units and currencies, and refer the interested reader on this topic to \cite{liu_rutian_semantic_2020}. So we shall later only consider measure attributes carrying actual values. 

\subsection{Literal functional dependencies and attribute graphs}
\label{sec:attribute-graphs}

Null markers in dimension attributes represent \emph{non applicable values}. This semantics is different from other interpretations where null values represent missing or unknown values and are considered as placeholders for non-null values. We consider null markers as regular values and apply the same literal equality semantics as in SQL unique constraints (see e.g.,\cite{d_ullman_database_nodate}): two attribute values $t_1.\A$ and $t_2.\A$ are \emph{literally equal}, denoted by $t_1.\A \equiv t_2.\A$, iff $t_1.\A = t_2.\A$ or both values are null markers. Observe that $t_1.\A = t_2.\A$ implies $t_1.\A \equiv t_2.\A$ but  the opposite is not true. Literal equality naturally extends to sets of attributes and leads to the notion of \emph{Literal Functional Dependencies (LFD)} \cite{badia_functional_2014}.  Let $\X$ and $\Y$ be two sets of attributes in a schema $\S$, an LFD $\X\mapsto \Y$ holds for some table $\XXT$ over $\S$ iff for any two tuples $t_1,t_2$ of $\XXT$, when $t_1.\X \equiv t_2.\X $ then $t_1.\Y \equiv t_2.\Y$. Note that if $\X$ does not contain any nullable attribute, 
the LFD $\X \mapsto \Y$ is equivalent to the \emph{Functional Dependency with Nulls (NFD)} $\X \to \Y$  \cite{atzeni_functional_1984}. A set of LFDs on a schema $\S$ expresses semantic properties constraining the possible \enquote{valid} tables over $\S$.

LFDs provide a formal system to define a set of logical and structural constraints over dimension tables. However, their practical use for characterizing a set of valid dimension tables is limited. The number of LFDs might rapidly increase for non-linear hierarchy types and the rule-based syntax does not exploit the hierarchical type structure to help  user in defining validity constraints. 
In \cite{liu2020discovering}, we thus introduced the notion of \emph{attribute graph}, which is a graph representation for LFDs in dimension tables, that characterizes all possible \enquote{valid} hierarchy instances of a dimension in a simple and natural way.

We provide an informal definition of attribute graphs through the following example. 

\begin{example}
\begin{figure}[htbp]
\centering
\includegraphics[width=0.65\linewidth]{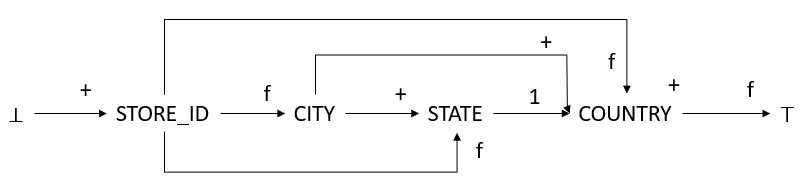}
\caption{Attribute graph of dimension \dimt{SALESORG}}
\label{fig:attgraph-salesorg}
\end{figure}

\Cref{fig:attgraph-salesorg} shows an attribute graph that is validated by dimension table \dimt{SALESORG}. First, the lower and upper bound attributes are respectively \attr{STORE\_ID} and \attr{COUNTRY}. In the attribute graph, they are therefore respectively connected to special nodes $\bot$ and $\top$. Second, for each pair of parent-child attributes in the attribute hierarchy of the dimension, there is a corresponding edge in the attribute graph. 
This yields edges: $(\attr{STORE\_ID}, \attr{CITY})$, $(\attr{CITY}, \attr{STATE})$ and $(\attr{STATE},\attr{COUNTRY})$. 
An additional edge is added between two attributes  when the attributes between them in the attribute hierarchy can have \emph{null} values and the higher attribute functional depends (literally or not) on the lower attribute).
This yields edges: $(\attr{STORE\_ID}, \attr{STATE})$, 
$(\attr{STORE\_ID}, \attr{COUNTRY})$, and $(\attr{CITY}, \attr{COUNTRY})$, because both \attr{CITY} and \attr{STATE} can have null values.
 Third, each edge is assigned a unique label encoding the presence of functional (label $\mathbf{1}$) or literal functional dependency constraints (label $\mathbf{f}$), or none of them ($\mathbf{+}$), between the connected attributes.

By convention, the labels of outgoing edges of $\bot$ are labelled as $\mathbf{+}$ and the labels of the incoming edges of $\top$ are labelled as $\mathbf{f}$.  
Attribute \attr{STORE\_ID} literally determines \attr{CITY}, \attr{STATE} and \attr{COUNTRY} and, therefore, the three  arcs $(\attr{STORE\_ID}, \attr{CITY})$,  $(\attr{STORE\_ID}, \attr{STATE})$ and $(\attr{STORE\_ID}, \attr{COUNTRY})$ are labeled by $\mathbf{f}$. The arc  $(\attr{CITY},\attr{STATE})$ is labeled by $\mathbf{+}$, which signifies that tuples with the same (possibly \emph{null}) value for \attr{CITY} can have different values for \attr{STATE}. Similarly, the arc $(\attr{CITY}, \attr{COUNTRY})$ is labeled by $\mathbf{+}$. Finally, the arc $(\attr{STATE},\attr{COUNTRY})$ is labeled by $\mathbf{1}$, since there exists a functional dependency from non-null $\attr{STATE}$ values to $\attr{COUNTRY}$, but not a literal functional dependency.

Similarly, we define the attribute graph of dimension \dimt{PROD} in \Cref{fig:attgraph-prod}. 

\begin{figure}[htbp]
\centering
\includegraphics[width=0.60\linewidth]{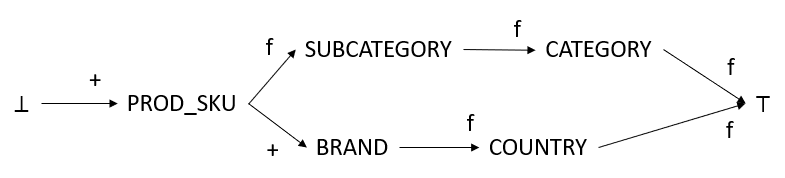}
\caption{Attribute graph of dimension \dimt{PROD}}
\label{fig:attgraph-prod}
\end{figure}

\end{example}

Attribute graphs must be defined by the designer of dimension tables. 
However, we showed in~\cite{liu2020discovering} that attribute graphs can also be automatically and efficiently computed from dimension tables or samples thereof. More details about the acquisition and maintenance of attribute graphs can be found in~\cite{liu_rutian_semantic_2020}. 

We also provided in~\cite{liu2020discovering} an efficient algorithm to compute the minimum set of dimension attributes (called \emph{dimension identifier}) that literally determines all other dimension attributes. Using the same properties of attribute graphs, we can determine if a set of dimensions attributes $U$ in a dimension table, literally determines a dimension attribute $B$ (\ie $U \mapsto \B$). 

\subsection{Analytic queries}
 \label{sec:red-query}
An \emph{interactive data analysis session} consists of a tree of interactive analytic queries having one or more (input) analytic tables as leaves and a single root which is the final result of the session. We assume that the result of any interactive data analysis session can be saved as an \emph{analytic view} whose definition is the tree of interactive queries that have been performed in the session. Hence, views can be reused to start a new interactive data analysis session. As usual, users can backtrack in their session and come back to a previous result. An example of an interactive data analysis session was given in \Cref{fig:sales-dem-usa_bis}. 

In this paper, we consider analytic queries consisting of unary operations
(filter, project, aggregation, pivot, and binary operations (union, difference, merge). All these operations, except pivot, are based on relational operations but their semantics is tailored to the case of analytic tables by restricting their usage depending on the type of attributes (dimension or measure) that are manipulated. 
Our set of operations includes the most common data transformation operations supported by self-service data preparation and BI tools~\cite{trifacta_web, paxata_web, data-preparation_sap_web, tableau_webpage, powerbi_web, qlik_web}. 
They also subsume the traditional interactive operations on a multidimensional (OLAP) cube, such as Roll-up, Drill-down, Slice, or Dice, as defined for instance in \cite{jensen_multidimensional_2010, olap_cube, gray_data_1996, sqltutorial_rollup}.  
In this section, we precisely define the semantics of these operations with a special attention to their manipulation of null values.

\mysubsubsection{Analytic filter queries}

The first single table analytic queries allow users to select a subset of tuples in the input table.

\begin{definition}[Filter query] \label{def:filter-red}
Let $\XXT(\S)$ be an analytic table.  
 We denote by $\Q(\XXT) = \FILTERQ_{\XXT}(P \mid \Y)$,  
an  \emph{analytic filter query} that returns all tuples in $\XXT$ satisfying a predicate $P$ on a set of attributes $\Y\subseteq \S$.
 \end{definition}
Observe that $P$ can be any well-formed Boolean predicate using negation, conjunction and disjunction over any subset of attributes in $\S$. We consider that predicate $P$ is a Boolean function which is  also defined for tuples with $null$ value attributes: except for literal equality, any comparison of an attribute value with a null marker evaluates to false. 
  
Analytic filter queries support operations on a multidimensional cube known as "slice" (selection by subset of values of a dimension) or "dice" (selection by subset of values of more than one dimension) \cite{olap_cube}. However, in our definition, a filter predicate can be expressed on any attribute. 

\begin{myexample}
Consider the table $\XXT(\S)$ in \Cref{tab:before-red-t1}. The result of  two filter queries  $\Q_1 = \FILTERQ_{ \XXT}(\{\A_1 =$\enquote*{$a_1$}$\} \mid \{\A_1\})$ and $\Q_2 = \FILTERQ_{ \XXT}(\{\A_2 \neq $\enquote*{$b_2$}$\} \mid \{\A_2\})$ are shown in \Cref{tab:fil-red-t} and \Cref{tab:non-fil-red-t}.

\begin{table}[htbp]
\caption{Filter queries}
\normalsize{
\begin{subtable}[b]{0.22\linewidth}
   \centering
    \setlength{\tabcolsep}{3pt}
  \begin{tabular}[t]{l|l|l|l|l|l}
    \cmidrule[1pt]{2-6}
    $ \XXT$ & $\A_1$ & $\A_2$ & $\A_3$ & $\M$ & $\N$ \\
    \midrule
    & $a_1$ & $b_1$ & $c_1$ & $x_1$ & $y_1$\\
    & $a_1$ & $b_1$ & $-$ & $x_2$ & $y_2$\\
    & $a_2$ & $b_1$ & $c_1$ & $x_3$ & $y_3$\\
    & $a_2$ & $b_2$ & $-$ & $x_4$ & $y_4$\\
    \cmidrule[0.8pt]{2-6}
    \end{tabular}
       \caption{Input table $\XXT$}
\label{tab:before-red-t1}
\end{subtable}
\hfill
\begin{subtable}[b]{0.28\linewidth}
   \centering
    \setlength{\tabcolsep}{3pt}
  \begin{tabular}[t]{l|l|l|l|l|l}
    \cmidrule[1pt]{2-6}
    $\Q_1$ & $\A_1$ & $\A_2$ & $\A_3$ & $\M$ & $\N$ \\
    \midrule
    & $a_1$ & $b_1$ & $c_1$ & $x_1$ & $y_1$\\
    & $a_1$ & $b_1$ & $-$ & $x_2$ & $y_2$\\
    \cmidrule[0.8pt]{2-6}
    \end{tabular}
       \caption{$\FILTERQ_{\XXT}(\{\A_1 =a_1\} \mid \{\A_1\})$}
    \label{tab:fil-red-t}
\end{subtable}
\hfill
    \begin{subtable}[b]{0.28\linewidth}
   \centering
    \setlength{\tabcolsep}{3pt}
  \begin{tabular}[t]{l|l|l|l|l|l}
    \cmidrule[1pt]{2-6}
    $\Q_2$ & $\A_1$ & $\A_2$ & $\A_3$ & $\M$ & $\N$ \\
    \midrule
    & $a_1$ & $b_1$ & $c_1$ & $x_1$ & $y_1$\\
    & $a_1$ & $b_1$ & $-$ & $x_2$ & $y_2$\\
    & $a_2$ & $b_1$ & $c_1$ & $x_3$ & $y_2$\\
    \cmidrule[0.8pt]{2-6}
    \end{tabular}
      \caption{$\FILTERQ_{\XXT}(\{\A_2 \neq b_2\} \mid \{\A_2\})$}
    \label{tab:non-fil-red-t}
\end{subtable}
}
\end{table}

\end{myexample}

\mysubsubsection{Analytic projection queries}

Projection can be used to remove measure attributes and add new calculated measure attributes. 

\begin{definition}[Analytic projection query]
\label{def:projection_query}
Let $\XXT(\S)$ be an analytic table with dimension attributes $\S_D\subseteq \S$. Let $\Y$ be a subset of $\S$ such that $\S_D \subseteq \Y \subseteq \S$. Let $f(\Z) \rightarrow A$ be an optional expression where $f(\Z)$ is an expression involving a set of attributes $\Z \subseteq \S$, constants, arithmetic operators and string operators, and $\A$ is a new name for a measure attribute that results from the calculation implied by $f(Z)$. We denote by $\Q(\XXT)=\PROJECTQ_{\XXT}(\Y, f(\Z) \rightarrow \A)$ (resp. $\Q(\XXT)=\PROJECTQ_{\XXT}(\Y)$ ) an \emph{analytic projection} which returns a table $\RRT$ with schema $\Y \cup \{A\}$ (resp. $\Y$), such that for every tuple $\XXT$ of $\XXT$, there exists a unique tuple $t'$ in $\RRT$ such that $t'.B = t.B$ for every $B \in \Y$, and $t'.\A= f(t.Z)$. 
\end{definition}

An analytic projection over an analytic table $\XXT(\S)$ is a special case of an \emph{extended projection}~\cite{d_ullman_database_nodate}. It can add a new measure attribute, whose value for each tuple is possibly computed from the values of other attributes of that tuple. 
The definition can easily be extended to a set of expressions $f(Z) \rightarrow A$. Note that expression $f(Z) \rightarrow A$ is optional in a projection query.

\begin{table}[htbp]
\caption{Analytic projection queries}
\normalsize{
\begin{subtable}[b]{0.22\linewidth}
   \centering
    \setlength{\tabcolsep}{3pt}
  \begin{tabular}[t]{l|l|l|l|l|l}
    \cmidrule[1pt]{2-6}
    $ \XXT$ & $\A_1$ & $\A_2$ & $\A_3$ & $\M$ & $\N$ \\
    \midrule
    & $a_1$ & $b_1$ & $c_1$ & $x_1$ & $y_1$\\
    & $a_1$ & $b_1$ & $-$ & $x_2$ & $y_2$\\
    & $a_2$ & $b_1$ & $c_1$ & $x_3$ & $y_3$\\
    & $a_2$ & $b_2$ & $-$ & $x_4$ & $y_4$\\
    \cmidrule[0.8pt]{2-6}
    \end{tabular}
       \caption{Input table $\XXT$}
\label{tab:before-red-t2}
\end{subtable}
\hfill
\begin{subtable}[b]{0.26\linewidth}
   \centering
    \setlength{\tabcolsep}{3pt}
  \begin{tabular}[t]{l|l|l|l|l}
    \cmidrule[1pt]{2-5}
    $\Q_3$ & $\A_1$ & $\A_2$ & $\A_3$ & $\M$ \\
    \midrule
    & $a_1$ & $b_1$ & $c_1$ & $x_1$ \\
    & $a_1$ & $b_1$ & $-$ & $x_2$ \\
    & $a_2$ & $b_1$ & $c_1$ & $x_3$\\
    & $a_2$ & $b_2$ & $-$ & $x_4$\\
    \cmidrule[0.8pt]{2-5}
    \end{tabular}
     \caption{$\PROJECTQ_{\XXT}(\{\A_1,\A_2,\A_3,\M\})$}
    \label{tab:fil-proj-t}
\end{subtable}
\hfill
\begin{subtable}[b]{0.40\linewidth}
   \centering
    \setlength{\tabcolsep}{3pt}
  \begin{tabular}[t]{l|l|l|l|l}
    \cmidrule[1pt]{2-5}
    $\Q_4$ & $\A_1$ & $\A_2$ & $\A_3$ & $\M'$ \\
    \midrule
    & $a_1$ & $b_1$ & $c_1$ & $x_1+y_1$ \\
    & $a_1$ & $b_1$ & $-$ & $x_2+y_2$ \\
    & $a_2$ & $b_1$ & $c_1$ & $x_3+y_3$\\
    & $a_2$ & $b_2$ & $-$ & $x_4+y_4$\\
    \cmidrule[0.8pt]{2-5}
    \end{tabular}
     \caption{$\PROJECTQ_{\XXT}(\{\A_1,\A_2,\A_3\},(\M+\N)\rightarrow \M')$}
    \label{tab:fil-proj-tf}
\end{subtable}
}
\end{table}

 \begin{myexample}
Reconsider the table $\XXT(\S)$ in \Cref{tab:before-red-t2}. 
\Cref{tab:fil-proj-t} and \Cref{tab:fil-proj-tf} show the result of two projections. The first projection $\PROJECTQ_{\XXT}(\{\A_1,\A_2,\A_3,\M\})$ simply keeps a subset of attributes of $\S$ whereas the second projection creates a new attribute $\M'$ which is the sum of $\M$ and $\N$. 
\end{myexample}
Projections must keep all dimension attributes of the original table. To remove dimension attributes, we introduce aggregate queries as explained next.

\mysubsubsection{Analytic aggregate queries}

Aggregate queries generally partition analytic tables along a subset of dimension attributes and aggregate the values of certain attribute in each partition. 
Analytic aggregate  queries support operations on a multidimensional cube known as "roll-up" (aggregation of data from a lower level to a higher level of granularity within a dimension hierarchy) or "dice" (grouping of data with respect to a subset of dimensions of a cube). 

\begin{definition}[Analytic aggregate query] 
\label{def:agg-red}
Let $\XXT(\S)$ be an analytic table with dimension attributes $\S_D\subseteq \S$, $\A$ be an aggregable attribute in $\S$, and $\AGGF$ be an aggregation function.
 We denote by $\Q(\XXT) = \AGGQ_\XXT(\AGGF(\A)\mid \X)$ where $\X \subseteq \S_D$, an \emph{analytic aggregate query} on table $\XXT$ that aggregates $\A$ using aggregation function $\AGGF$ with group-by attributes $\X$. We  say that $\XXT$ is \emph{aggregated along} $\A$ using $\AGGF$. The result contains one tuple for every tuple of distinct values of attributes in $\X$ including \emph{null} values (as for SQL group-by operations).
\end{definition}

The above definition can be easily generalized by replacing attribute $\A$ with a set of attributes. Unlike SQL Rollup \cite{gray_data_1996, sqltutorial_rollup}, note that our definition does not include tuples that represent subtotals in the query result. This facilitates the composition of aggregate queries, without having to deal with these special tuples, and better fits the purpose of interactive data analysis sessions. 

\begin{myexample}
Reconsider the table $\XXT(\S)$ in \Cref{tab:before-red-t3} with dimensional attributes $\A_1, \A_2$ from $\D_1$ and  $\A_3$ from dimension $\D_2$. 
 The result of $\Q_5 = \AGGQ_{ \XXT}( \SUM(\M) \mid \{\A_3\})$ is shown in \Cref{tab:agg-red-t}. Note that, as with SQL semantics, a group-by operator supports literal equality semantics for $null$ values.
The result of aggregate query $\Q_6 =\AGGQ_{ \XXT}( \SUM(\M) \mid \{\A_1, \A_3\})$  is shown in \Cref{tab:non-agg-red-t}.

\begin{table}[htbp]
\caption{Analytic aggregate queries}
\normalsize{
\begin{subtable}[b]{0.25\linewidth}
   \centering       
   \caption{Input table $\XXT$}
\label{tab:before-red-t3}
    \setlength{\tabcolsep}{3pt}
  \begin{tabular}[t]{l|l|l|l|l|l}
    \cmidrule[1pt]{2-6}
    $ \XXT$ & $\A_1$ & $\A_2$ & $\A_3$ & $\M$ & $\N$ \\
    \midrule
    & $a_1$ & $b_1$ & $c_1$ & $x_1$ & $y_1$\\
    & $a_1$ & $b_1$ & $-$ & $x_2$ & $y_2$\\
    & $a_2$ & $b_1$ & $c_1$ & $x_3$ & $y_3$\\
    & $a_2$ & $b_2$ & $-$ & $x_4$ & $y_4$\\
    \cmidrule[0.8pt]{2-6}
    \end{tabular}
\end{subtable}
\hfill
\begin{subtable}[b]{0.36\linewidth}
   \centering
   \caption{$\AGGQ_{ \XXT}( \SUM(\M) \mid \{\A_3\})$}
 \label{tab:agg-red-t}
    \setlength{\tabcolsep}{3pt}
    \begin{tabular}[t]{l|l|r}
    \cmidrule[1pt]{2-3}
    $\Q_5$ & $\A_3$ & $\SUM(\M)$ \\
    \midrule
    & $c_1$ &$x_1 + x_3$ \\
    & $-$ & $x_2 + x_4$ \\
    \cmidrule[0.8pt]{2-3}
   \end{tabular}
\end{subtable}
\hfill
\begin{subtable}[b]{0.36\linewidth}
   \centering
   \caption{$\AGGQ_{ \XXT}(\SUM(\M) \mid \{\A_1, \A_3\})$}
 \label{tab:non-agg-red-t}
    \setlength{\tabcolsep}{3pt}
    \begin{tabular}[t]{l|l|l|r}
    \cmidrule[1pt]{2-4}
    $\Q_6$ & $\A_1$ & $\A_3$ & $\SUM(\M)$ \\
    \midrule
    & $a_1$ &  $c_1$ & $x_1$\\
    & $a_1$ &  $-$ & $x_2$\\
    & $a_2$ &  $c_1$ & $x_3$\\
    & $a_2$ &  $-$ & $x_4$\\
    \cmidrule[0.8pt]{2-4}
   \end{tabular}
\end{subtable}
}   
\end{table}

\end{myexample}

\mysubsubsection{Analytic pivot queries}

Pivot queries also partition tables along a subset of dimension attributes. But instead of aggregating all values of a non partitioning attribute into a single  value for each partition, it generates a new attribute for each value. Analytic pivot queries are particularly useful in the data preparation phase of machine learning application scenarios like feature engineering \cite{zheng_feature_nodate,liu2020discovering}. They should not be mistaken with the OLAP cube pivot operation that keeps the schema of the input table unchanged.

\begin{definition}[Analytic pivot query] \label{def:pivot-red}
Let $\XXT(\S)$ be a fact table with dimension attributes $\S_D\subseteq \S$ and $\A$ be a measure attribute in $\S$. We denote by $\Q(\XXT) =  \PIVOTQ_\XXT(\A \mid \X)$, where $\X \subset \S_D$, an \emph{analytic pivot query} which pivots attribute $\A$ over $\X$. The result is a table $\RRT$ with all attributes in $\S_D - \X$ and an attribute $\A\_{v_i}$ for each value $v_i$ in the domain of $\XXT.\X$. 
The value $t.\A$ of each tuple $t\in \XXT$ such that $t.\X=v_i$ is a value in the attribute $\A\_{v_i}$ of the unique tuple $t'$ in $\RRT$ such that $t.(\S_D - \X)=t'.(\S_D - \X)$.
\end{definition}

The above definition can be easily generalized by replacing attribute $\A$ with a set of attributes. 

\begin{myexample}
\label{ex:pivotqueries}
Reconsider the table $\XXT(\S)$ in \Cref{tab:before-red-t4}. The result  of pivot query  $\Q_7 = \PIVOTQ_\XXT(\M \mid \A_1)$ that pivots attribute $M$ over $\A_1$ is shown in \Cref{tab:piv-red-t}. The schema of the resulting table $ \RRT$ contains all attributes in $\S - \A_1$ and two new attributes $\M\_a_1$ and $\M\_a_2$ for each value of $\XXT.\A_1$. 
The value $t.M$ of each tuple $t\in \XXT$ such that $t.\A_1=v$ is a value in the attribute $\M\_v$ of the unique tuple $t'$ in $\RRT$ such that $t.(\{\A_2, \A_3\})=t'.(\{\A_2, \A_3\})$.
The result of another pivot query  $\Q_8 = \PIVOTQ_\XXT(\M \mid \A_2)$  is shown in \Cref{tab:non-piv-red-t}.

\begin{table}[htbp]
\caption{Analytic pivot queries}
\normalsize{
\begin{subtable}[b]{0.25\linewidth}
   \centering
    \setlength{\tabcolsep}{3pt}
  \begin{tabular}[t]{l|l|l|l|l|l}
    \cmidrule[1pt]{2-6}
    $ \XXT$ & $\A_1$ & $\A_2$ & $\A_3$ & $\M$ & $\N$ \\
    \midrule
    & $a_1$ & $b_1$ & $c_1$ & $x_1$ & $y_1$\\
    & $a_1$ & $b_1$ & $-$ & $x_2$ & $y_2$\\
    & $a_2$ & $b_1$ & $c_1$ & $x_3$ & $y_3$\\
    & $a_2$ & $b_2$ & $-$ & $x_4$ & $y_4$\\
    \cmidrule[0.8pt]{2-6}
    \end{tabular}
       \caption{Input table $\XXT$}
\label{tab:before-red-t4}
\end{subtable}
\hfill
\begin{subtable}[b]{0.35\linewidth}
   \centering
    \setlength{\tabcolsep}{3pt}
  \begin{tabular}[t]{l|l|l|r|r}
    \cmidrule[1pt]{2-5}
    $\Q_7$ & $\A_2$ & $\A_3$ & $\M\_a_1$ & $\M\_a_2$ \\
    \midrule
    & $b_1$ & $c_1$ & $x_1$ & $x_3$ \\
    & $b_1$ & $-$ & $x_2$ & $-$ \\
    & $b_2$ & $-$ & $-$ & $x_4$ \\
    \cmidrule[0.8pt]{2-5}
    \end{tabular}
    \caption{$\PIVOTQ_\XXT(\M \mid \A_1)$}
    \label{tab:piv-red-t}
\end{subtable}
\hfill
    \begin{subtable}[b]{0.35\linewidth}
   \centering
    \setlength{\tabcolsep}{3pt}
  \begin{tabular}[t]{l|l|l|r|r}
    \cmidrule[1pt]{2-5}
    $\Q_8$ & $\A_1$ & $\A_3$ & $\M\_b_1$ & $\M\_b_2$ \\
    \midrule
    & $a_1$ & $c_1$ & $x_1$ & -\\
    & $a_1$ & $-$ & $x_2$ & -\\
    & $a_2$ & $c_1$ & $x_3$ & - \\
    & $a_2$ & $-$ & - & $x_4$\\
    \cmidrule[0.8pt]{2-5}
    \end{tabular}
    \caption{$\PIVOTQ_\XXT(\M \mid \A_2)$}
    \label{tab:non-piv-red-t}
\end{subtable}
}
\end{table}

\end{myexample}

\mysubsubsection{Analytic merge queries}

Analytic left-merge queries  combine the tuples of two analytic tables and correspond to natural left outer-join operations defined in the extended relational algebra with null values. Analytic left-merge queries  play an important role in so-called schema augmentation scenarios~\cite{liu2020discovering} and  can support the OLAP cube operation known as "drill-down" (the inverse of roll-up) by merging a fact table, that provides a higher-level of granularity within a dimension hierarchy for some measures, with a fact table that provides a lower level of granularity for the exact same measures. 

In our definition, we allow the merge of two fact tables on their common dimension attributes (which have the same names in the two fact tables) but we accept that the common attributes belong to different dimensions in each fact table. An example of such a merge operation was given in \cref{sec:introduction}, between \factt{SALES\_STORES} and \factt{DEM}, on common attributes \attr{CITY}, \attr{STATE}, \attr{COUNTRY}. These attributes belong to dimension \dimt{SALESORG} in \factt{SALES\_STORES} and dimension \dimt{REGION} in \factt{DEM}. 

\begin{definition}[Analytic left-merge query]
\label{def:merge_query}
Let $\Q = \pi_X(\XXT)$ where $\XXT=\XXT\leftouterjoin_{P_1 \land \cdots \land P_k} \YYT$, $\XXT(\S)$ and $\YYT(S')$ are two analytic tables, $\leftouterjoin$ is a \emph{left-outer join} operator, $P_i$ are join equality predicates over a set of (common) dimension attributes $\Y$, and $\pi_X$ is the duplicate elimination relational projection over a set of attributes $\X$ defined below. Then $\Q$ is a \emph{left-merge merge analytic query} if the following conditions hold: 
\begin{enumerate}
\item  \label{merge_item2} For each $\A_i\in \Y$, $\exists P_i$ such that $P_i=(\XXT.\A_i=\YYT.\A_i)\vee (\XXT.\A_i = null \wedge  \YYT.\A_i = null)$ ($null$ is a literal).

\item \label{merge_item1} If for each pair of attributes $\A_1,\A_2 \in \Y$ that belongs to both a dimension $D_1$ in $\XXT$  and a dimension $D_2$ in $\YYT$ ($D_1 \neq D_2$),  $\A_1$ and $\A_2$ are either connected with the same labelled paths in their respective attribute graphs or not connected by any path, then
$\X=\XXS\cup \YYS$  else $\X=\XXS \uplus \YYS$ ($\uplus$ denotes disjoint union, \ie union after renaming conflicting attributes). 
\end{enumerate}

In the following, we will abbreviate $\Q=\pi_X(\XXT\leftouterjoin_{P_1 \land \cdots \land P_k} \YYT))$ 
by  $\Q=\XXT \leftouterjoin_Y \YYT$, where $\Y$ is the set of join attributes, call it a \emph{left-merge query}, and refer to the result of $\Q$ as a \emph{merge table}.  
\end{definition}

Item~\ref{merge_item2} manages the join predicates in the merge query in the presence of nulls (we apply literal equality which is different from the SQL equality semantics for null values). The merge table preserves all rows in $\XXT$ (with possible row duplication) and contains all attributes in $\XXT$ and $\YYT$ (the merge query does not project out any attribute). 

Item~\ref{merge_item1} checks that, when two different dimensions are joined on their common attributes, the structure and properties of their respective hierarchies for the joined attributes are identical, that is, the attribute graphs (defined in \Cref{sec:attribute-graphs}), restricted to all common attributes and all paths connecting these attributes, are identical. When this is not the case (e.g., they differ on their hierarchical relationships or they have different functional dependencies), the merge query keeps the join attributes separately for each table and applies a disjoint union. 

Left-merge queries can also be generalized to a full-outer join ($\fullouterjoin{}$) between two tables, called an \emph{analytic full merge  query}, or restricted to a natural join ($\bowtie$), called an \emph{analytic strict merge query}.
Right-merge queries $\Q(\XXT,\YYT)=\XXT \rightouterjoin_{\Y} \YYT$ are equivalent to the symmetric left-merge queries $\Q(\YYT,\XXT)=\YYT \leftouterjoin_{\Y} \XXT$ on the switched tables. 
 
\begin{myexample}
Consider the fact tables $\XXT$ and $\YYT$ in \Cref{tab:merge-t1} and \Cref{tab:merge-t2}, respectively defined over dimension $\D_1$ (where $\A_1 \typeprec \A_2 \typeprec \A_3$) and dimension $\D_2$ (where $\A_2 \typeprec \A_3$). Suppose that in both dimensions, we have $\A_2 \mapsto A_3$, then by Item~\ref{merge_item1}, since attributes $\A_2$ and $\A_3$ are connected by the same labelled paths in their respective attribute graphs, they appear only once in the merge table, and the result of a left merge $\Q_{10}=\XXT\leftouterjoin \YYT$ is shown in \Cref{tab:leftmerge-red-t}. If the labelled paths between $\A_2$ and $\A_3$ were different in the attribute graphs of $D_1$ and $D_2$, all the dimension attributes of $\XXT$ and $\YYT$ will be kept separately in the result of the merge.

\begin{table}[htbp]
\caption{Analytic merge queries}
\normalsize{
\begin{subtable}[b]{0.28\linewidth}
   \centering
    \setlength{\tabcolsep}{3pt}
  \begin{tabular}[t]{l|l|l|l|l}
    \cmidrule[1pt]{2-5}
    $ \XXT$ & $\A_1$ & $\A_2$ & $\A_3$ & $\M$  \\
    \midrule
    & $a_1$ & $b_1$ & $c_1$ & $x_1$ \\
    & $a_1$ & $-$ & $c_2$ & $x_2$ \\
    & $a_2$ & $b_1$ & $c_3$ & $x_3$ \\
    & $a_3$ & $-$ & $c_2$ & $x_4$ \\
    \cmidrule[0.8pt]{2-5}
    \end{tabular}
       \caption{Input table $\XXT$}
\label{tab:merge-t1}
\end{subtable}
\hfill
\begin{subtable}[b]{0.24\linewidth}
   \centering
    \setlength{\tabcolsep}{3pt}
  \begin{tabular}[t]{l|l|l|l}
    \cmidrule[1pt]{2-4}
    $ \YYT$ &  $\A_2$ & $\A_3$ & $\N$  \\
    \midrule
     & $b_1$ & $c_1$ & $y_1$ \\
    & $b_1$ & $c_3$ & $y_2$ \\
     & $b_2$ & $c_4$ & $y_3$ \\
     & $b_3$ & $c_4$ & $y_4$ \\
    \cmidrule[0.8pt]{2-4}
    \end{tabular}
       \caption{Input table $\YYT$}
\label{tab:merge-t2}
\end{subtable}
\hfill
\begin{subtable}[b]{0.42\linewidth}
   \centering
    \setlength{\tabcolsep}{3pt}
   \begin{tabular}[t]{l|l|l|l|l|l}
    \cmidrule[1pt]{2-6}
    $ \Q_{10}$ & $\A_1$ & $\A_2$ & $\A_3$ & $\M$ & $\N$  \\
    \midrule
    & $a_1$ & $b_1$ & $c_1$ & $x_1$ & $y_1$\\
    & $a_1$ & $-$ & $c_2$ & $x_2$ & $-$\\
    & $a_2$ & $b_1$ & $c_3$ & $x_3$& $y_2$\\
    & $a_2$ & $-$ & $c_2$ & $x_4$  & $-$\\
    \cmidrule[0.8pt]{2-6}
    \end{tabular}
    \caption{$\XXT\leftouterjoin \YYT$}
    \label{tab:leftmerge-red-t}
\end{subtable}
}
\end{table}

\end{myexample}

\mysubsubsection{Analytic set queries}

Analytic tables are sets of tuples and can therefore be combined using set operations. However, compared to standard relational set operations, analytic set operations must respect additional constraints related to the separation between dimension and measure attributes and the condition that all measure attributes are determined by a subset of dimension attributes.

\begin{definition}[Analytic set queries]
\label{def:set_queries}
Let $\XXT$ and $\YYT$ be two analytic tables having the same schema with a set of dimension attributes $\Y$ (referring to the same dimensions):
\begin{itemize}
    \item If $\pi_\Y(\XXT)\cap \pi_\Y(\YYT)=\emptyset$, $\Q=\XXT \cup \YYT$ is a union analytic query where $\cup$ is the set union operator. 
    \item  $\Q=\XXT - \YYT$ is a difference analytic query where "$-$" is the set difference operator based on literal equality of attribute values.
 \end{itemize}
\end{definition}

Observe that analytic intersection $\XXT \cap \YYT$ is equivalent to $\XXT-(\XXT - \YYT)$.
Analytic union queries are useful to complement dimension or fact tables. 
A full merge query $\Q=\XXT \fullouterjoin_Y \YYT$ between two analytic tables $\XXT$ and $\YYT$ having exactly the same set of dimension attributes $\Y$, and such that $\pi_\Y(\XXT)\cap \pi_\Y(\YYT)=\emptyset$, expresses an \emph{analytic outer-union} between the two tables. 
Note that a \emph{data fusion} operation \cite{data_fusion_2009} can be expressed as a full merge query followed by an analytic projection. 

\begin{myexample}
\Cref{fig:analyticset} shows two tables and the result of two analytic set queries. Observe that the union $\XXT\cup \YYT$ is not defined since $\pi_{A_1,A_2,A_3}\XXT
\cap \pi_{A_1,A_2,A_3}\YYT\neq\emptyset$.

\begin{table}[htbp]
\caption{Analytic set queries}
\label{fig:analyticset}
\normalsize{
\begin{subtable}[b]{0.24\linewidth}
   \centering
    \setlength{\tabcolsep}{3pt}
  \begin{tabular}[t]{l|l|l|l|l}
    \cmidrule[1pt]{2-5}
    $ \XXT$ & $\A_1$ & $\A_2$ & $\A_3$ & $\M$ \\
    \midrule
    & $a_1$ & $b_1$ & $c_1$ & $x_1$ \\
    & $a_1$ & $b_1$ & $-$ & $x_2$ \\
    & $a_2$ & $b_1$ & $c_1$ & $x_3$ \\
    & $a_2$ & $b_2$ & $-$ & $x_4$\\
    \cmidrule[0.8pt]{2-5}
    \end{tabular}
       \caption{Input table $\XXT$}
\label{tab:tset1}
\end{subtable}
\hfill
\begin{subtable}[b]{0.24\linewidth}
   \centering
    \setlength{\tabcolsep}{3pt}
  \begin{tabular}[t]{l|l|l|l|l}
    \cmidrule[1pt]{2-5}
    $ \YYT$ & $\A_1$ & $\A_2$ & $\A_3$ & $\M$ \\
    \midrule
    & $a_1$ & $b_1$ & $c_1$ & $x_1$ \\
    & $a_1$ & $b_1$ & $-$ & $x_2$ \\
    & $a_1$ & $b_2$ & $c_1$ & $x_5$ \\
    & $a_2$ & $b_2$ & $c_1$ & $x_6$ \\
    & $a_2$ & $b_1$ & $-$ & $x_7$\\
    \cmidrule[0.8pt]{2-5}
    \end{tabular}
       \caption{Input table $\YYT$}
\label{tab:tset2}
\end{subtable}
\hfill
\begin{subtable}[b]{0.24\linewidth}
   \centering
    \setlength{\tabcolsep}{3pt}
  \begin{tabular}[t]{l|l|l|l|l}
    \cmidrule[1pt]{2-5}
    $ \XXT$ & $\A_1$ & $\A_2$ & $\A_3$ & $\M$ \\
    \midrule
    & $a_2$ & $b_1$ & $c_1$ & $x_3$ \\
    & $a_2$ & $b_2$ & $-$ & $x_4$\\
    \cmidrule[0.8pt]{2-5}
    \end{tabular}
       \caption{$\XXT-\YYT$}
\label{tab:setdiff1}
\end{subtable}
\hfill
\begin{subtable}[b]{0.24\linewidth}
   \centering
    \setlength{\tabcolsep}{3pt}
  \begin{tabular}[t]{l|l|l|l|l}
    \cmidrule[1pt]{2-5}
    $ \XXT$ & $\A_1$ & $\A_2$ & $\A_3$ & $\M$ \\
    \midrule
   & $a_1$ & $b_1$ & $c_1$ & $x_1$ \\
    & $a_1$ & $b_1$ & $-$ & $x_2$ \\
    & $a_2$ & $b_1$ & $c_1$ & $x_3$ \\
    & $a_2$ & $b_2$ & $-$ & $x_4$\\
    & $a_1$ & $b_2$ & $c_1$ & $x_5$ \\
    & $a_2$ & $b_2$ & $c_1$ & $x_6$ \\
    & $a_2$ & $b_1$ & $-$ & $x_7$\\
    \cmidrule[0.8pt]{2-5}
    \end{tabular}
       \caption{$(\XXT-\YYT)\cup T'$}
\label{tab:setdiffunion}
\end{subtable}
}
\end{table}

\end{myexample}


\section{Aggregability of Attributes in Analytic Tables}

\label{sec:aggregable-properties-def-prop}

An attribute of an analytic table does not necessarily aggregate with all aggregation functions along all dimension attributes. Describing when this aggregation is possible has been extensively studied  for statistical and OLAP databases (see \cite{mazon_survey_2009} for a survey). 
Focusing on function $\SUM$, \cite{kimball_data_2013}, \cite{horner_analysis_2004} and \cite{niemi_detecting_2014} proposed that the designer of a fact table declares the \emph{additivity} category of each measure: 
\emph{fully-additive} measures can be summed along any dimension, \emph{semi-additive} measures can be summed along some, but not all, dimensions, and \emph{non-additive} measures cannot be summed along any dimension. This approach has been implemented in several OLAP systems. 

Generalizing this approach, we introduce aggregable properties that enable a designer to declare for any attribute of an analytic table, which aggregation function is applicable and the set of dimension attributes along which this aggregation function can be computed. We introduce default rules that assist the designer of a table to define these properties. Finally, we show that aggregable properties can be automatically computed on the tables resulting from an analytic query, thereby saving the human effort to define them. 

\subsection{Aggregable properties of attributes}
\label{sec:aggprop-def}

If some attribute $\A$ is aggregable along a set of dimension attributes $\X$, then it is also aggregable along any subsets of $\X$. In the following, we denote by $\aggp_{\A} (\AGGF,\X)$ the \emph{aggregable property} of $\A$ and state that property $\aggp_{\A}(\AGGF,\X)$ holds in $\XXT$ if $\X$ is the maximal set of attributes along which $\A$ is aggregable using $\AGGF$ in $\XXT$. We now formally define aggregable properties $\aggp_\A(\AGGF,\X)$.


\begin{definition}[Aggregable Property]
\label{def:aggprop}
Let $\S_D$ be the set of dimension attributes in an analytic table $\XXT(\S)$, $\A$ be an aggregable attribute in $\S$ and $\AGGF$ be an aggregation function. 
\begin{itemize}
\item Let $\X_f \subseteq \S_D$ be the set all dimension attributes $\B$ such that any aggregation of $\A$ with $\AGGF$ along $\B$ is considered to be \emph{meaningless} by the user. We call $\X_f$ the set of \emph{forbidden} dimension attributes along which $\A$ cannot be aggregated using $\AGGF$. 
\item If $\A$ is a measure attribute, let $\X_d\subseteq \S_D$ be a minimal subset of dimension attributes such that $\X_d \mapsto \A$. Let $\X_d^+$ be the set of all dimension attributes $\B\in \S_D$ such that $\X_d \mapsto \B$.
We call $\X_d$ a determinant of $\A$ and $\X_d^+$ the closure of $\X_d$ in $\S_D$.
 \end{itemize}
Then the aggregable property $\aggp_\A(\AGGF,\X)$ holds in $\XXT$ for $\AGGF$ and $\X\subseteq \S_D-\X_f$, where:
\begin{enumerate}
\item \label{it:applicable} Function $\AGGF$ is \emph{applicable} to $\A$. \item \label{it:measuredep}
If $\A$ is a measure attribute then $\X = \X_d^+ - \X_f$.
\item \label{it:dimensiondep} If $\A$ is a dimension attribute then $\X = \S_D-\{\A\}-\X_f$.
\end{enumerate}
\end{definition}

\Cref{it:applicable} and the definition of the forbidden attributes $\X_f$ in \Cref{def:aggprop} cover the "information semantics" of an attribute $\A$ and restrict the functions and the dimensions for the aggregation of the attribute. 
Different categorizations have been proposed in the literature to determine the  aggregation functions which are \emph{applicable} to some measure attribute, 
such as a statistic classification of measurements \cite{stevens_theory_1946, niemi_detecting_2014,THANISCH2019116}, the attribute's aggregation behaviour \cite{pedersen_extending_1999, pedersen_foundation_2001}, or the compatibility between the type of dimensions and the type of a measure \cite{lenz_summarizability_1997}. 
These categorizations can also be used in our context to determine both the "applicability" of a function $\AGGF$ and the set $\X_f$ of forbidden dimensions for a given measure attribute.

\Cref{it:measuredep} covers the "logical semantics" defined by the literal functional dependencies between dimension and measure attributes.
 Essentially the closure $\X_d^+$ of the determinant $\X_d$ of $\A$ contains all dimension attributes which are "logically related to" measure attribute $\A$ through literal functional dependencies. First, it is easy to see that all  partitions generated by $\X_d^+$ have the same value for $\A$. Second, \Cref{it:measuredep}  considers that any aggregation of $\A$ along any subset of $\X_d^+$  is \emph{logically correct} and it is \emph{semantically meaningful} if it also respects the  applicability constraint (\Cref{it:applicable}) and excludes the attributes in $\X_f$. Symmetrically, all attributes that are not in $\X_d^+$ are considered as logically independent of measure $\A$ and must be preserved by the partitioning (i.e., appear in the GROUP BY clause of an SQL query).  
 
 Finally, \Cref{it:dimensiondep} mainly states that any dimension attribute can be aggregated along all other dimension attributes except those defined as "meaningless" in $\X_f$.
  This follows from the observation that all dimension attributes are considered to be descriptive  and the only aggregation  functions that can be applied are $\COUNT$ and $\COUNTDISTINCT$ (see also \Cref{tab:cat_agg} below). Then, we assume that there exists no "logical" constraint defined by LFDs when counting some values along any "semantically meaningful" set of attributes (see \Cref{exp:aggproperty-coca}).
 
 Observe that in \Cref{it:measuredep}, there may exist several 
 determinants $\X_d$ of $\A$ and each such determinants might define a different set of attributes $\X_i$ along which $\A$ can be aggregated using $\AGGF$. However, it is easy to show that if $\A$ can be aggregated along any subset of $\X_1$ and any subset of $\X_2$ using $\AGGF$, it also can be aggregated along any subset of the union $\X_1\cup \X_2$. 

In practice, the designer of a fact table is asked to indicate the set of semantically meaningless dimension attributes $\X_f$ and, in the case of a measure attribute only, a minimum set of logically correct dimension attributes $\X_d$ on which this attribute depends. The closure $\X_d^+$ is \emph{automatically obtained} using the attribute graphs of the dimensions. 
Thus, for each minimal set $\X_d$ provided by the user, \Cref{it:measuredep} of \Cref{def:aggprop} gives the corresponding aggregable property of a measure attribute. 

\begin{example}
\label{exp:aggproperty-coca}
Consider the fact table \factt{PRODUCT\_LIST} (\attr{PROD\_SKU}, \attr{COUNTRY}, \attr{BRAND}, \attr{YEAR},  \attr{QTY}) displayed in \Cref{tab:prod_list}. 
Suppose that the designer of the fact table indicates that the measure attribute \attr{QTY} literally depends on the minimal set of attributes $\X_d = \{\attr{PROD\_SKU}, \attr{YEAR}\}$, and that $\SUM$ aggregation is meaningful along any dimension attribute (i.e., $\X_f=\emptyset$). Since $\X_d$ does not determine any other attribute, 
by \Cref{it:measuredep} of the definition,
$\X_d$ is the largest set of attributes for which the aggregable property $\aggp_{\mattr{QTY}} (\SUM, \X_d)$ holds in \factt{PRODUCT\_LIST}. 
In other words, \mattr{QTY} can only be aggregated along attributes $\attr{PROD\_SKU}$ and $\attr{YEAR}$. 
For dimension attribute $\attr{PROD\_SKU}$, suppose that the designer indicates that $\X_f=\emptyset$, then $\aggp_{\mattr{PROD\_SKU}} (\AGGF, \X)$ holds for $\AGGF \in \{\COUNT, \COUNTDISTINCT\}$ and $\X=\{\attr{BRAND}, \attr{COUNTRY},\attr{YEAR}\}$. 

Next, consider the fact table \factt{STORE\_SALES} (\Cref{tab:state-summarizability-sales1}) and suppose that the designer of the fact table indicates that the measure attribute \attr{AMOUNT} depends on $\X_d = \{\attr{STORE\_ID}, \attr{YEAR}\}$ and $\X_f =\emptyset$ for function $\SUM$. Since \attr{STORE\_ID} literally determines the dimension attributes \attr{CITY}, \attr{STATE}, and \attr{COUNTRY}, by \Cref{it:measuredep} of the definition, \attr{AMOUNT} is aggregable along any subset of the dimension attributes of \factt{STORE\_SALES}. 

Finally, consider the fact table \factt{DEM} (Demographics) displayed in \Cref{tab:dem}. Summing up measure attribute \attr{POP} along dimension attribute \attr{YEAR} would clearly be incorrect, while it would be correct along any attribute of dimension \dimt{REGION}. Thus, the designer of the fact table should define $\X_f=\{\attr{YEAR}\}$ for $\SUM$ and \attr{POP}. After indicating that \attr{POP} depends on dimension attributes $\X_d=\{\attr{CITY}, \attr{COUNTRY}, \attr{YEAR}\}$, property $\aggp_{\mattr{POP}} (\SUM, \{\attr{CITY}, \attr{COUNTRY}\})$ can automatically be computed.

\begin{table}[htb]
        \caption{\factt{PRODUCT\_LIST}}
\label{tab:prod_list}
\normalsize{
    \centering
    \begin{tabular}{l|l|l|l|r}
        \toprule
         \attr{PROD\_SKU} & \attr{BRAND}& \attr{COUNTRY} & \attr{YEAR} & \attr{QTY} \\
         \midrule
         cz-tshirt-s & Coco Cola & USA & 2017 & 5 000 \\
         cz-tshirt-s & Zora & Spain & 2017 & 5 000 \\
         coco-33cl-can & Coco Cola & USA & 2017 & 10 000 \\
         \bottomrule
         \end{tabular}
         \\\ \\  Two aggregable properties: \\\ \\
 \begin{tabular}{rl}
          \emph{with user input}:& $\aggp_{\mattr{QTY}} (\AGGF, \{\attr{PROD\_SKU}, \attr{YEAR}\})$ for $\AGGF\in\{ \SUM, \COUNT, \AVG, ... \allowbreak \}$ \\
\emph{default}:& $\aggp_{\mattr{PROD\_SKU}} (\AGGF, \{\attr{BRAND}, \attr{COUNTRY}, \attr{YEAR}\})$ for $\AGGF\in\{ \COUNT,\allowbreak \COUNTDISTINCT\}$
    \end{tabular}
}
\end{table}

\end{example}

\subsection{Default rules for aggregable properties}
\label{sec:default-rules}
By inspecting the properties of measure attributes  and  aggregation functions, we define rules to obtain default aggregable properties for every aggregable attribute of every analytic table. 
The effort required from the designer of analytic tables is then to inspect and possibly correct the result produced by the application of the default rules, according to the known information semantics of attributes. With respect to \Cref{def:aggprop}, the only possible corrective actions taken by a designer consist of adding dimension attributes to the forbidden attribute set $\X_f$, or removing attributes from the determinant set $\X_d$ if $\X_d$ is not minimal. 

\paragraph{Default applicable functions:}
Existing methods that categorize measures to determine the applicability of an aggregation function rely on some external knowledge and require an analysis of every aggregable attribute of an analytic table. To reduce the user effort, we provide a default categorization into three categories of attributes $\NUM$ (numerical), $\DESC$ (descriptive/categorical) and $\STAT$ (statistical). These categories can automatically be  extracted from the schema  metadata: the two categories $\NUM$ and $\DESC$ are inferred from the (SQL) data type of attributes and the category $\STAT$ denotes a result from the use of some statistical function.  \Cref{tab:cat_agg}  describes the six common SQL aggregation functions applicable to each category, which will be used in the examples of this paper. We therefore use the attribute category of $\A$ to define which aggregation function $\AGGF$ is applicable to $\A$.  As mentioned before, a scale-based categorization of measure attributes could also be used (e.g., \cite{THANISCH2019116}).

\begin{table}[htb]
\centering
\caption{Categories of attributes and their properties}
\label{tab:cat_agg}
\normalsize{
\bgroup
\def\arraystretch{1.2}
\begin{tabular}{l|l}
    \toprule
    Attribute category& Properties\\
    \midrule
     $\NUM$&\labelitemi~~Numerical values\\
    &\labelitemi~~Applicable functions: \SUM, \AVG, \COUNT, \COUNTDISTINCT, \MIN, \MAX \\
    \hline
    $\DESC$&\labelitemi~~Descriptive or categorical values\\
    &\labelitemi~~Applicable functions: \COUNT, \COUNTDISTINCT\\
    \hline
    $\STAT$&\labelitemi~~Numerical statistical values\\
            &\labelitemi~~Applicable functions: \COUNT, \COUNTDISTINCT, \MIN, \MAX\\
  \bottomrule
\end{tabular}
\egroup
}
\end{table}

\paragraph{Default values of ~$X_d$ and $X_f$ for a measure attribute:}
If $\A$ is a measure attribute of $\XXT$ for which no minimal set of attributes that determines $\A$ has been defined by a user, then we use the default rule that $\A$ depends on all dimension attributes. This actually means that in \Cref{it:measuredep}, $\X_d$ contains the identifiers of all dimensions (automatically determined using the attribute graphs of the dimensions). 

We implicitly assume that $\X_d$ is minimal, which is a necessary condition in the definition of aggregable property. If $\A$ does not logically depend on some dimension, this  must be indicated by the designer of the fact table, and the corresponding dimension attributes are removed from $\X_d$. 

We assume by default that the set of meaningless attributes is empty ($\X_f = \emptyset$).
If there exists a "meaningless" aggregation along some dimensions (like in the fact table \factt{DEM} of the previous example), this should be indicated by the designer of the fact table, by adding the corresponding dimension attributes to $\X_f$.

\paragraph{Default  value of ~$X_f$ for a dimension attribute:}
As already mentioned before, if $\A$ is a dimension attribute,  we assume that its category is $\DESC$ to determine the applicable aggregation functions ($\COUNT$ and $\COUNTDISTINCT$). By definition, we also assume that all aggregations using  these two functions along any set of attributes (except $\A$) are correct.
As before, we also assume that there exist no meaningless aggregations, and we use the default rule that $\X_f = \emptyset$.

\paragraph{Important consequence}
We assure that each aggregable property $\aggp_\A(\AGGF,\X)$, with its determinant $\X_d$ and forbidden attribute set $\X_f$, is part of the metadata of attribute $\A$ in table $\XXT$. This is particularly needed when a user takes some action to either minimize the default value of $\X_d$ or add attributes to $\X_f$. Without keeping the values of $\X_d$ and 
$\X_f$, it would not be possible to infer them from the value of $\X$.

The default values and possible user actions are summarized in \Cref{tab:default_agg_prop}. 

\begin{table}[htb]
\centering
\caption{Default values of $\X_d$ and $\X_f$ and possible user actions}
\label{tab:default_agg_prop}
\normalsize{
\begin{tabular}{l|l|l}
    \toprule
    Attribute & Default values & Possible user action \\
    \midrule
    Measure &  
    $\X_d=$ fact identifier; 
    $\X_f = \emptyset$ 
    & remove attributes to minimize $\X_d$; 
    \\
    & & add  attributes to $\X_f$ \\
    \hline
    Dimension & $\X_f = \emptyset$ 
    & add  attributes to $\X_f$ \\
    \bottomrule
\end{tabular}
}
\end{table}

\begin{myexample}
Continuing with \Cref{exp:aggproperty-coca}, since attribute \mattr{QTY} in table \factt{PRODUCT\_LIST} is of category $\NUM$, we get from \Cref{tab:cat_agg} the list of applicable aggregation functions. Then, by default, the minimum set of attributes $\X_d$  that determines \mattr{QTY} will be the fact identifier of \factt{PRODUCT\_LIST}, $\X_d = \{\attr{PROD\_SKU}, \attr{BRAND}, \attr{YEAR}\}$. 
This set is however not minimal (\mattr{QTY} only depends on \{\attr{PROD\_SKU}, \attr{YEAR}\}) and the designer of the table should remove attribute \attr{BRAND} from $\X_d$. 
Finally, using the default rule that $\X_f = \emptyset$, we get the aggregable property displayed on the bottom of \Cref{tab:prod_list}.

The dimension attribute \attr{PROD\_SKU} is of category $\DESC$, which determines its applicable aggregation functions. Then, using the default rule for $\X_f$, we get the aggregable property displayed on the bottom of \Cref{tab:prod_list}. However, if the user considers that it makes no sense to count products along the time dimension, he might remove $\attr{YEAR}$ from the aggregable property by adding it to the set of forbidden attributes $\X_f$.
\end{myexample}

\subsection{Propagating aggregable properties}
\label{sec:prop-agg-prop}

\begin{figure}[htbp]
\centering
\includegraphics[width=0.60\linewidth]{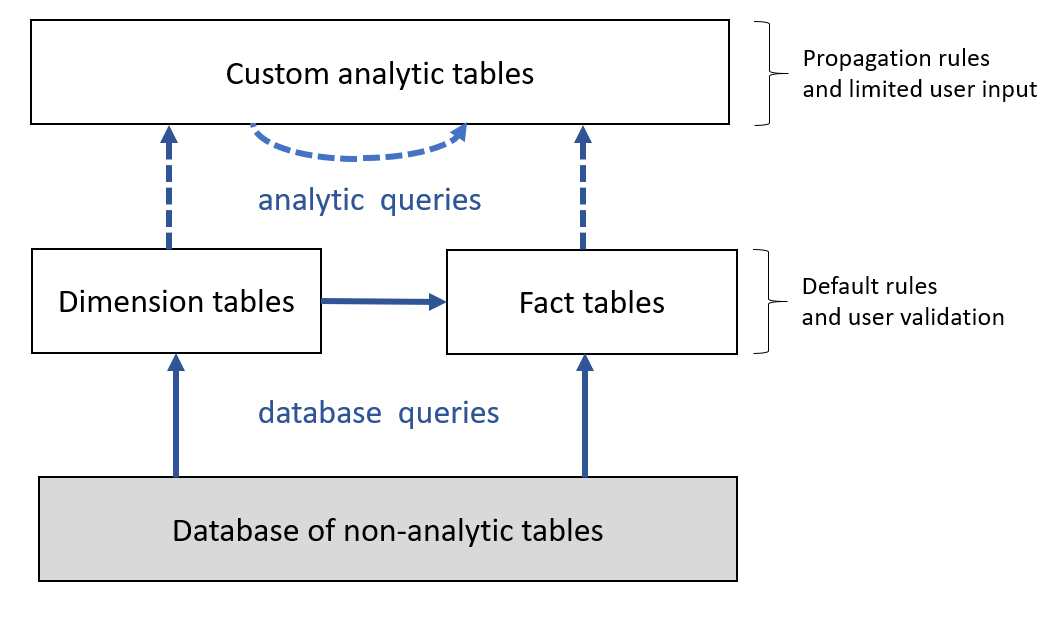}
\caption{Definition of aggregable properties}
\label{fig:agg-prop-process}
\end{figure}

We wish to limit the effort required from the designers of analytic tables, regarding the verification and possible correction of default rules, to the case of analytic tables that are defined from non-analytic tables. 
As shown in \Cref{fig:agg-prop-process}, this corresponds to dimension tables built from non-analytic tables, or fact tables built from dimension tables and non-analytic tables, using database queries (represented by bold arrows). These are the tables over which all other custom analytic tables are built, using self-service data preparation and BI tools.

For analytic tables that result from (analytic) queries  over analytic tables with aggregable properties    (represented by dashed arrows in \Cref{fig:agg-prop-process}), 
the following two sections  present \emph{propagation rules} to obtain the aggregable properties of their attributes. In most of the cases, these rules do not require any user input. 

\mysubsubsection{Propagating aggregable properties to the results of unary operations}
\label{sec:prop-aggprop-red}
 
To determine the aggregable property of some attribute $\A'$ in the result $\RRT=\Q(\XXT)$ of a  query $\Q$ over $\XXT$, we must first identify the aggregate functions which are applicable to $\A'$. This falls into one of the following cases:
\begin{enumerate}
    \item\label{it:propag_filter}If $\A'$ is also an attribute of $\XXT$ and $\AGGF$ is applicable to $\A'$ in $\XXT$ then $\AGGF$ is also applicable to $\A'$ in $\RRT$.
    \item\label{it:propag_pivot} If $\A'$ holds pivoted values of an attribute $\A$ of $\XXT$ and $\AGGF$ is applicable to $\A$ in $\XXT$, then $\AGGF$ is also applicable to $\A'$ in $\RRT$. 
    \item\label{propag_2} If $\A' = \AGGF(\A)$ is the result of applying some aggregation function $\AGGF$ over an attribute $\A$ in $\XXT$, then the aggregate functions that are applicable to $\A'$ are determined by the co-domain category of function $\AGGF$ using \Cref{tab:codomain_agg}.
    \item\label{it:project} If $\A'$ is a new attribute resulting from the evaluation of an expression $f(\Z) \rightarrow A'$ in $Q$, then the aggregation functions that are applicable to $\A'$ are determined by the category of $\A'$ (default or user-defined) using \Cref{tab:cat_agg}. 
\end{enumerate}

\begin{table}[htb]
\centering
\caption{Domain and co-domain categories for common SQL aggregation functions}
\label{tab:codomain_agg}
\normalsize{
\bgroup
\def\arraystretch{1.2}
\begin{tabular}{l|l|l}
    \toprule
    Functions & Domain category & Co-domain category \\
    \midrule
    $\SUM, \MIN, \MAX$ &  $\NUM$ & $\NUM$ \\
    $\MIN, \MAX$ &  $\STAT$ & $\STAT$ \\  
    $\COUNT, \COUNTDISTINCT$ & $\NUM, \DESC, \STAT$ & $\NUM$ \\
    $\AVG$ & $\NUM$ & $\STAT$\\
    \bottomrule
\end{tabular}
\egroup
}
\end{table}

Filter and pivot queries do not change the category of aggregable attributes of $\XXT$  that are in the result $\RRT$. Therefore, all functions that were applicable for attributes in $\XXT$, are still applicable to these attributes in the result of any filter or pivot query over $\XXT$. This is not true for aggregate and projection queries which might generate new attribute values of a different category than the aggregated  or projected attributes by applying a function. For example, while an attribute $\A$ of category $\NUM$ in $\XXT$ is still of category $\NUM$ in $\RRT$ when $\AGGF=\SUM(\A)$, the resulting attribute becomes of category $\STAT$ when $\AGGF=\AVG(A)$. This change is detected using the classification in \Cref{tab:codomain_agg}. 

The identification of all aggregation functions $\AGGF$ that are applicable to an attribute $\A'$ in the result $\RRT$ of a query is not sufficient for defining the aggregable properties of  $\A'$ that hold in $\RRT$.
We must also determine for each attribute $\A'$, the maximal subset of dimension attributes 
$\X'$ of $\RRT$ 
along which an aggregation using $\AGGF$ is correct. 
The propagation rules in \Cref{tab:prop_rules_unary} for determining $\X'$,
can be applied when $\A'$ is an attribute in the result of a filter, projection, pivot or aggregate query. The last column shows how to determine the new determinant  $\X'_d$ (for measure attributes) and the new forbidden attribute set $\X_f$ (for dimension and measure attributes) as well as the required user actions displayed in italics font (\emph{None} means no action required).

\begin{table}[htb]
\caption{Propagation rules for unary operations on $\XXT(S)$ returning table $\RRT(\S')$}
\label{tab:prop_rules_unary}
\normalsize{
\bgroup
\begin{normalsize}
\def\arraystretch{1.2}
\begin{tabular}{l|l|l}
    \toprule
    Unary query on $\XXT(\S)$ & 
    \makecell[tl]{Propagation rule for inferring the aggregable   properties of  attribute  \\ $\A'\in \RRS$  in the result $\RRT(\RRS)$ 
    }& 
    \makecell[tl]{
    \emph{User action}} \\
    \midrule
    $\FILTERQ_{\XXT}(P \mid \Y)$ 
    &  
    \makecell[tl]{
    attribute $\A' \in \RRS$ and $\aggp_{\A'}(\AGGF, \X)$ holds in $\XXT$:   \\
    \hspace{3mm} $\aggp_{\A'}(\AGGF, \X')$ holds in $\RRT$ with $\X'=\X$, $\X'_d=\X_d$ and $\X'_f=\X_f$.}  
    & 
    \makecell[tl]{
     \emph{None}}
    \\
    \midrule
    $\PROJECTQ_{\XXT}(\Y,f(\Z)\rightarrow \M)$ 
    & 
    \makecell[tl]{dimension attribute $\A' \in \Y$ and $\aggp_{\A'}(\AGGF, \X)$  holds in $\XXT$: \\
     \hspace{3mm} $\aggp_{\A'}(\AGGF, \X')$  holds in $\RRT$  with $\X'=\X$, $\X'_d=\X_d$ and $\X'_f=\X_f$.}
    & 
    \makecell[tl]{
    \emph{None}     }
    \\
    \cline{2-3}
    & \makecell[tl]{
    new measure attribute $\A'=\M$: \\
        \hspace{3mm} if $\AGGG$ can be applied on $\M$ as defined in \Cref{tab:cat_agg}\\
        \hspace{3mm} then $\aggp_{\A'}(\AGGG, \X')$  holds in $\RRT$
            where $\X'$ is defined by the  rules of 
         \\ \hspace{9mm} \Cref{tab:default_agg_prop} with 
          $\X'_d=$ determinant of $\Z$ and   $\X'_f=\emptyset$.
         }
    & 
    \makecell[tl]{
   \\\ \\\emph{Minimize $\X'_d$} \\
   \emph{Complete $\X'_f$ }}
     \\
    \midrule
    $\PIVOTQ_{\XXT}(\A \mid \Y)$ 
    & 
    \makecell[tl]{
    attribute $\A'\in \RRS-\{\A\}$ and $\aggp_{\A'}(\AGGF, \X)$ holds  in $\XXT$: \\
    \hspace{3mm} if $\X_d\cap\Y=\emptyset$\\
    \hspace{3mm} then $\aggp_{\A'}(\AGGF, \X')$  holds in $\RRT$ with $\X'=\X-\Y$, $\X'_d=\X_d$  and 
    \\\hspace{9mm}    $\X'_f=\X_f-\Y$.
    }
    & 
    \makecell[tl]{
    \\\ \\\emph{None}}
    \\   \cline{3-3}
    & 
     \makecell[tl]{
     \hspace{3mm} else $\aggp_{\A'}(\AGGF, \X')$  holds in $\RRT$ where $\X'$ is defined by the rules of 
     \\  \hspace{9mm} \Cref{tab:default_agg_prop} with $\X'_d=$ fact identifier and    $\X'_f=\X_f-\Y$.
        }
     &
    \makecell[tl]{ 
    \emph{Minimize $\X'_d$}}
    \\
    \cline{2-3}
    & 
    \makecell[tl]{
    new pivot attribute $\A'\in \RRS$ and $\aggp_{\A}(\AGGF, \X)$ holds in $\XXT$:  \\
     \hspace{3mm} if $\X_d\not\subseteq\Y$ \\
     \hspace{3mm} then $\aggp_{\A'}(\AGGF, \X')$ holds in $\RRT$ with $\X'=\X-\Y$, $\X'_d=\X_d-\Y$    \\
     \hspace{9mm} and $\X'_f=\X_f-\Y$}
    & 
    \makecell[tl]{
    \\\ \\
    \emph{None}}
    \\[8mm]
    \cline{3-3}
    & 
    \makecell[tl]{
     \hspace{3mm} else 
     $\aggp_{\A'}(\AGGF, \X')$ holds in $\RRT$ 
     where $\X'$ is defined by the  rules of\\
     \hspace{9mm} \Cref{tab:default_agg_prop} with $\X'_d$ = fact identifier and     $\X'_f=\X_f-\Y$
     }
    & 
    \makecell[tl]{
    \emph{Minimize $\X'_d$}}
    \\
    \midrule
    $\AGGQ_{\XXT}(\AGGF(\A) \mid \Y)$ 
    & 
    \makecell[tl]{
    dimension attribute $\A' \in \Y$ and $\aggp_{\A'}(\AGGF, \X)$ holds in $\XXT$:  \\
     \hspace{3mm} $\aggp_{\A'}(\AGGF,\X')$ holds in $\RRT$ with $\X'= \X \cap \Y$ and $\X'_f=\X_f\cap \Y$} 
     & 
     \makecell[tl]{ 
     \emph{None}}
     \\
    \cline{2-3}
    & 
    \makecell[tl]{new measure attribute $\A'=\AGGF(\A)$ and $\aggp_{\A}(\AGGF, \X)$ holds in $\XXT$:  \\
     \hspace{3mm} if $\AGGG$ can be applied on $\A'$ as defined in \Cref{tab:cat_agg}\\ 
     \hspace{3mm} then $\aggp_{\AGGF(\A)}(\AGGG, \X')$ holds in $\RRT$ with $\X'=\Y$, $\X'_d=$ fact identifier \\
     \hspace{9mm} and     $\X'_f=\emptyset$.
     }
    &  
     \makecell[tl]{
     \\\ \\\emph{Minimize $\X'_d$}
     \\
     \emph{Complete $\X'_f$}} \\
    \bottomrule
\end{tabular}
\end{normalsize}
\egroup
}
\end{table}


\begin{proposition}[Propagation rules for filter, project and pivot]
\label{def:propagate-fil-piv-agg-prop}
Let $\RRT(\RRS)=\Q(\XXT)$ be the result of a filter, project or pivot query $\Q$ over an analytic table $\XXT(\S)$, $\A'$ be an attribute of $\RRT$, and $\S_D$ be the set of dimension attributes in $\XXT$. 
Then the propagation rules of \Cref{tab:prop_rules_unary} for filter, project and pivot are correct. 
\end{proposition}
\begin{myproof}
Suppose that whenever $\A' \in T$ then $\aggp_{\A'}(\AGGF, \X)$ holds in $\XXT$, for $\X \subseteq \S_D$.
\begin{enumerate}
    \item \label{rule:filter}
Filter queries:  Let $\RRT = \FILTERQ_{\XXT}(P \mid \Y)$. Then $\RRT$ is a subset of $\XXT$ and all conditions, and in particular the literal functional dependencies, in the \Cref{def:aggprop} of for $\aggp_{\A'}(\AGGF, \X)$ still hold for $\A'$ and we obtain $\X'_d=\X_d$ and $\X'_f=\X_f$. 

\item \label{rule:projection}
Projection queries: Let $\RRT=\PROJECTQ_{\XXT}(\Y,f(\Z)\rightarrow \M)$. 

\begin{itemize}
    \item $\A'\in\Y$: Since, by definition of projection,  $\Y$ contains all dimension attributes of $\XXT$, $\RRT$ also contains all dimension attributes of $\XXT$ (and possibly some other measure attributes). Therefore,  all conditions in \Cref{def:aggprop}  still hold for all measure and dimension attributes $\A' \in \Y$ and we obtain $\X'_d=\X_d$ and $\X'_f=\X_f$. 
   \item For new measure attribute $\M$,    we have to show that $\X'_d$ must be a determinant of $\M$: Since $\X'_d$ is a determinant of $\Z$ and $\M$ is the result of a function applied to attributes $\Z$, by transitivity, $\X'_d$  is also a determinant of $\M$.
\end{itemize}

\item \label{rule:pivot} 
Pivot queries: Let $\RRT =  \PIVOTQ_\XXT(\A \mid \Y)$. 
\begin{itemize}
    \item $\A'$ in $\S-\Y-\{\A\}$ and $\aggp_{\A'}(\AGGF, \X)$ holds in $\XXT$: If $\X_d\cap\Y=\emptyset$, by definition of pivot, each tuple $t\in \XXT$ is mapped to a tuple $t'\in \RRT$ where $t.(\X_d\cup\A')=t'.(\X_d\cup\A')$ and we obtain $\X'_d=\X_d$ is a determinant of $\A'$. If $\X_d\cap\Y\not=\emptyset$, we cannot conclude that the remaining attribute set in $\X_d-\Y$ is still a determinant of $\A'$ and we must apply the default rules to find $\X'_d$. However, observe that $\A'$ existed in the input table and we can conclude that all remaining forbidden attributes $\X'_f=\X_f-\Y$ are still forbidden.

\item  $\A'$ is a new attribute that holds pivoted values of $\A$  and $\aggp_{\A}(\AGGF, \X)$ holds in $\XXT$: 
If $\X_d$ is a determinant of $\A$ in $\XXT$ and $\X_d\not\subseteq\Y$, we can conclude that the "remaining" attributes $\X'_d=\X_d-\Y$ are a determinant of $\A'$. Suppose that there are two tuples $t_1$ and $t_2$ in $\RRT$ which have the same value for $\X_d-\Y$, but different values for attribute $\A'$: $t_1.\A'\neq t_2.\A'$. By definition of pivot, these two tuples are the result of two distinct tuples $t'_1$ and $t'_2$ in $\XXT$ where  $t_1.\A'=t'_1.\A'$, $t_2.\A'=t'_2.\A'$, $t'_1.\Y=t'_2.\Y$,  and $t'_1.(\X_d-\Y)=t'_2.(\X_d-\Y)$. We get $t.\X_d= t'.\X_d$ and $t'_1.\A'\neq t'_2.\A'$ which is in contradiction with $\X_d$ is a determinant of $\A'$. 
 If $\X_d\subseteq\Y$ we have to recompute the determinant set of all pivoted attributes. All attributes which were forbidden for $\A$ are also forbidden for $\A'$ and we obtain $\X'_f=\X_f-\Y$.
\end{itemize}

\end{enumerate}
\end{myproof}

We next define the following proposition for attributes in the result of an aggregate query. 
\begin{proposition}[Propagation rule for aggregation]
\label{def:propagate-agg-prop-try}
Let $\XXT(\S)$ be an analytic table with dimension attributes $\S_D\subseteq \S$, and $\aggp_{\A}(\AGGF, \X)$ be an aggregable property that holds in $\XXT$ with determinant $\X_d$ and forbidden set $\X_f$.
Let $\RRT=\AGGQ_{\XXT}(\AGGF(\A) \mid \Y)$ be a valid aggregate query (i.e., $\S_D - \X \subseteq \Y$).
Then the propagation rule of \Cref{tab:prop_rules_unary} for aggregation is correct. 
\end{proposition}
\begin{myproof}
We prove each case of attribute $\A'$:
\begin{enumerate}
    \item Every attribute $ \A' \in \Y$ is a dimension attribute in $\XXT$ and $\RRT$. For every aggregable property  $\aggp_{\A'}(\AGGF, \X)$ that holds in $\XXT$,  we must only  determine $\X'$ and $\X'_f$ for $\aggp_{\A'}(\AGGF, \X')$ in $\RRT$.  Since $\A'$ is aggregable along $\X$, it is also aggregable along the subset of remaining attributes $\X' = \X \cap \Y$ and all remaining forbidden attributes $\X'_f=\X_f\cap \Y$ are still forbidden.  We conclude that $\aggp_{\A'}(\AGGF, \X')$ holds in $\RRT$ where $\X' = \X \cap \Y$ and  $\X'_f = \X_f \cap \Y$.
    
    \item  We show that $\aggp_{\AGGF(\A)}(\AGGG, \X')$ holds for new attribute $\AGGF(\A)$ in $\RRT$ with $\X' = \X_d^{'+}-\X'_f=\Y$. By the assumption in \Cref{it:applicable} of \Cref{def:aggprop}, $\AGGG$ is applicable to $\AGGF(\A)$.
    By applying the default rules of \Cref{tab:default_agg_prop} $\X'_d$ is the fact identifier of $\RRT$ and determines $\AGGF(\A)$ as well as all attributes in $\Y$ ($\Y=\X_d^{'+}$ is the closure of $\X_d$). $\X'_f$ is by default empty. $\X'_d$ and $\X'_f$ and must be validated by the user by removing incorrect attributes from $\X'_d$ and adding meaningless attributes to $\X_f$.

\end{enumerate}
\end{myproof}

\begin{myexample}
\label{ex:prop-filter-pivot}
\label{exp:summarizable-att}

Consider fact table \factt{PRODUCT\_LIST} in \Cref{tab:prod-list2-repeat3} and attribute \attr{QTY}. As seen in \Cref{exp:aggproperty-coca}, 
$\aggp_{\attr{QTY}}(\SUM \mid \X)$ holds in \factt{PRODUCT\_LIST} for $\X = \{\attr{PROD\_SKU}, \attr{YEAR}\}$, $\X_d=\{\attr{PROD\_SKU}, \attr{YEAR}\}$ and $\X_f=\emptyset$. 

\begin{table}[htb]
    \caption{Table \factt{PRODUCT\_LIST}}
    \label{tab:prod-list2-repeat3}
    \centering
    \begin{tabular}{l|l|l|l|r}
        \toprule
         \attr{PROD\_SKU} & \attr{BRAND}& \attr{COUNTRY} & \attr{YEAR} & \attr{QTY} \\
         \midrule
         cz-tshirt-s & Coco Cola & USA & 2017 & 5 000 \\
         cz-tshirt-s & Coco Cola & USA & 2018 & 7 000 \\         
         cz-tshirt-s & Zora & Spain & 2017 & 5 000 \\
         cz-tshirt-s & Zora & Spain & 2018 & 7 000 \\
         coco-can-33cl & Coco Cola & USA & 2017 & 10 000 \\
         \bottomrule
    \end{tabular}
\end{table}

First, let $\RRT =  \FILTERQ_{\factt{PRODUCT\_LIST}}(\{\attr{YEAR} = \text{\enquote*{2017}}\})$. By \Cref{tab:prop_rules_unary}, 
$\aggp_{\attr{QTY}}(\SUM \mid \X)$  still holds in $\RRT$. 
Next, let $\RRT = \PIVOTQ_{\factt{PRODUCT\_LIST}}(\attr{QTY} \mid \attr{BRAND})$ be a query producing two new attributes $\attr{QTY\_CocoCola}$ and $\attr{QTY\_Zora}$ with values from attribute $\attr{QTY}$. 
Then, by \Cref{tab:prop_rules_unary}, since $\X_\D \not \subseteq Y$, 
both aggregable properties $\aggp_{\attr{QTY\_CocoCola}}(\SUM \mid \X')$ and $\aggp_{\attr{QTY\_Zora}}(\SUM \mid \X')$  hold 
in $\RRT$ where $\X' = \X - \{\attr{BRAND}\} = \X=\{\attr{PROD\_SKU}, \attr{YEAR}\}$. 
Finally, let $\RRT=\AGGQ_{\factt{PRODUCT\_LIST}}(\SUM(\attr{QTY}) \mid \Y)$ with $\Y=\{\attr{BRAND}, \attr{YEAR}\}$. Function $\SUM$ returns a value of category $\NUM$, 
so by \Cref{tab:prop_rules_unary}, 
$\aggp_{\SUM(\attr{QTY})}(\AGGG \mid \X')$ holds in table $\RRT$ with $\X'= \Y = \{\attr{BRAND}, \attr{YEAR}\}$  and $\AGGG\in \{\SUM, \AVG, \allowbreak \COUNT, \allowbreak \COUNTDISTINCT, \MIN, \allowbreak \MAX\}$. By default, $\X'_d=\{\attr{BRAND}, \attr{YEAR}\}$ is the fact identifier of $\RRT$ and $\X_f$ is empty.

Let us now consider attribute \attr{PROD\_SKU}. As seen in \Cref{exp:aggproperty-coca},   properties $\aggp_{\attr{PROD\_SKU}}(\COUNT \mid \X)$ and  $\aggp_{\attr{PROD\_SKU}}(\COUNTDISTINCT \mid \X)$ hold in \factt{PRODUCT\_LIST} for $\X = \{\attr{BRAND}, \attr{COUNTRY}, \attr{YEAR}\}$. 
Let $\RRT=\AGGQ_{\factt{PRODUCT\_LIST}}(\AGGF(\attr{PROD\_SKU}) \mid \{\attr{BRAND}, \attr{YEAR}\})$, with $\AGGF=\COUNT$ or $\AGGF=\COUNTDISTINCT$. 
 
Both of these functions return values of category $\NUM$. 
So by \Cref{tab:prop_rules_unary}, 
$\aggp_{\AGGF(\attr{PROD\_SKU})}(\AGGG \mid \X')$ holds in table $\RRT$ for $\AGGG\in \{\SUM, \AVG, \COUNT, \COUNTDISTINCT,\allowbreak \MIN,\allowbreak \MAX\}$ and $\X'= \X \cap \{\attr{BRAND}, \attr{YEAR}\} = \{\attr{BRAND}, \attr{YEAR}\}$. 
\end{myexample}

\mysubsubsection{Propagating aggregable properties to 
the results of binary operations}
\label{sec:prop-aggprop-natcomp}

We now consider the problem of determining the aggregable properties of the attributes in the result of binary merge queries and binary set queries (union, difference). The propagation rules are summarized in \Cref{tab:prop_rules_binary}

\begin{table}[htb]
\centering
\caption{Propagation rules for binary operations}
\label{tab:prop_rules_binary}
\bgroup
\def\arraystretch{1.5}
\begin{normalsize}
\begin{tabular}{l|l|l}
    \toprule
    \makecell[tl]{
    Binary query on\\
    $\XXT(\XXS)$ and $\YYT(\YYS)$
    }
    & \makecell[tl]{Propagation rule for inferring the aggregable properties of attribute $\A'\in \XXS$ \\  in the result $\RRT(\S_r)$
    }& \makecell[tl]{
    \emph{User action}
    } \\
    \midrule
    \makecell[tl]{
    $\RRT = \XXT \leftouterjoin_{\Y} \YYT$\\
    $\RRT = \XXT \rightouterjoin_{\Y} \YYT$}
    & 
    \makecell[tl]{
        dimension attribute 
        $\A'\in \XXS_D$
        and  $\aggp_{\A'}(\AGGF, \X)$ holds in $\XXT$: \\ 
        \hspace{3mm} 
        $\aggp_{\A'}(\AGGF, \X')$ holds in $\RRT$ 
        with $\X'=\X\cup(\YYS_D-\Y) - \X'_f$ and $\X'_f=\X_f$.}
     & \makecell[tl]{
     \emph{Complete $\X'_f$}}
    \\
    \cline{2-3}
     \makecell[tl]{
     $\RRT = \XXT \fullouterjoin_{\Y} \YYT$\\
    $\RRT = \XXT \bowtie_{\Y} \YYT$}
    &
    \makecell[tl]{
        measure attribute 
        $\A'\in \XXS-\XXS_D$ 
        and $\aggp_{\A'}(\AGGF, \X)$  holds in $\XXT$  : \\
        \hspace{3mm} $\aggp_{\A'}(\AGGF, \X')$ holds in $\RRT$ 
        with $\X'=\X$, $\X'_d=\X_d$ and $\X'_f=\X_f$
        } 
     & 
     \makecell[tl]{
     \emph{Complete $\X'_f$}}\\
    \midrule
    \makecell[tl]{ 
    $\RRT=\XXT \cup \YYT$}
    & \makecell[tl]{ 
        dimension attr. $\A' \in \RRS$ 
        and $\aggp_{\A'}(\AGGF, \X)$  holds in $\XXT$ and $\YYT$  \\
        \hspace{3mm} $\aggp_{\A'}(\AGGF, \X')$ holds in $\RRT$ with $\X'=\X$, $\X'_d=\X_d$ and $\X'_f=\X_f$.}
    & \makecell[tl]{
    \emph{None}}
    \\[5mm]
    \cline{2-3}    
    & \makecell[tl]{ 
         measure attr. $\A' \in \RRS$ 
        and $\aggp_{\A'}(\AGGF, \X)$ holds in $\XXT$ and $\YYT$:  \\
        \hspace{3mm} if $\X_d \mapsto \A'$ holds in $\RRT$ \\
        \hspace{3mm} then $\aggp_{\A'}(\AGGF, \X')$ holds in $\RRT$ with $\X'=\X$, $\X'_d=\X_d$ and $\X'_f=\X_f$. 
        }
    & \makecell[tl]{
    \\\ 
    \emph{None}}
    \\
    \cline{3-3}    
    & \makecell[tl]{ 
        \hspace{3mm} else $\aggp_{\A'}(\AGGF, \X')$ holds in $\RRT$   where $\X'$ is defined by the  rules of
        \\ \hspace{9mm} 
        \Cref{tab:default_agg_prop} with $\X'_d=$ fact identifier 
     and     $\X'_f=\X_f$.
     }
    & \makecell[tl]{
     \emph{Minimize $\X'_d$}} \\    
    \midrule
    \makecell[tl]{ 
    $\RRT=\XXT - \YYT$}
    & \makecell[tl]{ 
    attribute $\A' \in \RRS$ 
        and $\aggp_{\A'}(\AGGF, \X)$ holds in $\XXT$ and $\YYT$ :  \\
        \hspace{3mm} 
        $\aggp_{\A'}(\AGGF, \X')$ holds in $\RRT$ with $\X'= X^{'+}_d - X'_f$         where $\X^{'+}_d$ is the set of 
        \\    \hspace{6mm}
        attributes in $\XXS_D\cup \YYS_D$ literally determined  by $\X'_d$ with $\X'_d=\X_d$
        \\  \hspace{6mm}
           and $\X'_f=\X_f$.}
        
    & \makecell[tl]{
    \emph{None}} \\
    
    \bottomrule
\end{tabular}
\end{normalsize}
\egroup
\end{table}

\begin{proposition}[propagation rule for merge]
\label{prop:propagation_aggregable_merge}
Let $\XXT(\XXS)$ and $\YYT(\YYS)$ be two analytic tables with dimension attributes $\XXS_D\subseteq \XXS$ and $\YYS_D\subseteq \YYS$ respectively.
Let $\RRT(\RRS)$ be the result of a merge query between  $\XXT$ and $\YYT$ over a set of common dimension attributes $\Y \subseteq \XXS_D\cap\YYS_D$ and let $\S^r_D=\XXS_D\cup \YYS_D$ be the dimension attributes in $\RRT$.  
Let $\A' \in \RRS$ be an attribute of $\XXT$ with aggregable property $\aggp_{\A'} (\AGGF,\X)$ holding in $\XXT$. 

Then the propagation rules of \Cref{tab:prop_rules_binary} for merge queries are correct. 

\end{proposition}

\begin{myproof} 
We proceed with each case of attribute $\A'$:

\begin{enumerate}
\item If $\A'$ is a dimension attribute, by \Cref{def:aggprop},  
$\X=\XXS_D -\{\A'\}-\X_f$ and since $\{\A'\} \cup \X_f \subseteq \XXS_D$ (all forbidden attributes are in $\XXS_D$), 
we can add all attributes of $\YYS_D$ which are not in $\XXS_D$ to $\X'$:  $\X' = \X \cup (\YYS_D -\Y)$.  
The user must add all new meaningless attributes in $\YYS_D$ to $\X'_{f}=\X_f$. 

\item If $\A'$ is a measure attribute,
$\A'$ is an attribute of table $\XXT$  but not of table $\YYT$. Let $\X = \X_d^+ - \X_f$, as in \Cref{def:aggprop}. 
We show that the LFD $\X_d \mapsto \A'$ is still valid in $\RRT$. Suppose that there exist two tuples $t$ and $t'$ in $\RRT$ such that $t.X_d \equiv t'.X_d$ and $t.\A' \not \equiv t'.\A'$. 
We show that, for each merge operation, the projection of these two tuples on $\XXS$ would also exist in $\XXT$, which contradicts that LFD $\X_d \mapsto \A'$ holds in $\XXT$: 
\begin{itemize}
    \item If $\RRT = \XXT \leftouterjoin_{\Y} \YYT$ (left merge): 
    by definition of $\leftouterjoin_{\Y}$ any projection on $\XXS$ of a tuple $\XXT$ in $\RRT$ is also a tuple in $\XXT$ (similar to filter queries). 
    \item If $\RRT = \XXT \rightouterjoin_{\Y} \YYT$ (right merge):
    any projection on $\XXS$ of a tuple $\XXT$ in $\RRT$ is either a tuple in $\XXT$ or a tuple that does not exist in $\XXT$ and has a null value for each attribute in $\XXS - \Y$. In the latter case, the LFD is preserved in $\RRT$ because all these tuples also have a null value on $\A$ and there is no tuple in $\XXT$ that has null values on $\X_d$ and a non-null value for $\A$.  
    \item If $\RRT = \XXT \fullouterjoin_{\Y} \YYT$ (full merge): The proof as a combination of the proofs for left merge and right merge (any projection on $\XXS$ of a tuple $\XXT$ in $\RRT$ is either a tuple in $\XXT$ or a tuple that does not exist in $\XXT$ and has a null value for each attribute in $\XXS - \Y$). 
    \item If $\RRT = \XXT \bowtie_{\Y} \YYT$ (strict merge): 
    any projection on $\XXS$ of a tuple $\XXT$ in $\RRT$ is also a tuple in $\XXT$.
\end{itemize}

Then $\X_d$ is still a minimum set of dimension attributes in $\RRT$ which literally determines $\A'$ (otherwise it would not be minimum in $\XXT$) and, by \Cref{def:aggprop},  $\X'$ contains all attributes determined by $\X_d$ in $\RRT$ (closure of $\X_d$ in $\RRT$). Similarly, all meaningless attributes in $\X_f$ remain meaningless and the user can add new meaningless attributes from $\YYS-\Y$. 
\end{enumerate}
\end{myproof}

In \Cref{tab:prop_rules_binary}, in the case of a merge query, 
$\X'_f = \X_f$ by default 
and for a measure attribute, the attribute graphs of each dimension are used to compute the closure $\X_d^+$ over $\YYS_D$.
In the propagation rule, we only considered the case of an attribute $\A$ in $\XXT$, but the same result would apply for an attribute $\A$ in $\YYT$ due to the symmetry of the merge operations. 

\begin{myexample}
In \Cref{tab:propagate_merge}, 
table \factt{STORE\_SALES\_YEARLY} 
is defined over dimensions \dimt{SALESORG} and \dimt{TIME}, table \factt{DEM2} is defined over \dimt{REGION} (see \Cref{tab:region}) and \dimt{TIME}, and  
table \factt{STORE\_SALES\_DEM} is the result of the left-merge query:
\begin{align*}
\factt{STORE\_SALES\_DEM}=\factt{STORE\_SALES\_YEARLY}\leftouterjoin_{\attr{CITY}, \attr{YEAR}} \factt{DEM2}
\end{align*}

\sloppy
We first consider attribute \attr{Amount}. Suppose that $\aggp_{AMOUNT}(\SUM, \X)$ holds in \factt{STORE\_SALES\_YEARLY} for $\X=\{\attr{Store\_Id},\allowbreak \attr{City},  \attr{Year}\}$, with $\X_d=\{\attr{Store\_Id}, \attr{Year}\}$  
and $\X_f=\emptyset$.
Then, 
$\Z= \{\attr{CITY}, \allowbreak  \attr{STATE}, \attr{COUNTRY}\}$ is the set of attributes determined by $\X_d$ in \factt{DEM2}. 
Then, $\X^{'+}_d= \X^+_d \cup \Z = \{\attr{Store\_Id}, \attr{Year}, \attr{CITY}, \attr{STATE}, \attr{COUNTRY}\}$ is the set of attributes determined by $\X_d$ in \factt{STORE\_SALES\_DEM}.
So, by the propagation rule of \Cref{tab:prop_rules_binary}, 
aggregable property $\aggp_{\attr{AMOUNT}}(\SUM, \X')$ holds in \factt{STORE\_SALES\_DEM} where $\X'= \X_d^{'+}-\X'_f = \X_d^{'+}$.

\begin{table}[htbp]
\caption{Example of a left-merge query}
\label{tab:propagate_merge}
\normalsize{
    \begin{subtable}[b]{.49\linewidth}
    \centering
     \caption{\factt{STORE\_SALES\_YEARLY}}
    \label{tab:store_sales_yearly}
    \begin{tabular}[t]{l|l|l|r}
      \toprule
    \attr{Store\_Id} & \attr{City} &  \attr{Year} & \attr{Amount} \\
    \midrule
    Oh\_01 & Dublin &  2017 & 3.2 \\
    Ca\_01 & Dublin &  2017 & 5.3 \\
    Oh\_01 & Dublin &  2018 & 8.2 \\
    Ca\_01 & Dublin &  2018 & 6.3 \\
    Pa\_01 & Paris &  2017 & 45.1 \\
      \bottomrule
    \end{tabular}
\end{subtable}
\hfill
\begin{subtable}[b]{0.5\linewidth}
\centering
\caption{\factt{DEM2}}
\label{tab:region2}
    \begin{tabular}[t]{l|l|l|l|r}
      \toprule
       \attr{City} & \attr{State} & \attr{Country} & \attr{YEAR} & \attr{UNEMP} \\
    \midrule
      Dublin & Ohio & USA & 2017 & 4.2 \\
      Dublin & California & USA & 2017 & 3.1 \\
      Palo Alto & California & USA & 2017 & 2.1 \\
      Paris & - & France & 2017 & 11.9\\
      Dublin & - & Ireland & 2017 & 6.7 \\
      \bottomrule
    \end{tabular}
\end{subtable}

\vspace{5mm}
\begin{subtable}[b]{.8\linewidth}
     \centering
      \caption{\factt{STORE\_SALES\_DEM} (left merge of \factt{STORE\_SALES\_YEARLY} and \factt{DEM2})}
     \label{tab:store_sales_dem}
     \begin{tabular}[t]{l|l|l|l|l|l|l|r}
       \toprule
     \attr{Store\_Id} & \attr{City} & \attr{State} & \attr{Country} & \attr{Year} & \attr{Amount} & \attr{UNEMP} \\
     \midrule
     Oh\_01 & Dublin & Ohio & USA  & 2017 & 3.2 & 4.2 \\
     Oh\_01 & Dublin & California & USA  & 2017 & 3.2 & 3.1 \\
     Oh\_01 & Dublin & - & Ireland & 2017 & 3.2 & 6.7 \\
     Ca\_01 & Dublin & California & USA  & 2017 & 5.3 & 3.1 \\
     Ca\_01 & Dublin & Ohio & USA  & 2017 & 5.3 & 4.2 \\
     Ca\_01 & Dublin & - & Ireland & 2017 & 5.3 & 6.7 \\
     Oh\_01 & Dublin & - & -  & 2018 & 8.2 & - \\
     Ca\_01 & Dublin & - & -  & 2018 & 6.3 & - \\
     Pa\_01 & Paris & - & France & 2017 & 45.1 & 11.9 \\
       \bottomrule
     \end{tabular}
     \end{subtable}
 \hfill 
}
\end{table}

Observe that the aggregable property computed by the propagation rule for merge queries does not guarantee that any  aggregation query  of \attr{AMOUNT} produces the same result when it is applied on \factt{STORE\_SALES\_YEARLY}  or on \factt{STORE\_SALES\_DEM}. We will show in \Cref{sec:prop_summarize} how to refine the propagation rule to guarantee this summarizability property.  

We now consider attribute \attr{UNEMP} in \factt{DEM2} of category $\STAT$. Suppose that  $\aggp_{\attr{UNEMP}}(\AGGF, \X)$ holds in \factt{DEM2} for $\X =  \{\attr{Year},\allowbreak \attr{CITY}, \allowbreak \attr{STATE}, \allowbreak \attr{COUNTRY}\})$ and $\AGGF \in \{\COUNT, \COUNTDISTINCT, \MIN, \MAX \}$ with $\X_d = X$.
We use the propagation rule of \Cref{tab:prop_rules_binary} on the (equivalent) right-merge of \factt{DEM2} with \factt{STORE\_SALES\_YEARLY}. No new attribute of \factt{STORE\_SALES\_YEARLY} is determined by $\X_d$, so $\X_d^{'+}=\X_d$. Hence, by the propagation rule, $\aggp_{\attr{UNEMP}}(\AGGF, \X')$ holds in \factt{STORE\_SALES\_DEM} for $\X'= \X_d^{'+} - \X'_f = \X_d = \X$.
\end{myexample}

The next proposition states which aggregable properties hold for attribute $\A'$ in the union  $\RRT=\XXT\cup \YYT$ and the set difference $\RRT=\XXT- \YYT$, knowing the aggregable properties of $\A'$ in tables $\XXT$ and $\YYT$. 


\begin{proposition}[propagation rules for union and difference]
\label{prop:propagation_aggregable_set}
Let 
$\XXT(\XXS)$ and  $\YYT(\XXS)$ be two tables over the same schema $\XXS$ having a set of dimension attributes $\X$. Let $\attr{\A'} \in \XXS$ be an attribute with aggregable property $\aggp_{\A'} (\AGGF,\X)$ \emph{holding in both tables $\XXT$ and $\YYT$}. 
Then the propagation rules of \Cref{tab:prop_rules_binary} for set union and difference are correct. 
\end{proposition}

\begin{myproof}
We distinguish each case where $\A'$ is a dimension attribute or a measure attribute in $\RRT(\RRS)$:
\begin{enumerate}
    \item If $\A'$ is a dimension attribute: let $\X' = \S_D - \{A\} - \X_f$, where $\S_D$ is the set of dimension attributes in $\S$. Since $\RRS = \S$ for difference and union, and $\X_f$ is defined for a given set of attributes (independently of a specific table), $\X'$ does not change in the aggregable property of $\RRT$ for $\A'$.
    \item If $\A'$ is a measure attribute: let $\Z = \X_d^+ - \X_f$. Then $\X_d \mapsto \A'$ in each table $\XXT$ and $\YYT$. We analyze each query case:
    \begin{itemize}
        \item Difference: By definition of analytic difference queries, $\RRT \subseteq \XXT$, thus $\X_d \mapsto \A'$ also holds in $\RRT=\XXT-\YYT$.
        \item Union: 
            there could be two tuples that have the same values on their $\X$ attributes, so we must check that  $\X_d \mapsto A$ holds in $\RRT=\XXT\cup \YYT$. Otherwise, we apply the default rules for initializing $\X'_d$.
    \end{itemize} 
    Finally, $\X_f$ must be the same in $\RRT$ since it only depends on the schema.
\end{enumerate}
\end{myproof}

Propagation rules for all operations, except union, are immediate to compute because they only involve the manipulation of metadata properties. In the case of union, when a measure depends on a subset of the dimensions of a fact table, the propagation rule requires a uniqueness test (to check the LFD $\X_d \mapsto A$) on the result of the union. Since the uniqueness test involves hierarchical dimension attributes only, it can be performed efficiently using specific data structures used to represent fact tables in main-memory (see \cite{brunel_supporting_2015, brunel_index-assisted_2016}). 




\section{Summarizability of aggregable attributes}


\label{sec:summarize-reduction}
In this section, we consider  the properties of attributes that characterize the equivalence between computing an aggregated value of an attribute from a table $\XXT$ and computing the same aggregated value from the result of a query $\Q$ over $\XXT$.
In \Cref{sec:distributive_functions}, we first define the property of \emph{summarizable attributes} in the case when $\Q$ is an aggregate query, which corresponds to the traditional notion of summarizability addressed by previous work. In \Cref{sec:prop_summarize}, we then extend our propagation rules to compute aggregable properties such that aggregate queries can only be expressed over summarizable attributes. Finally, in \Cref{sec:generalized_summarizability}, we introduce the new property of \emph{G-summarizability} of attributes in the case when $\Q$ is any analytic query. 
In \Cref{sec:control-g-sum}, we then again extend our propagation rules to compute  aggregable properties such that aggregation queries are expressed over G-summarizable attributes only.  

\subsection{Summarizable attributes and distributive functions}
\label{sec:distributive_functions}

\begin{figure}[htbp]
\centering
\includegraphics[width=0.5\linewidth]{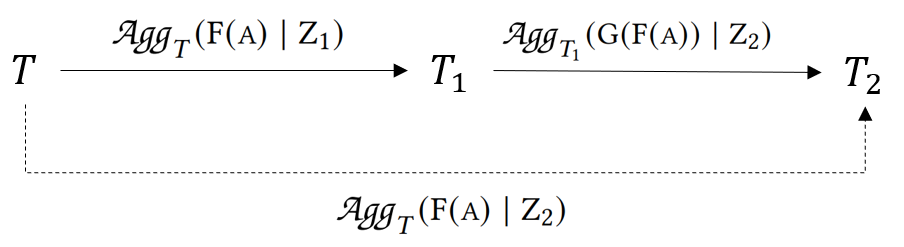}
\caption{Summarizable attribute $A$ with respect to $Z_1$ and $\AGGF$ using $\AGGG$}
\label{fig:summ-att-fig1}
\end{figure}

\Cref{fig:summ-att-fig1} illustrates the definition of summarizable attributes.
It depicts that when some  attribute $\A$ of table $\XXT$ is aggregated with some function $\AGGF$ for each partition of attributes $\Z_2$, it is possible to obtain the same result, by first aggregating $\A$ for each partition of attributes $\Z_1$ where $Z_2 \subset Z_1$, using function $\AGGF$, and then further aggregating $\A$ for each partition $\Z_2$, using either the same function $\AGGF$ or a different function $\AGGG$. We shall say that $\A$ is summarizable with respect to grouping set $\Z_1$ and function $\AGGF$, using function $\AGGG$.

The \Cref{def:summarizable-attr} below formalizes \Cref{fig:summ-att-fig1} and, as shown later in \Cref{sec:related-work}, subsumes the definition of summarizability addressed in previous work. 
For simplicity, we use the expression \emph{valid aggregate query} for a query $\AGGQ_{\XXT}(\AGGF(\A) \mid \Z)$ such that there exists an aggregable property $\aggp_{\A}(\AGGF, \X)$ that holds in $\XXT$, and $Z$ contains the dimension attributes of $\XXT$ that are not in $\X$.

\label{sec:summarize-distributive}
\begin{definition}[Summarizable attribute]
\label{def:summarizable-attr}
Let $\XXT(\S)$ be an analytic table, $\A$ be an aggregable attribute of $\XXT$, and 
$\RT = \AGGQ_{\XXT}(\AGGF(\A) \mid \Z_1)$ be a valid aggregate query over $\XXT$. 
If for \emph{any} subset $\Z_2  \subset \Z_1$, there exists an applicable aggregate function $\AGGG$ such that the equation 
\begin{align}
\label{eq:summarizable-attr}
\AGGQ_{\RT}(\AGGG(\AGGF(\A)) \mid \Z_2) = \AGGQ_{\XXT}(\AGGF(\A) \mid \Z_2)
\end{align}
holds, then $\A$ is said to be \emph{summarizable in $\XXT$ with respect to grouping set $\Z_1$ and function $\AGGF$ using function $\AGGG$}. 
\end{definition}

Attribute summarizability is  related to the notion of distributive aggregation functions, also called decomposable aggregation functions in \cite{decomp-agg-functions}.

\begin{definition}[Distributive aggregation function] 
\label{def:distributivefct}
Let $\AGGF$ be an aggregation function applicable to a set of domain values $\V$ and $P=\{\V_1, \ldots, \V_n\}$, $n\geq 1$ be a partitioning of $\V$. 
If there exists an aggregate function $\AGGG$ such that
$\AGGF(\V_1 \cup \ldots \cup \V_n )=\AGGG(\AGGF(\V_1 )\cup \ldots \cup \AGGF(\V_n))$ then $\AGGF$ is said to be \emph{distributive on partitioning $P$} using function $\AGGG$. 
\end{definition}
If  $\AGGF$ is distributive using function $\AGGG$ on any partitioning of domain $\V$, we say that $\AGGF$ is distributive using function $\AGGG$ over domain $\V$.
If  $\AGGF$ is distributive using function $\AGGG$ over any domain $\V$, we say that $\AGGF$ is distributive using function $\AGGG$, and if $\AGGF=\AGGG$, we simply say that $\AGGF$ is distributive.

It is easy to show that functions $\SUM, \MIN$ and $\MAX$ are  distributive, function $\COUNT$ is distributive using function $\SUM$ whereas function $\COUNTDISTINCT$ 
is distributive using function $\SUM$ only on partitionings where the same value does not appear in two distinct partitions. Finally, function $\AVG$ is distributive over all domains $V$ containing only two elements or where all elements are identical.

\begin{example}
For $V=\{1,2,2,3\}$ we have $\COUNTDISTINCT(\V)=3$. Then $\COUNTDISTINCT$ is  distributive using $\SUM$ on  partitioning $P'=\{\{1,2,2\}, \{3\}\}$: $$\SUM(\COUNTDISTINCT(\{1,2,2\}),\COUNTDISTINCT\{3\}))=3$$
However, $\COUNTDISTINCT$ is not distributive using $\SUM$ on partitioning $P=\{\{1,2\}, \{2,3\}\}$:
$$\SUM(\COUNTDISTINCT(\{1,2\}),\COUNTDISTINCT(\{2,3\})=\SUM(2,3)=4$$
\end{example}

We say that $\AGGF$ is \emph{distributive using function $\AGGG$ on attribute $\A$ of table $\XXT$ with partitioning attributes $\Z$} if $\AGGF$ is distributive using function $\AGGG$ on all partitions of the values of
$\A$ in $\XXT$ defined by $\Z$ and any subset of $\Z$. 
The following proposition relates the definition of distributive functions to the notion of summarizable attributes.

\begin{proposition}[Function distributivity and attribute summarizability]
\label{prop:summarizabile-att}
Let $\XXT(\S)$ be an analytic table with dimension attributes $\S_D\subseteq \S$ and an aggregable attribute $\A$ such that $\aggp_{\A}(\AGGF, \X)$ holds in $\XXT$. 
If $\AGGF$ is distributive using function $\AGGG$ on attribute $\A$ in table $\XXT$ with partitioning attributes $\Z \supseteq \S_D - \X$
then $\A$ is summarizable with respect to grouping set $\Z$ and function $\AGGF$ using function $\AGGG$.
\end{proposition}

\begin{myproof}
Suppose that $\aggp_{\A}(\AGGF, \X)$ holds in $\XXT$ and $\RT = \AGGQ_{\XXT}(\AGGF(\A) \mid \Z_1)$ and $\AGGF$ is \emph{distributive using function $\AGGG$ on attribute $\A$ of table $\XXT$ with partitioning attributes $\Z_1$}. To prove that $\A$ is summarizable in $\XXT$ with respect to grouping set $\Z_1$ and $\AGGF$ using function $\AGGG$, we prove that for any subset $\Z_2 \subset \Z_1$, the following equation holds:
\begin{equation}
    \label{eq:proof-sum1}
   \AGGQ_{\XXT}(\AGGF(\A) \mid \Z_2) = \AGGQ_{\RT}(\AGGG(\AGGF(\A)) \mid \Z_2) 
\end{equation}
First, it is obvious that both tables $\XXT$ and  $\RT$ contain the same $\Z_2$ values and therefore, the result tables in \cref{eq:proof-sum1} contain the same  tuples with distinct $\Z_2$ values. 
We now  show that for each pair of tuples $t\in \AGGQ_{\XXT}(\AGGF(\A) \mid \Z_2)$ and $t'\in \AGGQ_{\RT}(\AGGG(\AGGF(\A)) \mid \Z_2)$ where $t.\Z_2=t'.\Z_2$, we have $t.\AGGF(\A)=t'.\AGGG(\AGGF(\A))$. Let $x=t.\Z_2=t'.\Z_2$ and $\XXT^{x}=\sigma_{\Z_2=t.\Z_2}(\XXT)$ and $\RT^{x}=\sigma_{\Z_2=t.\Z_2}(\RT)$ be the partitions of $\XXT$ and $\RT$ on attributes $\Z_2$ corresponding to the tuples used to compute $t.\AGGF(\A)$. For each tuple $t'_i\in \RT^{x}$ there also exists a partition $\XXT^{y_i}=\sigma_{\Z_1=y_i}(\XXT)$ of $\XXT$ where $y_i=t'_i.\Z_1$ and $t_i'.\AGGF(\A)=\AGGF(\pi_{\A}(\XXT^{y_i}))$. 
All tuples $t'_i$ have the same $\Z_2$ value $x=t.\Z_2$ and are  aggregated to tuple $t'$ whose value for attribute $t'.\AGGG(\AGGF(\A))=\AGGG(\AGGF(\pi_{A}(\XXT^{y_1}) \cup \ldots \cup \AGGF(\pi_{\A}(\XXT^{y_n})$. Since $\AGGF$ is distributive using function $\AGGG$ and $\XXT^{x} = \XXT^{y_1} \cup \ldots \cup \XXT^{y_n}$, we obtain $\AGGG(\AGGF(\pi_{\A}(\XXT^{y_1}) \cup \ldots \cup \AGGF(\pi_{\A}(\XXT^{y_n}))=\AGGF(\pi_{\A}(\XXT^{y_1})\cup \ldots \cup \pi_{\A}(\XXT^{y_n}))=\AGGF(\pi_{\A}(\XXT^{x}))$. We conclude $t.\Z_2=t'.\Z_2$ and $t.\AGGF(\A)=t'.\AGGG(\AGGF(\A))$. 
\end{myproof}

\sloppy
\begin{myexample}
\label{exp:summarizable-att2}
Function $\COUNT$ is distributive using function $\SUM$. Therefore, \attr{PROD\_SKU} in table \factt{PRODUCT\_LIST} (\Cref{tab:prod_list} on \Cpageref{tab:prod_list}) is summarizable with respect to grouping set $\Z=\{\attr{BRAND}, \attr{COUNTRY}, \attr{YEAR}\}$ and $\COUNT$ using function $\SUM$.  Thus, if $\Z_2 = \{\attr{COUNTRY}, \attr{YEAR}\}$ and $\RT = \AGGQ_{PRODUCT\_LIST}(\COUNT(\attr{PROD\_SKU}) \mid \Z)$, the following equation folds:
$$\AGGQ_{PRODUCT\_LIST}(\COUNT(\attr{PROD\_SKU}) \mid \Z_2) = \AGGQ_{\RT}(\SUM(\COUNT(\attr{PROD\_SKU})) \mid \Z_2)$$
However, as explained before, $\COUNTDISTINCT$ is only distributive using $\SUM$ if no pair of partitions share the same value. This is not the case (there exist two partitions of $Z$ with the same product ''cz-tshirt-s''), so attribute \attr{PROD\_SKU} is not summarizable with respect to grouping set $Z$ and function $\COUNTDISTINCT$. 
\end{myexample}
%
Function distributivity is a sufficient but not a necessary condition for summarizability. This is illustrated in the following proposition, which defines a sufficient condition for summarizability with $\COUNTDISTINCT$ and $\SUM$.
\begin{proposition}[Summarizability with $\COUNTDISTINCT$ and $\SUM$]
\label{prop:countditinct-distributive}
\label{prop:countdistinct-summarizable}
Let $\XXT(\S)$ be an analytic table with a set of dimension attributes $\S_{\D}$ and an aggregable attribute $\A$.
Let $\RT = \AGGQ_{\XXT}(\COUNTDISTINCT(\A) \mid \Z_1)$ be a valid aggregate query over $T$, where $\Z_1\subseteq \S_{\D}$.
If $\Z_2 \subset \Z_1$ and the literal functional dependency $\Z_2\cup \{\A\} \mapsto \Z_1$ holds in  $\XXT$, the following equation is true:
\begin{equation}
    \label{eq:proof-countdis}
    \AGGQ_{\RT}(\SUM(\COUNTDISTINCT(\A)) \mid \Z_2) = \AGGQ_{\XXT}(\COUNTDISTINCT(\A) \mid \Z_2)
\end{equation}
We say that attribute $\A$ (in $\XXT$) is summarizable with respect to grouping set $\Z_1$ and $\COUNTDISTINCT$ using function $\SUM$ \emph{with partitioning attributes $\Z_2$}. 
\end{proposition}

\begin{myproof}
The previous proposition mainly states that $\A$ is summarizable with respect to $\Z_1$ and $\COUNTDISTINCT$ using function $\SUM$ with partitioning attributes $\Z_2$ if all tuples $\XXT$ in some partition $\XXT^{x}\subseteq \XXT$ defined by attributes $\Z_2\subseteq \Z_1$ which have the same value for attribute $t.\A$ are assigned to the \emph{same partition} $\XXT^{y}\subseteq \XXT$ defined by attributes $\Z_1$. This avoids the double counting of distinct $\A$ values when taking the $\SUM$ of $\COUNTDISTINCT$ over the partitions generated by attributes $\Z_1$. 

We first show by contradiction that when $\Z_2\cup \{\A\} \mapsto \Z_1-\Z_2$ holds in  $\XXT$, all tuples $\XXT$ in some partition $\XXT^{x}\subseteq\XXT$ of $\XXT$ generated by attributes $\Z_2$ with the same value for attribute $t.\A$ are assigned to the same partition $\XXT^{y}\subseteq \XXT$ generated by attributes $\Z_1$.

Let $\XXT^{x}$ be a partition of $\XXT$ which contains all tuples $\XXT$ such that $t.\Z_2=x$.
Since $\Z_2\subset \Z_1$, $\XXT^{x}$ is the union of a set of partitions $\XXT^{y}_0 \ldots, \XXT^{y}_n$, $n \geq 0$ of $\XXT$ defined by attributes $\Z_1$.
Suppose that there exist two tuples $t\in\XXT^{y}_i$ and $t'\in \XXT^{y}_j$ where $i\neq j$ and $t.\A=t'.\A$. Then, we have $t.\Z_2=t'.\Z_2=x$, $t.\A=t'.\A$ and, since $i\neq j$,  $t.\Z_1\neq t'.\Z_1$ (two different partitions generated by $\Z_1$ contain the same values for $\Z_2$ and $\A$). This is in contradiction with $\Z_2\cup \{\A\} \mapsto \Z_1-\Z_2$. Then, if $d_i$ is the number of distinct $\A$ values in some partition $\XXT^{x}_i\subseteq \XXT$, we can easily show that $\sum_{i=0}^{n} d_i$ is the number of distinct $\A$ values in partition $V$. 
\end{myproof}
Observe that if \cref{eq:proof-countdis} holds for any subset $\Z_2\subset \Z_1$,  $\A$  is summarizable with respect to grouping set $\Z_1$ and $\COUNTDISTINCT$ using function $\SUM$ (\Cref{def:summarizable-attr}).

\begin{myexample}
Let $\RT = \AGGQ_{\factt{PRODUCT\_LIST}}(\COUNTDISTINCT(\attr{PROD\_SKU}) \mid \Z_1)$ where $\Z_1=\{\attr{BRAND}, \attr{COUNTRY}\}$. Attribute \attr{PROD\_SKU} is not summarizable with respect to grouping set $Z_1$ and function $\COUNTDISTINCT$ using $\SUM$. However, if $\attr{PROD\_SKU} \mapsto \attr{COUNTRY}$ holds in table \factt{PRODUCT\_LIST}, then for $\Z_2=\{\attr{BRAND}\}$ and $\A=\attr{PROD\_SKU}$, we have $\{\attr{PROD\_SKU}, \attr{BRAND}\} \mapsto \{\attr{BRAND}, \attr{COUNTRY}\}$. Therefore, $\attr{PROD\_SKU}$ is summarizable with respect to $\Z_1$ and $\COUNTDISTINCT$ using function $\SUM$ with partitioning attribute $\Z_2=\{\attr{BRAND}\}$.

\end{myexample}

\subsection{Controlling attribute summarizability using aggregable properties}
\label{sec:prop_summarize}

\begin{figure}[htbp]
\centering
\includegraphics[width=0.6\linewidth]{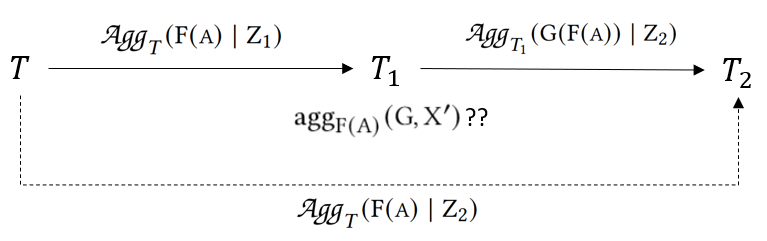}
\caption{Aggregable property that controls the summarizability of $A$ in $\XXT$ with respect to grouping set $Z_1$}
\label{fig:summ-att-fig2}
\end{figure}

Given the result of an aggregate query $\RT =\AGGQ_{\XXT}(\AGGF(\A) \mid \Z_1)$ over some attribute $\A$, 
we want to control the possible aggregations of attribute $\AGGF(\A)$ depending on the summarizability of $\A$. We use aggregable properties for that purpose, as shown on \Cref{fig:summ-att-fig2}. We want to automatically compute the subset of dimension attributes $\X' \subseteq \Z_1$ of $\RT$ such that $\aggp_{\AGGF(\A)}(\AGGG,\X')$ holds in $\RT$, 
for some function $\AGGG$ that is applicable to $\AGGF(\A)$,
and which guarantees the summarizability of $\A$ in $\RT$ for any $\Z_2$ such that 
$\Z_1 -\X' \subseteq \Z_2$. 
Our rationale is therefore to 
refine the  propagation rules introduced in \Cref{sec:prop-agg-prop} to take into account the summarizability correctness criteria. 

We next define the notion of summarizability preserving aggregable property which formalizes \Cref{fig:summ-att-fig2}.   
\begin{definition}[Summarizability preserving aggregable property]
\label{def:summarizability_preserving_aggregable_property}
Let $\XXT(\S)$ be an analytic table and  
$\RT=\AGGQ_{\XXT}(\AGGF(\A) \mid \Z_1)$ be the result of a valid aggregate  query. We say that $\aggp_{\AGGF(\A)}(\AGGG,\X')$ \emph{preserves the summarizability of $\A$ with respect to grouping set $\Z_1$} 
if for any subset $\Z_2$ such that $\Z_1-\X'\subseteq \Z_2$, attribute $\A$ is summarizable in $\XXT$ with respect to grouping set $\Z_2$ and function $\AGGF$ using $\AGGG$. 
\end{definition}
 
The next proposition uses \Cnameref{prop:summarizabile-att} and \Cnameref{prop:countdistinct-summarizable} to refine the previous propagation rule for aggregation in \Cref{tab:prop_rules_unary} so that summarizability preserving aggregable properties are inferred. The refined rule for the case when $\A' = \AGGF(A)$ is displayed in \Cref{tab:refined-prop_rules_aggregate}.

\begin{table}[htb]
\caption{Propagation rule for aggregate operation preserving summarizability}
\label{tab:refined-prop_rules_aggregate}
\normalsize{
\bgroup
\begin{normalsize}
\def\arraystretch{1.2}
\begin{tabular}{l|l|l}
    \toprule
    Query on $\XXT(\S)$ & 
    \makecell[tl]{New propagation rule for inferring the aggregable properties of new measure  \\ attribute $\A'$ in the result $\RRT(\RRS)$
    }& 
    \makecell[tl]{
    \emph{User action}} \\
    \midrule

    $\AGGQ_{\XXT}(\AGGF(\A) \mid \Y)$ 
    & 
    
    \makecell[tl]{new measure attribute $\A'=\AGGF(\A)$ and $\aggp_{\A}(\AGGF, \X)$ holds in $\XXT$:  \\
     \hspace{3mm} if $\AGGG$ can be applied on $\A'$ (\Cref{tab:cat_agg}) and  $\AGGF$ is distributive using $\AGGG$ \\
     \hspace{3mm} then $\aggp_{\AGGF(\A)}(\AGGG, \X')$ holds in $\RRT$ with $\X'= \X \cap \Y$,  $\X'_d=$ fact identifier  in $\T_r$ \\ 
     \hspace{9mm} and     $\X'_f=\emptyset$. 
     }
     &  
     \makecell[tl]{
     \emph{Minimize $\X'_d$ }
     \\
     \emph{Complete $\X'_f$}} \\[10mm]
     &
     \makecell[tl]{
     \hspace{3mm} if $\AGGF=\COUNTDISTINCT$ and $\X'$ is a \emph{maximal} subset of $\X \cap \Y$ \\ 
     \hspace{9mm}   such that  $(\Y-\X')\cup \{A\}\mapsto \Y$ holds in $\XXT$\\ 
     \hspace{3mm}  then $\aggp_{\AGGF(\A)}(\SUM, \X')$ holds in $\RRT$ with $\X'_d=$ fact identifier in $\T_r$ and $\X'_f=\emptyset$.}
    &  
     \makecell[tl]{
     \emph{Minimize $\X'_d$ }
     \\
     \emph{Complete $\X'_f$}} \\[10mm]
    \bottomrule
\end{tabular}
\end{normalsize}
\egroup
}
\end{table}

\begin{proposition}[Propagation of aggregable properties with summarizability preservation]
\label{def:propagate-agg-prop}
Let $\XXT(\S)$ be an analytic table with dimension attributes $\S_D\subseteq \S$ and let $\RRT=\AGGQ_{\XXT}(\AGGF(\A) \mid \Y)$ be the result of a valid aggregate  query. 
Then the aggregable properties inferred by the rule of \Cref{tab:refined-prop_rules_aggregate} when $\A' = F(A)$ preserve the summarizability of $\A$ with respect to grouping set $\Y$.
\end{proposition}
\begin{myproof}  By \Cnameref{def:summarizability_preserving_aggregable_property}, we have to show that for any subset $\Z_2\subseteq \Y$ such that $\Y-\X'\subseteq \Z_2$, attribute $\A$ is summarizable in $\XXT$ with respect to $\Z_2$ and function $\AGGF$ using $\AGGG$. We examine both cases of
\Cref{def:propagate-agg-prop}: 
\begin{itemize}
\item 
$\AGGG$ can be applied on $\A'$ as defined in \Cref{tab:cat_agg} and $\AGGF$ is distributive using $\AGGG$: 
Since $\AGGF$ is distributive using function $\AGGG$, it is also distributive on attribute $\A$ with partitioning attributes $\Y$. Then, by \Cnameref{prop:summarizabile-att}, $\A$  is  summarizable with respect to $\Y$ and $\AGGF$ using function $\AGGG$, and \Cref{eq:summarizable-attr} in \Cnameref{def:summarizable-attr} holds for any subset $\Z_2\subseteq \Y$.
\item 
$\AGGF=\COUNTDISTINCT$ and $\X'$ is a \emph{maximal} subset of $\X \cap \Y$ such that $(\Y-\X')\cup \{A\}\mapsto \Y$ holds in $\XXT$: 
By \Cnameref{prop:countditinct-distributive}, it is sufficient to show that $\Z_2\cup\{\A\}\mapsto \Y$ for all $\Z_2$ where $\Y-\X' \subseteq \Z_2 \subset \X \cap \Y$. 
Since $(\Y-\X')\cup \{\A\}\mapsto \Y$ and $\Y-\X' \subseteq \Z_2$ we also have $\Z_2\cup \{\A\}\mapsto \Y$. 
\end{itemize}
\end{myproof}

For the second condition in \Cref{tab:refined-prop_rules_aggregate}, observe that there might exist several maximal subsets of attributes $\X'_i$. The process to compute these subsets is quite simple. Each maximal subset $\X'_i$ corresponds to a \emph{minimal subset} of attributes $K_i=\Y-\X'_i\subset \Y$ such that $K_i\cup\{A\}$ determines all attributes of $\Y$. These sets $\K_i$ can easily be computed using the attribute graphs of the corresponding dimensions to obtain $\X'_i = (\Y \cap \X) - \K_i$.

\begin{myexample}
\label{ex:def-rule-agg2}
Aggregable property $\aggp_{\attr{PROD\_SKU}}(\COUNT \mid \X)$ holds in table \factt{PRODUCT\_LIST} (\Cref{tab:prod-list2-repeat2})
for $\X = \{\attr{BRAND}, \attr{COUNTRY}, \attr{YEAR}\}$. 
Let $\RRT = \AGGQ_{\factt{PRODUCT\_LIST}}(\COUNT(\attr{PROD\_SKU}) \mid \Y)$, where $\Y = \{\attr{BRAND},\attr{COUNTRY},\attr{YEAR}\}$.

By the first condition in \Cref{tab:refined-prop_rules_aggregate} and distributivity of $\COUNT$ using $\SUM$, the aggregable property $\aggp_{\COUNT(\attr{PROD\_SKU})} (\SUM\mid \X')$, where $\X'= \X \cap \Y = \{\attr{BRAND}, \attr{COUNTRY}, \attr{YEAR}\}$, preserves the summarizability of \attr{PROD\_SKU} with respect to grouping set $\Y$.

\end{myexample}

\begin{myexample}
\label{ex:def-rule-agg3}
Property $\aggp_{\attr{PROD\_SKU}}(\COUNTDISTINCT \mid \X)$ holds in table \factt{PRODUCT\_LIST} for $\X = \{\attr{BRAND}, \allowbreak \attr{COUNTRY}, \attr{YEAR}\}$. 
Let $\T_r=\AGGQ_{\factt{PRODUCT\_LIST}}(\COUNTDISTINCT(\attr{PROD\_SKU}) \mid \Y)$ where $\Y = \{\attr{BRAND}, \allowbreak \attr{COUNTRY}, \attr{YEAR}\}$. 
By the second condition in \Cref{tab:refined-prop_rules_aggregate}, we must compute the maximum subset $\X'$ of $\X \cap \Y$ such that 
$(\Y-\X')\cup \{\A\}\mapsto \Y$. We use the method explained before. 
By the attribute graphs of dimension \dimt{Time} and \dimt{PROD}, the only LFD which holds among the attributes of $\Y$ is $\attr{BRAND}\mapsto \attr{COUNTRY}$.
Thus, there is a single minimal set $K= \{\attr{BRAND}, \attr{YEAR}\}$ such that $K \cup \{\attr{PROD\_SKU}\}$ determines all  attributes in $\Y$. 
We obtain that $\X'= \Y - K = \{\attr{COUNTRY}\}$. Hence, 
$\aggp_{\COUNTDISTINCT(\attr{PROD\_SKU})} (\SUM\mid \{\attr{COUNTRY}\})$ 
preserves the summarizability of \attr{PROD\_SKU} with respect to grouping set $\Y$.

\end{myexample}

Aggregable properties provide "explanations" for end users of which aggregate queries preserve the summarizability condition of the aggregated attribute in a given stage of the data analysis session. In the previous example, the aggregable property $\aggp_{\COUNTDISTINCT(\attr{PROD\_SKU})} (\SUM\mid \{\attr{COUNTRY}\})$ of $\T_r$ explains that
table $\T_r$ can be used to count the number of distinct products per brand \emph{and} year by taking the sum of $\COUNTDISTINCT(\attr{PROD\_SKU})$ along \attr{COUNTRY}. However, $\T_r$ cannot be used to obtain the number of distinct products by brand \emph{or} by year. In this case, the user must "backtrack" in the interactive data analysis session to the table \factt{PRODUCT\_LIST} to obtain this number.

\subsection{Generalized attribute summarizability}

\label{sec:generalized_summarizability}
In the previous sections, as illustrated in \Cref{fig:summ-att-fig1}, we defined an attribute $\A$ in some table $\XXT$ to be summarizable with respect to some aggregate query $\Q(\XXT)=\AGGQ_{\factt{T}}(\AGGF(\A) \mid \Z_1)$ and function $\AGGF$ using $\AGGG$, if for any query $\Q'(\XXT)=\AGGQ_{\factt{\XXT}}(\AGGF(\A) \mid \Z_2)$ aggregating $\A$ along a subset $\Z_2$ of $\Z_1$, there exists an \emph{equivalent} aggregate query $\Q''(\RT)=\AGGQ_{\factt{T_1}}(\AGGG(\AGGF(\A)) \mid \Z_2)$ on the result $\RT$ of $\Q(\XXT)$. \Cref{def:gen-summ-attr-unary} generalizes this notion to any analytic query $\Q$ 
as follows. 

\begin{definition}[Generalized summarizable attribute]
\label{def:gen-summ-attr-unary}
Let $\XXT(\S)$ be an analytic table that is the input of an analytic query $\Q$ returning a table $\RT(\S_1)$.
Let $\A$ be an aggregable attribute of both $\XXT$ and $\RT$ and $\Z$ be a subset of the dimension attributes of $\S \cap \S_1$. 
If for any two valid aggregate queries $\Q'(\XXT)=\AGGQ_{\XXT}(\AGGF(\A) \mid \Z')$ and $\Q'(\RT)=\AGGQ_{\RT}(\AGGF(\A) \mid \Z')$
such that $\Z \subseteq \Z'$, and any two tuples $t_1 \in \Q'(\XXT)$ and $t_2 \in \Q'(\RT)$, we have:
$$t_1.\Z' \equiv t_2.\Z' \Rightarrow t_1.\AGGF(\A) \equiv t_2.\AGGF(\A)$$
then $\A$ is said to be \emph{G-summarizable} in $\XXT$ with respect to query $\Q$, grouping set $\Z$ and function $\AGGF$.
\end{definition}

\begin{figure}[htbp]
\begin{center}
\includegraphics[width=0.6\linewidth]{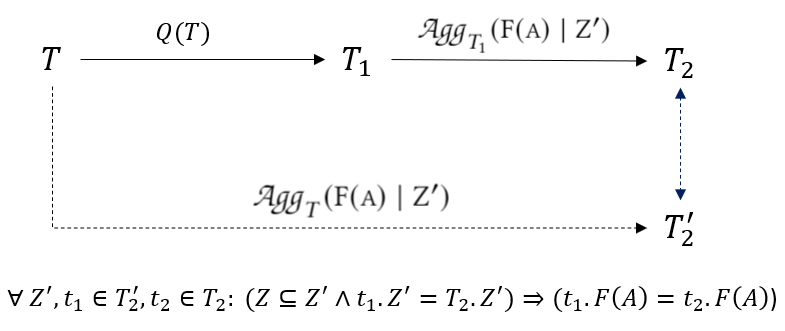}
\end{center}
\caption{G-summarizable attribute $A$ in $T$ with respect to $Q$, grouping set $Z$ and function $F$}
\label{fig:gen-summ-att2}
\end{figure}

The above definition is illustrated in \Cref{fig:gen-summ-att2}. 
We make a few observations. First, $\T_2$ and $\YYT_2$ are not necessarily equal, \ie $\T_2$ might contain tuples that are not in $\YYT_2$ and vice versa. Second, in \Cnameref{sec:prop-agg-prop}, we established the propagation rules to compute the aggregable properties on $\A$ that hold in $\RT$ for $\AGGF$, which are used in the definition to determine which aggregate queries $\Q'$ on $\RT$ are valid. 
Third, an implicit assumption is that $\A$ 
is an attribute that exists in both $\XXT$ and $\RT$. Therefore, in the case of an aggregate or a pivot query $\Q$, $\A$ must be a dimension attribute (since measure attributes of $\XXT$ do not exist anymore in $\RT$ - they have either been aggregated, pivoted or eliminated). Finally, grouping set $\Z$ defines a set of attributes that must be contained in $\Z'$ and implicitly restricts the attributes along which aggregation can be done in $\Q'$. 
\bernd{Add: Thus, the notions of summarizability and G-summarizability are complementary...}

Before formalizing sufficient conditions for G-summarizability, we present a few  examples.

\begin{example}
Consider a fact table $\XXT$ (in \Cref{tab:filter-non-coverage2}) defined over two dimension $\D_1$ and $\D_2$ with dimension attributes $\A_1, \A_2, \A_3$ from dimension $\D_1$ (where $\A_1 \typeprec A_2 \typeprec A_3$) and $\B_1, \B_2$ from dimension $\D_2$ (where $B_1 \typeprec B_2$). 
We shall say that within table $\XXT$, $\A_3$ is the \emph{highest} attribute of $\D_1$ while $B_2$ is the \emph{highest} attribute of $D_2$.  
\begin{table}[htbp]
\normalsize
\caption{G-summarizability in $\XXT$ with respect to a filter query}
\label{tab:filter-non-coverage2}
\begin{subtable}[b]{0.3\linewidth}
    \centering
    \setlength{\tabcolsep}{3pt}
    \begin{tabular}[t]{l|l|l|l|l|l|l}
   \cmidrule[1pt]{2-7}
    $\XXT$& $\A_1$ & $\A_2$ & $\A_3$ & $\B_1$ &$\B_2$ & $\M$\\
    \midrule
    $t_0$ & $a_1$ & $b_1$ & $c_1$ & $f_1$ & $e_1$ & $x_1$\\
    $t_1$ & $a_2$ & $b_2$ & $c_1$ & $f_1$ & $e_1$ & $x_2$\\
    $t_2$ & $a_3$ & $b_1$ & $c_1$ & $f_2$ & $e_1$ & $x_3$\\
    $t_3$ & $a_2$ & $b_1$ & $c_2$ & $f_2$ & $e_1$ & $x_4$\\
    \cmidrule[0.8pt]{2-7}
   \end{tabular}
\end{subtable}
\hfill
\begin{subtable}[b]{0.3\linewidth}
    \centering
    \setlength{\tabcolsep}{3pt}
    \begin{tabular}[t]{r|l|l|l|l|l|l}
   \cmidrule[1pt]{2-7}
    $\Q_1(\XXT)$ & $\A_1$ & $\A_2$ & $\A_3$ & $\B_1$ &$\B_2$ & $\M$ \\
    \midrule
    $t_0$ & $a_1$ & $b_1$ & $c_1$ & $f_1$ & $e_1$ & $x_1$\\
    $t_1$ & $a_2$ & $b_2$ & $c_1$ & $f_1$ & $e_1$ & $x_2$\\
    $t_2$ & $a_3$ & $b_1$ & $c_1$ & $f_2$ & $e_1$ & $x_3$\\
    \cmidrule[0.8pt]{2-7}
   \end{tabular}
\end{subtable}
\hfill
\begin{subtable}[b]{0.3\linewidth}
    \centering
    \setlength{\tabcolsep}{3pt}
    \begin{tabular}[t]{r|l|l|l|l|l|l}
   \cmidrule[1pt]{2-7}
    $\Q_2(\XXT)$ & $\A_1$ & $\A_2$ & $\A_3$ & $\B_1$ &$\B_2$ & $\M$ \\
    \midrule
    $t_2$ & $a_3$ & $b_1$ & $c_1$ & $f_2$ & $e_1$ & $x_3$\\
    $t_3$ & $a_2$ & $b_1$ & $c_2$ & $f_2$ & $e_1$ & $x_4$\\
    \cmidrule[0.8pt]{2-7}
   \end{tabular}
\end{subtable}
\end{table}

Take query $\Q_1(\XXT)  = \FILTERQ_{ \XXT}(\{\A_3 = c_1 \} \mid \{\A_3\})$ whose result table $\RT$ is displayed in \Cref{tab:filter-non-coverage2}. 
If we take a grouping set $\Z = \{\A_3\}$, then for each partition $\XXT^p = \sigma_{\A_3 = p}(\XXT)$ of $\XXT$ we either have $\Q_1(\XXT^p)=\XXT^x$ or $\Q_1(\XXT^p)=\emptyset$. In our example, we have:  
$\Q_1(\XXT^{c_1})=\XXT^{c_1}$ and $\Q_1(\XXT^{c_2})=\emptyset$. Thus,  
for any subset $\Z'$ of dimension attributes of $\XXT$ containing $\A_3$, we 
either have $\Pi_{\Z'}(\Q_1(\XXT^{p})) = \Pi_{\Z'}(\XXT^{p})$ or $\Pi_{\Z'}(\Q_1(\XXT^{p}))$ is empty, 
where $\Pi$ is a projection without duplicate elimination.   
Therefore, any valid aggregation query with grouping attributes $\Z'$ containing $\A_3$  returns, for each partition of $\RT$ defined by $\Z'$, the same result as for the corresponding partition of $\XXT$ defined by $\Z'$.
Hence, any attribute $\A$ 
is G-summarizable in $\XXT$ with respect to query $\Q_1$, grouping set $\Z = \{A_3\}$ and any  function $\AGGF$ applicable to $\A$ in $\XXT$. 
Observe that this is not the case for any other grouping set $Z$ that does not contain attribute $\A_3$. For instance, if $\Z = \{\A_2\}$ then for partition $\XXT^{b_1}$,  $\Q_1(\XXT^{b_1}) \neq \XXT^{b_1}$.  

However, the previous reasoning does not apply if $\XXT$ is filtered on a measure attribute $\attr{M}$, like in query $\Q_2(\XXT) = \FILTERQ_{ \XXT}(\{\M \in \{x_3, x_4\}\} \mid \{\M\})$. Since a measure attribute cannot belong to the grouping set of an aggregate query, we cannot define a partitioning set $\Z=\{\M\}$. 
Then, to identify a set of dimension attributes $\Z$ such that all attributes $\A$ are G-summarizable in $\XXT$ with respect to query $\Q_2$, grouping set $\Z$ and an applicable function $\AGGF$, we must find a set of dimension attributes $\Z$ that partitions $\XXT$ into $\Q_2(\XXT)$ and $\XXT - \Q_2(T)$. In our example, $\Z=\{\B_1\}$ would be a possible solution that cannot be easily found. 
\end{example}

\begin{example}
\label{ex:merge-coverage}
Consider now the fact tables $\XXT$ and $\YYT$ defined over the same dimensions $\D_1$ and $D_2$ as before, as shown below in \Cref{tab:merge-non-coverage}. 
Take the left-merge query $\Q(\XXT,\YYT)=\XXT\leftouterjoin_{\Y} \YYT = T_1$ where $\Y=\{A_1, A_2, B_1, B_2\}$. Any attribute $\A$ of $\XXT$ 
is G-summarizable in $\XXT$ with respect to $\Q$, grouping set $\Z = \emptyset$, and any function $\AGGF$ applicable to $\A$, because the duplicate preserving projection of $\RT$ on the attributes of $\XXT$ is equal to table $\XXT$.

Let us look at the G-summarizability of attributes in $\YYT$ with respect to $\Q$ and $\AGGF$. First, grouping set $\Z=\{\A_2, \B_2\}$ containing the "highest" attributes in $\Y$ defines two partitions of $\YYT$.
We can see that partition $\YYT^{b_1, e_1}$ is different from the duplicate-preserving projection of $\RT^{b_1, e_1}$ on the attributes of $\YYT$ (tuples $t_3$, $t_4$ of $\YYT$ have no corresponding tuples in $\RT$). So, any valid aggregation query over a partitioning on $Z$ would violate the G-summarizability property.
Indeed, the only grouping set $\Z$ for which we have the equality of non-empty partitions is $\Z=\{A_1, A_2, B_1, B_2\}$. 
We shall see later that if a valid aggregation can be expressed over $Y$ in $T'$ then $Z$ can be equal to $Y$.

\begin{table}[htbp]
\normalsize
\caption{G-summarizability in $\XXT$ with respect to a left-merge query}
\label{tab:merge-non-coverage}
\begin{subtable}[b]{0.2\linewidth}
    \centering
    \setlength{\tabcolsep}{3pt}
    \begin{tabular}[t]{l|l|l|l|l|l}
   \cmidrule[1pt]{2-6}
    $\XXT$& $\A_1$ &$\A_2$ & $\B_1$ &$\B_2$ & $\M$\\
    \midrule
    $t_0$ & $a_1$ &$b_1$ & $f_2$ & $e_1$ & $x_1$\\
    $t_1$ & $a_4$ & $b_2$ & $f_4$ & $e_2$ & $x_2$\\
    \cmidrule[0.8pt]{2-6}
   \end{tabular}
\end{subtable}
\hfill
\begin{subtable}[b]{0.3\linewidth}
    \centering
    \setlength{\tabcolsep}{3pt}
    \begin{tabular}[t]{l|l|l|l|l|l|r}
   \cmidrule[1pt]{2-7}
    $T'$ & $\A_1$ & $\A_2$ & $\A_3$ & $\B_1$ &$\B_2$ & $\M'$\\
    \midrule
    $t_3$ & $a_1$ & $b_1$ & $c_1$ & $f_1$ & $e_1$ & $y_1$\\
    $t_4$ & $a_2$ & $b_1$ & $c_1$ & $f_2$ & $e_1$ & $y_3$\\
    $t_5$ & $a_1$ & $b_1$ & $c_1$ & $f_2$ & $e_3$ & $y_4$\\
    $t_6$ & $a_1$ & $b_1$ & $c_1$ & $f_2$ & $e_1$ & $y_5$\\
    \cmidrule[0.8pt]{2-7}
   \end{tabular}
\end{subtable}
\hfill
\begin{subtable}[b]{0.40\linewidth}
    \centering
    \setlength{\tabcolsep}{3pt}
    \begin{tabular}[t]{r|l|l|l|l|l|l|r}
   \cmidrule[1pt]{2-8}
    $\XXT \leftouterjoin \YYT$ & $\A_1$ & $\A_2$ & $\A_3$ & $\B_1$ &$\B_2$ & $\M$ & $\M'$\\
    \midrule
    $t_{10}$ & $a_1$ & $b_1$ & $c_1$ & $f_2$ & $e_1$ & $x_1$ & $y_5$\\
    $t_{11}$ & $a_4$ & $b_2$ & - & $f_4$ & $e_2$ & $x_2$ & - \\
    \cmidrule[0.8pt]{2-8}
   \end{tabular}
\end{subtable}
\end{table}

Now take the same left-merge query $\Q$ as before applied to the tables displayed in \Cref{tab:merge-non-coverage2}.
For grouping set $\Z = \{\A_2, \B_2\}$, the partition of $T'$ with values $(b_1, e_1)$ has a corresponding identical partition in $T_1$ after a duplicate-preserving projection on the attributes of $T'$. The partition of $T'$ with values $(b_1, e_3)$ has a corresponding empty partition in $T_1$. However, the partition of $T_1$ with values $(b_2, e_1)$ has one extra tuple with respect to the corresponding partition in $T'$ because tuple $t_7$ is matched by two tuples of $T$ and its attribute values appear duplicated in $T_1$ (in tuples $t_{12}$ and $t_{13}$). 
Hence, for attributes of $T'$, G-summarizability in $\YYT$ with respect to $Q$ and grouping set $\Z = \{\A_2, \B_2\}$ must be restricted to functions that are insensitive to duplicates (i.e., \COUNTDISTINCT, \MIN, \MAX). 

Finally, any attribute of $\XXT$ is G-summarizable in $\XXT$ with respect to $\Q$ and grouping set $\Z = \emptyset$ because the duplicate preserving projection of $\RT$ on the attributes of $\XXT$ is equal to  $\XXT$. Thus, $T_1$ is aggregable over any partitioning of attributes of $T$.

\begin{table}[htbp]
\normalsize
\caption{G-summarizability in $T'$ with respect to a left-merge query}
\label{tab:merge-non-coverage2}
\begin{subtable}[b]{0.3\linewidth}
    \centering
    \setlength{\tabcolsep}{3pt}
    \begin{tabular}[t]{l|l|l|l|l|l|r}
   \cmidrule[1pt]{2-7}
    $T$ & $\A_1$ & $\A_2$ & $\A_3$ & $\B_1$ &$\B_2$ & $\M$\\
    \midrule
    $t_0$ & $a_1$ & $b_1$ & $c_1$ & $f_1$ & $e_1$ & $x_1$\\
    $t_1$ & $a_2$ & $b_1$ & $c_1$ & $f_1$ & $e_1$ & $x_2$\\
    $t_2$ & $a_3$ & $b_2$ & $c_1$ & $f_2$ & $e_1$ & $x_3$\\
    $t_3$ & $a_3$ & $b_2$ & $c_2$ & $f_2$ & $e_1$ & $x_4$\\
    $t_4$ & $a_4$ & $b_2$ & $c_1$ & $f_2$ & $e_1$ & $x_5$\\
    \cmidrule[0.8pt]{2-7}
   \end{tabular}
\end{subtable}
\hfill
\begin{subtable}[b]{0.2\linewidth}
    \centering
    \setlength{\tabcolsep}{3pt}
    \begin{tabular}[t]{l|l|l|l|l|l}
   \cmidrule[1pt]{2-6}
    $\YYT$& $\A_1$ &$\A_2$ & $\B_1$ &$\B_2$ & $\M'$\\
    \midrule
    $t_5$ & $a_1$ &$b_1$ & $f_1$ & $e_1$ & $y_1$\\
    $t_6$ & $a_2$ & $b_1$ & $f_1$ & $e_1$ & $y_2$\\
    $t_7$ & $a_3$ & $b_2$ & $f_2$ & $e_1$ & $y_3$\\
    $t_8$ & $a_2$ & $b_1$ & $f_3$ & $e_3$ & $y_4$\\
    \cmidrule[0.8pt]{2-6}
   \end{tabular}
\end{subtable}
\hfill
\begin{subtable}[b]{0.40\linewidth}
    \centering
    \setlength{\tabcolsep}{3pt}
    \begin{tabular}[t]{r|l|l|l|l|l|l|r}
   \cmidrule[1pt]{2-8}
    $\XXT \leftouterjoin \YYT$ & $\A_1$ & $\A_2$ & $\A_3$ & $\B_1$ &$\B_2$ & $\M$ & $\M'$\\
    \midrule
    $t_{10}$ & $a_1$ & $b_1$ & $c_1$ & $f_1$ & $e_1$ & $x_1$ & $y_1$\\
    $t_{11}$ & $a_2$ & $b_1$ & $c_1$ & $f_1$ & $e_1$ & $x_2$ & $y_2$\\
    $t_{12}$ & $a_3$ & $b_2$ & $c_1$ & $f_2$ & $e_1$ & $x_3$ & $y_3$ \\
    $t_{13}$ & $a_3$ & $b_2$ & $c_2$ & $f_2$ & $e_1$ & $x_4$ & $y_3$ \\
    $t_{14}$ & $a_4$ & $b_2$ & $c_1$ & $f_2$ & $e_1$ & $x_5$ & - \\
    \cmidrule[0.8pt]{2-8}
   \end{tabular}
\end{subtable}
\end{table}
\end{example}

\begin{proposition} [Queries satisfying G-summarizability]
\label{prop:gen-summ}
Let $\Q$ be a unary or binary analytic query with some input table $\XXT(\XXS)$ returning a table $\RT(\S_1)$ and $\XXS_D$ be the dimension attributes in $\XXS \cap \S_1$.
Let $\Z$ be a subset of $\XXS_D$, and $\A'$ be an attribute in $\XXS\cap\S_1$ such that $\aggp_{\A'}(\AGGF, \X)$ and $\aggp_{\A'}(\AGGF, \X_1)$ hold in $\XXT$ and $\RT$ respectively.
Then, \emph{the attribute $\A'$ is G-summarizable in $\XXT$ with respect to query $\Q$, grouping set $\Z$, and function $\AGGF$} in the following cases:

\noindent\emph{Unary queries:}
\begin{enumerate}
    \item  
    $\Q=\FILTERQ_{\XXT}(P \mid \Y)$,  $\Y \subseteq \XXS_D$, $\A'\in \XXS$
    and $\Z=\XXS_D - \X_1 \cup \Y$.
    \item  
    $\Q=\PROJECTQ_{\XXT}(\Y, f(\Z') \rightarrow \M)$, $\A'\in \Y$ 
    and $\Z = \XXS_D - \X_1$.
    
    \item  
    $\Q=\AGGQ_{\XXT}(\AGGG(\A) \mid \Y)$,  $\A'\in \Y$,  
    $\Z = \Y - \X_1$ and $\AGGF \in \{\MIN,\allowbreak \MAX,\allowbreak \COUNTDISTINCT\}$.
        
    \item  $\Q=\PIVOTQ_{\XXT}(\A \mid \Y)$, $\A'\in \XXS - \Y - \{\A\}$, 
    $\Z =\XXS_D- \X_1-\Y- \{\A\})$  and $\AGGF \in \{\MIN,\allowbreak \MAX,\allowbreak \COUNTDISTINCT\}$.
\end{enumerate}

\noindent\emph{Merge queries:} 
In the following, let $\Ytop\subseteq \Y$ denote the subset of "highest" attributes in the set of join attributes $\Y$.
\begin{enumerate}
    \item  $\Q=\XXT  \leftouterjoin_\Y \YYT $,  $\A'\in\S$, 
    $\Z = \S_D - \X_1$  and  if $\Y \not \mapsto \RRS$ then $\AGGF \in \{\MIN,\allowbreak \MAX, \allowbreak \COUNTDISTINCT \}$.
    
    \item 
    $\Q=\XXT  \rightouterjoin_\Y \YYT$ or $\Q=\XXT  \fullouterjoin_\Y \YYT$: 
    \begin{enumerate}
    \item If $\A'\in\Y$ and for all non-empty partitions $\YYT^y=\sigma_{\Ytop=y}(\YYT)$  of $\YYT$,  the corresponding partition $\XXT^y=\sigma_{\Ytop=y}(\XXT)$ of $\XXT$ is empty or $\pi_\Y(\XXT^y)$ is \textit{equal} to $\pi_\Y(\YYT^y)$, then  $\Z=\S_D-\X_1\cup \Ytop$

    \item If $\A'\in\S-\Y$ and for all non-empty partitions $\YYT^y=\sigma_{\Y=y}(\YYT)$  of $\YYT$  and corresponding partitions $\XXT^y=\sigma_{\Y=y}(\XXT)$ of $\XXT$, $\pi_\Y(\XXT^y)$ is a \textit{subset of}  $\pi_\Y(\YYT^y)$, then $\Z=\S_D-\X_1$.

    \item  Otherwise, $\Z=(\S_D-\X_1)\cup \Y$. 
    \end{enumerate}
    
    In addition, for all cases, if $\Y \not \mapsto \RRS$ then $\AGGF \in \{\COUNTDISTINCT, \MIN, \MAX\}$.
    
   \item 
    $\Q=\XXT  \bowtie_\Y \YYT$, $\A'\in\S$:  if for all non-empty partitions $\YYT^y=\sigma_{\Ytop=y}(\YYT)$  and corresponding partitions $\XXT^y=\sigma_{\Ytop=y}(\XXT)$,  $\pi_\Y(\YYT^y)$ contains $\pi_{\Y}(\XXT^y)$ then $\Z=\S_D-\X_1\cup \Ytop$  else $\Z=\S_D-\X_1\cup \Y$. 
    
    In addition, if $\Y \not \mapsto \RRS$ then $\AGGF \in \{\COUNTDISTINCT, \MIN, \MAX\}$.
   
\end{enumerate}

\noindent\emph{Set queries:} 
In the following, let $\Ytop\subseteq \XXS_D$ denote the set of "highest" attributes in the set of dimension attributes $\XXS_D\subseteq \XXS$ of $\XXT$ and $\YYT$.
\begin{enumerate}

    \item 
    $\Q=\XXT  \cup \YYT $,  $\A'\in\S$ and if $\pi_{\Ytop}(\XXT) \cap \pi_{\Ytop}(\YYT)=\emptyset$, then $\Z = \XXS_D - \X_1\cup \Ytop$.
    

    \item 
    $\Q=\XXT  - \YYT$,  $\A'\in\S$ and if all partitions  $\sigma_{\Ytop=y}(\XXT)$ are equal to or disjoint with $\sigma_{\Ytop=y}(\YYT)$, then
    $\Z=\XXS_D-\X_1\cup \Ytop$.
    
\end{enumerate}
\end{proposition}

\begin{myproof}
\noindent\emph{Unary queries:}
We shall use symbol $\Pi$ to denote the duplicate preserving projection and $\pi$ do denote duplicate eliminating projection. For each case of a unary query $\Q$ on some table $\XXT$ producing a table $\RT$, we first prove that, for any partition $\XXT^x=\sigma_{\Z=x}(\XXT)$ of $\XXT$ and $\RT^x=\sigma_{\Z=x}(\RT)$ of $\RT$, we have: (1) $\pi_{\Z,\A'}(\RT^x) = \pi_{\Z,\A'}(T^x)$ (both partitions are equal modulo duplicates), or (2) $\T^x$ is empty, or (3) $\RT^x$ is empty. We call the previous condition the \emph{G-summarizability condition on $\XXT$ and $\RT$ for $\Z$ and $\A'$}. If this condition holds, we can show that $\A'$ is G-summarizale in $\XXT$ with respect to $\Q$, grouping set $\Z$ and $\AGGF$ as follows:
\begin{itemize}
    \item In case (1), if $\Z\mapsto \A'$, then $\Pi_{\Z, \A'}(\RT^x) = \Pi_{\Z, \A'}(T^x)$ (both partitions are identical including duplicates) and it is obvious that any query, that aggregates $\A'$ using $\AGGF$ grouped by $\Z'$ containing all attributes in $\Z$, produces the same result on $\XXT$ and $\RT$ and the conditions for G-summarizability are fulfilled. Otherwise, if $\Z\not\mapsto \A'$, $\AGGF$ must be restricted to aggregation functions that are not sensitive to duplicates ($\AGGF \in \{\COUNTDISTINCT, \MIN, \MAX\}$). 
    \item In cases (2) and (3), the aggregated value does not exist, respectively, in the input table $\XXT$ or the result table $\RT$. This is also sufficient for satisfying the G-summarizability property.
\end{itemize}
Then, since $\aggp_{\A'}(\AGGF, \X_1)$ holds in $\RT$, any query that aggregates $\A'$ using $\AGGF$ grouped by $\Z'$ containing all dimension attributes in $\RRS\cap \XXS_D - \X_1$ is valid. Thus it is sufficient to show that $\Z$ contains all dimension attributes in $\RRS\cap \XXS_D - \X_1$. We call this condition the \emph{aggregable property condition}.

\begin{enumerate}
   
    \item Analytic filter $\RT=\FILTERQ_{\XXT}(P \mid \Y)$: By the condition $\Y\subseteq \XXS_D$, all attributes in $\Y$ are  dimension attributes. By  $\Z = \XXS_D - \X_1 \cup \Y$ we have $\Y\subseteq \Z$. Then for any non-empty partition $\XXT^x=\sigma_{\Z=x}(\XXT)$ of $\XXT$ we can show that if $P$ is true for some tuple in $\T^x$, it is true for \emph{all} tuples in $\T^x$: $\Z$ contains all attributes of filtering predicate $P$ and we can show that for all $\Z=x$ either  $\Z=x \Rightarrow P(\Y)$ or $\Z=x \Rightarrow \neg P(\Y)$. Therefore, the corresponding partition   $T_1^x=\sigma_{\Z=x}(\FILTERQ_{\XXT}(P \mid \Y))=\FILTERQ_{\XXT}(P\wedge \Z=x \mid \Z)$ is either empty or equal to $\XXT^x$.
    Hence, for any $\XXT^x$ and attribute $\A'$ in $\XXS$, we either have $\Pi_{\Z, \A'}(\RT^x) = \Pi_{\Z, \A'}(\XXT^x)$ (both partitions are identical with duplicates) or $\RT^x$ is empty. Finally, the aggregable property condition holds: $\RRS\cap \XXS_D - \X_1=\XXS_D - \X_1\subseteq \Z=\XXS_D - \X_1 \cup \Y$.
 
     \item Analytic projection $\RT=\PROJECTQ_{\XXT}(\Y, f(\Z') \rightarrow \M)$: By definition of analytic projection, for any subset $\X \subseteq \Y$, we have $\Pi_{\X}(\RT) = \Pi_{\X}(\XXT)$. Since $\Z  = \XXS_D - \X_1 \subseteq \XXS_D \subseteq  \Y$ and $\A'\in\Y$, we have $\Pi_{\Z,\A'}(\RT) = \Pi_{\Z,\A'}(\XXT)$. 
     Finally, the aggregable property condition holds: $\RRS\cap \XXS_D - \X_1=\XXS_D - \X_1\subseteq \Z=\XXS_D - \X_1$.

    \item Analytic aggregate $\RT=\AGGQ_{\XXT}(\AGGG(\A) \mid \Y)$:  
    By definition of aggregate queries, for any $\X \subseteq \Y$  and partition $\XXT^x$, we have $\pi_{\X}(\RT^x) = \pi_{\X,\A'}(\XXT^x)$ (duplicate eliminating projection). Since $\Z\subseteq \Y$ and $\A'\in \Y$, we then have $\pi_{\Z,\A}(\RT^x) = \pi_{\Z,\A}(\XXT^x)$. However, $\XXT^x$ generally contains several tuples that are merged into a single tuple in $\RT$. Therefore, $\AGGF$ must be restricted to functions that are not sensitive to duplicates ($\AGGF \in \{\COUNTDISTINCT, \MIN, \MAX\}$). 
     Finally,  the aggregable property condition holds: $\RRS\cap \XXS_D - \X_1=\Y - \X_1\subseteq \Z=\Y - \X_1$. 
    
    \item Analytic pivot $\RT=\PIVOTQ_{\XXT}(\A' \mid \Y)$: We apply similar arguments as for aggregate queries on the remaining attributes $\Y'=\XXS- \Y- \{\A\}$ in $\RT$. By definition of pivot queries, for any subset $\X \subseteq \Y'$ and partition $\XXT^x$, we have $\pi_{\X}(\RT^x) = \pi_{\X}(\XXT^x)$ (duplicate eliminating projection) and  $\pi_{\Z,\A'}(\RT^x) = \pi_{\Z,\A'}(\XXT^x)$ in particular for $\Z\subseteq \Y'$ and $\A'\in \Y'$. However, in the general case, $\XXT^x$  contains several tuples that are merged into a single tuple in $\RT$ and $\AGGF$ must be restricted to functions that are not sensitive to duplicates ($\AGGF \in \{\COUNTDISTINCT, \MIN, \MAX\}$). 
    Finally, the aggregable property condition holds: $\RRS\cap \XXS_D - \X_1=\XXS_D-\Y - \X_1 - \{\A\}\subseteq \Z=\XXS_D- \X_1-\Y- \{\A\}$.
  
 \end{enumerate}

\noindent\emph{Merge queries:}
For merge queries $\Q$ over two tables $\XXT$ and $\YYT$ producing a table $\RT$, we also check the G-summarizability condition and the aggregable property condition as for unary queries:
\begin{itemize}
    \item For any couple of partitions $\XXT^x=\sigma_{\Z=x}(\XXT)$ and $\RT^x=\sigma_{\Z=x}(\RT)$, either at least one of the two partitions is empty or $\pi_{\S}(\XXT^x)$ is equal to $\pi_{\S}(\RT^x)$. 
    \item $\Z$ contains all dimension attributes in $\RRS\cap \XXS_D - \X_1$.
\end{itemize} 
\begin{enumerate}
    \item Left-merge query $\XXT(\S) \leftouterjoin_\Y \YYT(\S')$:  
    By definition, $\pi_\S(\XXT) = \pi_\S(\RT)$ (each tuple of $\XXT$ produces one or more tuples in $\RT$ and vice versa). Thus, for any $\Z\cup\{\A'\} \subseteq \S_d$ we also have $\pi_{\Z,\A'}(\XXT) = \pi_{\Z,\A'}(\RT)$. By definition, a tuple in $\XXT$ can only appear twice in $\Pi_{\Z}(\RT)$ if $\Y \not \mapsto \S'$. If that is the case, $\AGGF$ must be restricted to functions that are not sensitive to duplicates: $\AGGF \in \{\COUNTDISTINCT, \MIN, \MAX\}$.  
    Finally, the aggregable property condition holds: $\RRS\cap \XXS_D - \X_1=\XXS_D - \X_1\subseteq \Z=\XXS_D - \X_1$.

    \item Right-merge query $\XXT(\S) \rightouterjoin_\Y \YYT(\S')$ or full-merge query $\Q=\XXT(\S) \fullouterjoin_\Y \YYT(\S')$: 
    
    \begin{enumerate}
     \item $\A' \in \Y$: We assume that for all non-empty partitions $\YYT^y=\sigma_{\Ytop=y}(\YYT)$ of $\YYT$, the corresponding partition $\XXT^y=\sigma_{\Ytop=y}(\XXT)$ is either empty or $\pi_Y(\XXT^y)$ is equal to $\pi_{\Y}(\sigma_{\Ytop=y}(\YYT))$. From this assumption and the definition of right-outer join, it directly follows that for any non-empty partition $\RT^y=\sigma_{\Ytop=y}(\RT)$ of $\RT$ and corresponding partition $\XXT^y$,  $\XXT^y=\emptyset$  or $\pi_{\S}(\XXT^y)=\pi_{\S}(\RT^y)$. Then, for all $\X\supseteq \Ytop$, all non-empty partitions $\pi_{\X}(\sigma_{\X=x}(\XXT))$ are equal to $\pi_{\X}(\sigma_{\X=x}(\RT))$ and since, by definition of $\Z$, $\Z\cup \{\A'\}\supseteq \Ytop$, the previous condition also holds for $\X=\Z\cup\{\A'\}$. 
     Therefore the G-summarizability condition holds on $\XXT$ and $\RT$ for $\Z$ and $\A'$.
     Finally, the aggregable property condition also holds: $\RRS\cap \XXS_D - \X_1=\XXS_D - \X_1\subseteq \Z=\XXS_D - \X_1 \cup \Y^{top}$.
     
     \item $\A' \in \S-\Y$: We assume that for all non-empty partitions, $\YYT^y=\sigma_{\Ytop=y}(\YYT)$, and corresponding partitions $\XXT^y=\sigma_{\Ytop=y}(\XXT)$, $\pi_Y(\XXT^y)$ is a subset of $\pi_{\Y}(\sigma_{\Ytop=y}(\YYT))$.  From this assumption and the definition of right-outer join, it directly follows that for any non-empty partition $\RT^y=\sigma_{\Ytop=y}(\RT)$ and corresponding partition $\XXT^y$, $\pi_{\S-\Y}(\RT^y)-\pi_{\S-\Y}(\XXT^y)$ only contains $null$ values. Then, for all $\X\subseteq \S_D$ and $\A' \in \S-\Y$, all non-empty partitions $\pi_{\X,\A'}(\sigma_{\X=x}(\XXT))$ are equal to $\pi_{\X}(\sigma_{\X=x\wedge \A'\neq null}(\RT))$ and since, by definition of $\Z$, $\Z\subseteq \S_D$, the previous condition also holds for $\X=\Z\cup\{\A'\}$. Therefore, the G-summarizability condition holds on $\XXT$ and $\RT$ for $\Z$ and $\A'$.
     Finally, the aggregable property condition also holds: $\RRS\cap \XXS_D - \X_1=\XXS_D - \X_1\subseteq \Z=\XXS_D - \X_1 \cup \Y$.
     
     \item Otherwise: From the definition of right-merge and full-merge, it directly follows that for any non-empty partition $\RT^y=\sigma_{\Y=y}(\RT)$, the corresponding partition $\XXT^y$ is either empty or equal to $\pi_{\S}(\RT^y)$. Then, for all $\X\supseteq \Y$, all non-empty partitions $\pi_{\X}(\sigma_{\X=x}(\XXT))$ are equal to $\pi_{\X}(\sigma_{\X=x}(\RT))$ and since, by definition of $\Z$, $\Z\cup \{\A'\}\supseteq \Y$, the previous condition also holds for $\X=\Z\cup\{\A'\}$. Therefore, the G-summarizability condition holds for $\XXT$ and $\RT$ for $\Z$ and $\A'$.
      Finally, the aggregable property condition also holds: $\RRS\cap \XXS_D - \X_1=\XXS_D - \X_1\subseteq \Z=\XXS_D - \X_1 \cup \Y$.
     \end{enumerate}
     
   \item $\Q=\XXT(\S) \bowtie_\Y \YYT(\S')$: 
    
    \begin{enumerate}
    \item "if" part: We assume  that for all non-empty partitions $\YYT^y=\sigma_{\Ytop=y}(\YYT)$, the corresponding partition $\XXT^y=\sigma_{\Ytop=y}(\XXT)$ is contained in  $\pi_{\Y}(\YYT^y)$. From this assumption and the definition of inner join, it directly follows that any non-empty partition $\RT^y=\sigma_{\Ytop=y}(\RT)$ in the result is equal to the corresponding partition $\XXT^y$: $\pi_{\S}(\XXT^y)=\pi_{\S}(\RT^y)$. Then, for all $\X\supseteq \Ytop$, all non-empty partitions $\pi_{\X}(\sigma_{\X=x}(\XXT))$ are equal to $\pi_{\X}(\sigma_{\X=x}(\RT))$ and since, by definition of $\Z$, $ \Z\cup \{\A'\}\supseteq \Ytop$, the previous condition also holds for $X=\Z\cup\{\A'\}$. Therefore the G-summarizability condition holds on $\XXT$ and $\RT$ for grouping set $\Z$ and attribute $\A'$.
  Finally, the aggregable property condition also holds: $\RRS\cap \XXS_D - \X_1=\XXS_D - \X_1\subseteq \Z=\XXS_D - \X_1 \cup \Y^{top}$.
 
  \item  "else" part: From the definition of inner join, it directly follows that for any non-empty partition $\RT^y=\sigma_{\Y=y}(\RT)$, the corresponding partition $\XXT^y$ is equal to $\pi_{\S}(\RT^y)$. Then, for all $\X\supseteq \Y$, all non-empty partitions $\pi_{\X}(\sigma_{\X=x}(\RT))$ are equal to $\pi_{\X}(\sigma_{\X=x}(\XXT))$ and since, by definition of $\Z$, $\Z\cup \{\A'\} \supseteq \Y$, the previous condition also holds for $\X=\Z\cup\{\A'\}$. Therefore the G-summarizability condition holds on $\XXT$ and $\RT$  for groupîng set $\Z$ and $\A'$.
    Finally, the aggregable property condition also holds: $\RRS\cap \XXS_D - \X_1=\XXS_D - \X_1\subseteq \Z=\XXS_D - \X_1 \cup \Y$.
    \end{enumerate}
\end{enumerate}

\noindent\emph{Set queries:} 
In the following, let $\Ytop\subseteq \XXS_D$ denote the set of "highest" attributes in the set of dimension attributes $\XXS_D\subseteq \XXS$ of $\XXT$ and $\YYT$.
\begin{enumerate}
  
    \item     $\Q=\XXT(\S) \cup \YYT(\S)$: By the assumption $\pi_{\Ytop}(\XXT) \cap \pi_{\Ytop}(\YYT)=\emptyset$ and the definition of union, it follows that for any non-empty partition $\RT^y=\sigma_{\Ytop=y}(\RT)$ in the result, the corresponding partition $\XXT^y$ is either empty or $\XXT^y$ is equal to $\RT^y$. Then, for all $\X\supseteq \Ytop$, all non-empty partitions $\pi_{\X}(\sigma_{\X=x}(\XXT))$ are equal to $\pi_{\X}(\sigma_{\X=x}(\RT))$ and since, by definition of $\Z$, $\Z \supseteq \Ytop$, the previous condition also holds for $\Z=\S_D-\X_1\cup \Ytop$. Finally, the aggregable property condition also holds: $\RRS\cap \XXS_D - \X_1=\XXS_D - \X_1\subseteq \Z=\XXS_D - \X_1 \cup \Y^{top}$.
    
    \item     $\Q=\XXT(\S) - \YYT(\S)$: By assumption, all partitions  $\sigma_{\Ytop=y}(\XXT)$ are equal to or disjoint with $\sigma_{\Ytop=y}(\YYT)$. Then, by the definition of set-difference, it follows that for any non-empty partition $\RT^y=\sigma_{\Ytop=y}(\RT)$ in the result, the corresponding partition $\XXT^y$ is either empty or $\XXT^y$ is equal to $\RT^y$. Then, for all $\X\supseteq \Ytop$, all non-empty partitions $\pi_{\X}(\sigma_{\X=x}(\XXT))$ are equal to $\pi_{\X}(\sigma_{\X=x}(\RT))$ and since, by definition of $\Z$, $\Z\supseteq \Ytop$, the previous condition also holds for $\X=\Z\cup\{\A'\}$. Therefore, the G-summarizability condition holds on $\XXT$ and $\RT$ for $\Z$ and $\A'$.
Finally, the aggregable property condition also holds: $\RRS\cap \XXS_D - \X_1=\XXS_D - \X_1\subseteq \Z=\XXS_D - \X_1 \cup \Y^{top}$.

\end{enumerate}
\end{myproof}

Observe that for merge queries and set queries we choose the "highest" dimension $\Ytop$ as candidates for checking the G-summarizability conditions. In fact, we might check this condition for any subset $\Y'$ of attributes from $\Y$ or $\XXS_D$ instead of $\Ytop$, and identify the minimal candidates for which these conditions hold. There are two main reasons for only choosing $\Ytop$. First, checking the G-summarizability condition for a subset of attributes mainly corresponds to comparing the size of partitions in two different tables obtained by two aggregate queries. This basic operation is costly and the systematic exploration of all attribute subsets $\Y'$ might, even with efficient pruning techniques, take too much time in an interactive data exploration session. Secondly, the choice of the highest attributes $\Ytop$ is based on the realistic hypothesis that the majority of analytic queries aggregate values along these attributes and other lower attributes.

\subsection{Controlling G-summarizability using aggregable properties}
\label{sec:control-g-sum}
 
The following proposition refines the propagation rules for aggregable properties of \Cref{sec:prop-agg-prop},  using the results of \Cref{prop:gen-summ}, to guarantee the G-summarizability of attributes.

\begin{proposition} [aggregable properties with G-summarizability for unary queries]
\label{cor:gen-summ}
Let $\Q$ be a unary analytic query with some input table $\XXT(\XXS)$ returning a table $\RRT(\S')$, and $\S_D$ be all the dimension attributes of $\XXS \cap \S'$. 
Let $\A'$ be an attribute in $\XXS \cap \S'$ such that $\aggp_{\A'}(\AGGF, \X)$ holds in $\XXT$.
Then, in the cases of queries $\Q$ of \Cref{tab:prop_rules_unary_gsum}, the aggregable property $\aggp_{\A'}(\AGGF, \X')$ holds in $\RRT$ and is such that for all $\Z$ where $\S_D - \X' \subseteq \Z$, $\A'$ is G-summarizable in $\XXT$ with respect to query $\Q$, grouping set $\Z$, and function $\AGGF$.
\end{proposition}

\begin{table}[htb]
\caption{Propagation rules for unary operations on $T(S)$ preserving G-summarizability}
\label{tab:prop_rules_unary_gsum}
\normalsize{
\bgroup
\begin{normalsize}
\def\arraystretch{1.2}
\begin{tabular}{l|l|l}
    \toprule
    Query on $\XXT(\S)$ & 
    \makecell[tl]{Propagation rule for inferring the aggregable    properties  for \\
    attributes $\A'\in \XXS \cap \RRS$ of the result $\RRT(\RRS)$ 
    } & 
    \makecell[tl]{
    \emph{User action}} \\
    \midrule
    $\FILTERQ_{\XXT}(P \mid \Y)$ 
    &  
    \makecell[tl]{
    attribute $\A' \in \RRS$, $\Y \subseteq S_D$ and $\aggp_{\A'}(\AGGF, \X)$ holds in $\XXT$:   \\
    \hspace{3mm} $\aggp_{\A'}(\AGGF, \X')$ holds in $\RRT$ with $\X'=\X - \Y$, $\X'_d$ = fact identifier\\
    \hspace{6mm}   and $\X'_f=\X_f - \Y$
    }  
    & 
    \makecell[tl]{ 
     \emph{Minimize $\X'_d$}
     }
    \\
    \midrule
    $\PROJECTQ_{\XXT}(\Y,f(\Z)\rightarrow \M)$ 
    & 
    \makecell[tl]{dimension attribute $\A' \in \Y$ and $\aggp_{\A'}(\AGGF, \X)$  holds in $\XXT$: \\
     \hspace{3mm} $\aggp_{\A'}(\AGGF, \X')$  holds in $\RRT$  with $\X'=\X$, $\X'_d=\X_d$ and $\X'_f=\X_f$.}
    & 
    \makecell[tl]{
    \emph{None}     }
    \\     
    \midrule
    $\PIVOTQ_{\XXT}(\A \mid \Y)$ 
    & 
    \makecell[tl]{
    attribute $\A'\in \RRS-\{\A\}$ and $\aggp_{\A'}(\AGGF, \X)$ holds  in $\XXT$: \\
    \hspace{3mm} if $\X_d\cap\Y=\emptyset$\\
    \hspace{3mm} then $\aggp_{\A'}(\AGGF, \X')$  holds in $\RRT$ with $\X'=\X-\Y$, $\X'_d=\X_d$  \\
    \hspace{9mm} and   $\X'_f=\X_f-\Y$ \\}
    & 
    \makecell[tl]{
    \\\ \\\emph{None}}
    \\  \cline{3-3}
    & 
     \makecell[tl]{
     \hspace{3mm} else $\aggp_{\A'}(\AGGF, \X')$  holds in $\RRT$ with $\X'$ \\   
        \hspace{9mm}  as defined by the rules of \Cref{tab:default_agg_prop} with $\X'_d=$ fact identifier\\
    \hspace{9mm}  and
    $\X'_f=\X_f-\Y$\\
    \hspace{3mm} and $\AGGF \in \{\MIN,\allowbreak \MAX,\allowbreak \COUNTDISTINCT\}$ }
     &
    \makecell[tl]{
    \\ \ \\
    \emph{Minimize $\X'_d$}}
 
   \\  \midrule
    $\AGGQ_{\XXT}(\AGGG(\A) \mid \Y)$ 
    & 
    \makecell[tl]{
    dimension attribute $\A' \in \Y$ and $\aggp_{\A'}(\AGGF, \X)$ holds in $\XXT$:  \\
     \hspace{3mm} $\aggp_{\A'}(\AGGF,\X')$ holds in $\RRT$ with $\X'= \X \cap \Y$ and $\X'_f=\X_f\cap \Y$. \\
     \hspace{3mm} and $\AGGF \in \{\MIN,\allowbreak \MAX,\allowbreak \COUNTDISTINCT\}$} 
     & 
     \makecell[tl]{
     \emph{None}}
     \\
    \bottomrule
\end{tabular}
\end{normalsize}
\egroup
}
\end{table}

\begin{myproof}
The proof mainly consists in defining the "new" $\X'$ as the "complement" of $\Z$ as defined in \Cref{prop:gen-summ} where $\X'$ is replaced by its definition in \Cref{tab:prop_rules_unary}. For example, for filter queries,  since $\Z=(\XXS_D-\X')\cup \Y$ and $\X'=\X$, we obtain $\Z=\XXS_D-(\X - \Y)$ and its complement $\X'=\RRS-\Z=\X - \Y$. For aggregation queries, $\Z=\Y-\X_1$ and $\X_1=\Y\cap \X$ and we obtain $\Z=\Y-(\Y\cap\X)$ and its complement $\X'=\Y-\Z=\Y\cap\X$.
\end{myproof}

We make the following observations on the rules of \Cref{tab:prop_rules_unary_gsum}. First, the rule for Project is unchanged with respect to \Cref{tab:prop_rules_unary}.
Second, when $Q$ is an aggregate query $\Q=\AGGQ_{\XXT}(\AGGG(\B) \mid \Y)$, the aggregable property for attribute $\AGGG(\B)$ is computed using the rule of \Cref{tab:prop_rules_unary} to guarantee the summarizability of attribute $B$. For attributes of $\Y$, the only refinement to the rule in \Cref{tab:prop_rules_unary} is to restrict the scope of $\AGGF$. The same observation applies to the refined propagation rule for a pivot query.  

\begin{proposition}[aggregable properties with G-summarizability for binary queries]
\label{prop:propagation_aggregable_binary} 
Let $\XXT(\XXS)$ and $\YYT(\YYS)$ be two analytic tables with dimension attributes $\XXS_D\subseteq \XXS$ and $\YYS_D\subseteq \YYS$ respectively, $\XXS_D^{top}p$ denote the highest attributes in $\XXS_D$ and $\Ytop$ denote the highest attributes in $\Y=\XXS_D\cap \YYS_D$.
Let $\RRT(\RRS)$ be the result of a binary query between  $\XXT$ and $\YYT$ and
 $\A' \in \RRS\cap \XXS$ be an attribute of $\XXT$ with aggregable property $\aggp_{\A'} (\AGGF,\X)$ holding in $\XXT$.  
Then, for all queries $\Q$ satisfying the conditions of \Cref{tab:prop_rules_binary_gsum}, the aggregable property $\aggp_\A(\AGGF, \X')$ holds in $\RRT$ and is such that for all $\Z$ where $\S_D - \X' \subseteq \Z$, $\A'$ is G-summarizable in $\XXT$ with respect to query $\Q$, grouping set $\Z$, and function $\AGGF$.
\end{proposition}

\begin{myproof}
As for unary queries, the proof mainly consists in defining the "new" $\X'$ as
the "complement" of $\Z$ as defined in \Cref{prop:gen-summ} where $\X'$ is replaced by its definition in \Cref{tab:prop_rules_unary}. 
For dimension attributes $\A'$, we can show that we always obtain a new $\X'$ which is equal to the old $\X'$ defined in \Cref{tab:prop_rules_unary}.
For example, for right-merge and full-merge queries, if $\A'$ is a dimension attribute in $\XXS_D$,  $\X'=\X\cup\YYS_D-\Y-\X'_f$ and in the first case where $\Z=(\XXS_D-\X')\cup \Ytop$, we obtain $\Z=\XXS_D-(\X\cup\YYS_D-\Y-\X'_f)\cup\Ytop=\XXS_D-(\X\cup\YYS_D-\Y-\X'_f-\Ytop)=\XXS_D-(\X\cup\YYS_D-\Y-\X'_f)$ which we also obtain in the second case: $\Z=(\XXS_D-\X')\cup \Y=\XXS_D-(\X\cup\YYS_D-\Y-\X'_f-\Y)=\XXS_D-(\X\cup\YYS_D-\Y-\X'_f)$.

For measure attributes, $\Z=\S_D-\X'$ and $\X'=\X$, we obtain the new $\X'=\X$.
\end{myproof}

We make the following observations on the rules of \Cref{tab:prop_rules_binary_gsum} and \Cref{tab:prop_rules_binary_gsum2}.
First, all rules refine the conditions and actions of the propagation rules of \Cref{tab:prop_rules_binary} by taking into account the restrictions described in \Cref{prop:gen-summ}.
Second, in the case of a right-merge, full-merge, union and difference query, 
it is possible to search for any subset $\Y'\subseteq \Y$ instead of $\Ytop$ for which the conditions on $\Ytop$ hold. If no such $\Y'$ is found then the set of attributes $\Y$ must be removed from $\X'$ (we illustrated that in \Cref{ex:merge-coverage}). 
Third, note that the propagation rule for right-merge assumes that $\A'$ is an exclusive attribute of $\XXT$. If $\A'$ is also in $\YYT$ then its aggregable property is computed using the propagation rule for left-merge.

\bgroup
\def\arraystretch{1.5}
\begin{normalsize}
\begin{tabularx}{\linewidth}{l|l|l}
\caption{Propagation rules for merge operations with G-summarizability}
\label{tab:prop_rules_binary_gsum}
    \\\toprule
    \makecell[tl]{Merge query on \\ $\XXT(\XXS)$ and $\YYT(\YYS)$
    } 
    & \makecell[tl]{Propagation rule for inferring the aggregable properties of \\ attributes $\A'\in \XXS$ in the result $\RRT(\S_r)$  } 
    & \makecell[tl]{  \emph{User action}} \\
    \midrule
    \endhead
    &
    \makecell[tl]{
    if $\Y \not \mapsto \S'$ then $\AGGF \in \{\COUNTDISTINCT, \MIN, \MAX\}$ 
    }
    & 
    \\*
    \cline{2-3}    
    \makecell[tl]{
    $\RRT = \XXT \leftouterjoin_{\Y} \YYT$}
    & 
    \makecell[tl]{
        dimension attribute $\A' \in \XXS_D$ and  $\aggp_{\A'}(\AGGF, \X)$ holds in $\XXT$: \\
        \hspace{3mm}  $\aggp_{\A'}(\AGGF, \X')$ holds in $\RRT$ with $\X'=\X \cup (\YYS_D - \X'_f)$ and $\X'_f=\X_f$   }
     & \makecell[tl]{\emph{Complete $\X'_f$}}  \\*
    \cline{2-3} 
    &
    \makecell[tl]{
        measure attribute $\A' \in \XXS-\XXS_D$ and $\aggp_{\A'}(\AGGF, \X)$  holds in $\XXT$  : \\ 
        \hspace{3mm} $\aggp_{\A'}(\AGGF, \X')$ holds in $\RRT$ with $\X'=\X$, $\X'_d=\X_d$ and $\X'_f=\X_f$.} 
     & 
     \makecell[tl]{ \emph{Complete $\X'_f$}}\\
     \midrule
     \pagebreak
    &
    \makecell[tl]{
    if $\Y \not \mapsto \S'$  then $\AGGF \in \{\COUNTDISTINCT, \MIN, \MAX\}$ 
    }
    &  \\*
    \cline{2-3}    
    \makecell[tl]{
    $\RRT = \XXT \rightouterjoin_{\Y} \YYT$ \\
    $\RRT = \XXT \fullouterjoin_{\Y} \YYT$
    }
    & 
    \makecell[tl]{
        dimension attribute $\A' \in \Y$ and  $\aggp_{\A'}(\AGGF, \X)$ holds in $\XXT$: \\
         \hspace{3mm} if $\pi_\Y(\sigma_{\Ytop=y}(\XXT))=\pi_\Y(\sigma_{\Ytop=y}(\YYT))$ or $\pi_\Y(\sigma_{\Ytop=y}(\XXT))=\emptyset$ \\  \hspace{9mm} for all non-empty partitions $\sigma_{\Ytop=y}(\YYT)$ : \\
        \hspace{3mm} then $\aggp_{\A'}(\AGGF, \X')$ holds in $\RRT$ with $\X' = \X \cup (\YYS_D - \X'_f) -\Ytop$ \\\hspace{9mm}  and $\X'_f=\X_f$.    \\
        \hspace{3mm} else $\aggp_{\A'}(\AGGF, \X')$ holds in $\RRT$ with $\X' = \X \cup (\YYS_D - \X'_f)-\Y$ \\\hspace{9mm} and $\X'_f=\X_f$.
        }
        &
        \makecell[tl]{ \emph{Complete $\X'_f$}}   \\ 
    \cline{2-3} 
    &
    \makecell[tl]{
    dimension attribute $\A' \in \S_D-\Y$ and  $\aggp_{\A'}(\AGGF, \X)$ holds in $\XXT$: \\
        \hspace{3mm} if $\pi_\Y(\sigma_{\Ytop=y}(\XXT))\subseteq\pi_\Y(\sigma_{\Ytop=y}(\YYT))$  or $\pi_\Y(\sigma_{\Ytop=y}(\XXT))=\emptyset$ \\ \hspace{9mm} for all non-empty $\sigma_{\Ytop=y}(\YYT)\neq \emptyset$: \\
        \hspace{3mm} then $\aggp_{\A'}(\AGGF, \X')$ holds in $\RRT$ with $\X' = \X \cup (\YYS_D - \X'_f)$ \\
        \hspace{3mm} else $\aggp_{\A'}(\AGGF, \X')$ holds in $\RRT$ with $\X' = \X \cup (\YYS_D - \X'_f)-\Y$ 
        \\\hspace{9mm} 
        and $\X'_f=\X_f$.
    }
     & 
     \makecell[tl]{
     \emph{Complete $\X'_f$}} \\
    \cline{2-3}
    &
    \makecell[tl]{
        measure attribute $\A' \in \XXS-\XXS_D$ and $\aggp_{\A'}(\AGGF, \X)$  holds in $\XXT$  :  \\
        \hspace{3mm} if $\pi_\Y(\sigma_{\Ytop=y}(\XXT))\subseteq\pi_\Y(\sigma_{\Ytop=y}(\YYT))$  or $\pi_\Y(\sigma_{\Ytop=y}(\XXT))=\emptyset$ \\  \hspace{9mm} for all non-empty $\sigma_{\Ytop=y}(\YYT)\neq \emptyset$: \\
        \hspace{3mm} then $\aggp_{\A'}(\AGGF, \X')$ holds in $\RRT$ with $\X' = \X$, $\X'_d=\X_d$ and $\X'_f=\X_f$.}
    & 
     \makecell[tl]{ \\\ \\ \emph{Complete $\X'_f$}\\ }  \\ 
     \cline{3-3}   &
    \makecell[tl]{
        \hspace{3mm} else $\aggp_{\A'}(\AGGF, \X')$ holds in $\RRT$ with $\X' = \X -\Y$ and $\X'_f=\X_f$.
    } 
     & 
     \makecell[tl]{ Recompute $\X'_d$}   \\
     \midrule
    &
    \makecell[tl]{ if $\Y \not \mapsto \S'$ then $\AGGF \in \{\COUNTDISTINCT, \MIN, \MAX\}$   }
    & 
    \\
    \cline{2-3}    
    \makecell[tl]{ $\RRT = \XXT \bowtie_{\Y} \YYT$}
    & 
    \makecell[tl]{
        dimension attribute $\A' \in \XXS_D$ and  $\aggp_{\A'}(\AGGF, \X)$ holds in $\XXT$: \\
        \hspace{3mm} if $\pi_\Y(\sigma_{\Ytop=y}(\XXT))\subseteq\pi_\Y(\sigma_{\Ytop=y}(\YYT))$ or $\pi_\Y(\sigma_{\Ytop=y}(\XXT))=\emptyset$ \\  \hspace{9mm} for all non-empty partitions $\sigma_{\Ytop=y}(\YYT)\neq \emptyset$: \\
        \hspace{3mm} then $\aggp_{\A'}(\AGGF, \X')$ holds in $\RRT$ with $\X' = \X \cup (\YYS_D - \X'_f)-\Ytop$  \\\hspace{9mm} and $\X'_f=\X_f$.\\
        \hspace{3mm} else $\aggp_{\A'}(\AGGF, \X')$ holds in $\RRT$ with $\X' = \X \cup (\YYS_D - \X'_f)-\Y$  \\\hspace{9mm} and $\X'_f=\X_f$.
        }
     & \makecell[tl]{ \emph{Complete $\X'_f$}}
    \\
    \cline{2-3}
    &
    \makecell[tl]{
        measure attribute $\A' \in \XXS-\XXS_D$ and $\aggp_{\A'}(\AGGF, \X)$  holds in $\XXT$  : \\ 
        \hspace{3mm} if $\pi_\Y(\sigma_{\Ytop=y}(\XXT))\subseteq\pi_\Y(\sigma_{\Ytop=y}(\YYT))$ or $\pi_\Y(\sigma_{\Ytop=y}(\XXT))=\emptyset$ \\  \hspace{9mm} for all non-empty  partitions $\sigma_{\Ytop=y}(\YYT)\neq \emptyset$: \\
        \hspace{3mm} then $\aggp_{\A'}(\AGGF, \X')$ holds in $\RRT$ with $\X' = \X -\Ytop$ and $\X'_f=\X_f$.\\
        \hspace{3mm} else $\aggp_{\A'}(\AGGF, \X')$ holds in $\RRT$ with $\X' = \X -\Y$ and $\X'_f=\X_f$
        }
     & 
     \makecell[tl]{\emph{Recompute $\X'_d$}\\     \emph{Complete $\X'_f$}}
    \\\bottomrule
\end{tabularx}
\end{normalsize}
\egroup

\subsection{Wrapping up results on summarizability}
To wrap up our results on summarizability and G-summarizability, we illustrate them using the motivating example presented in the introduction of this paper. We then discuss some directions for future work around the generation of explanations associated with the result of an analytic query.  

\bgroup
\def\arraystretch{1.5}
\begin{normalsize}
\begin{tabularx}{\linewidth}{l|l|l}
\caption{Propagation rules for set operations with G-summarizability}
\label{tab:prop_rules_binary_gsum2}
    \\\toprule
    \makecell[tl]{Set query on \\ $\XXT(\XXS)$ and $\YYT(\YYS)$
    } 
    & \makecell[tl]{Propagation rule for inferring the aggregable properties of \\ attributes $\A'\in \XXS$ in the result $\RRT(\S_r)$  } 
    & \makecell[tl]{  \emph{User action}} \\
    \midrule
    \endhead

    \makecell[tl]{ 
    $\RRT=\XXT \cup \YYT$}
    & \makecell[tl]{ 
         dimension attribute $\A' \in \XXS_D$  and $\aggp_{\A'}(\AGGF, \X)$  holds in $\XXT$ and $\YYT$:  \\
        \hspace{3mm} if $\pi_{\S_D^{top}}(\XXT)\cap \pi_{\Ytop}(\YYT)=\emptyset$\\
        \hspace{3mm} then $\aggp_{\A'}(\AGGF, \X')$ holds in $\RRT$ with $\X'=\X-\Ytop$ and $\X'_f=\X_f$}
    & \makecell[tl]{
        \\\ \\\emph{Recompute $\X'_d$}
    }
    \\* 
    \cline{2-3}    
    & \makecell[tl]{ 
         measure attribute $\A' \in \S$  and $\aggp_{\A'}(\AGGF, \X)$ holds in $\XXT$ and $\YYT$:  \\
        \hspace{3mm} if  $\pi_{\Ytop}(\XXT)\cap \pi_{\Ytop}(\YYT)=\emptyset$  \\
        \hspace{3mm} then if $\X_d\mapsto \A'$ holds in $\RRT$\\
        \hspace{9mm} then $\aggp_{\A'}(\AGGF, \X')$ holds in $\RRT$ with $\X'=\X-\Ytop$ and $\X'_f=\X_f$. }
    & \makecell[tl]{
         \\\ \\\ \\\emph{Recompute $\X'_d$}
        }
    \\*
    \cline{2-3}    
    & \makecell[tl]{ 
        \hspace{9mm} else $\aggp_{\A'}(\AGGF, \X')$ holds in $\RRT$ with $\X'=\X-\Ytop$,  \\
        \hspace{15mm} $\X'_d=\X_d$ and $\X'_f=\X_f$
     }
    & \makecell[tl]{
    \\ \emph{Minimize $\X'_d$}
    }
    \\\midrule    
    \makecell[tl]{ 
    $\RRT=\XXT - \YYT$}
    & \makecell[tl]{ 
   $\A' \in \RRS$ 
        and $\aggp_{\A'}(\AGGF, \X)$ holds in $\XXT$ and $\YYT$ :  \\
        \hspace{3mm}  if all partitions $\sigma_{\Ytop}(\XXT)$ are equal to or disjoint with $\sigma_{\Ytop}(\YYT)$\\
        \hspace{3mm} then $\aggp_{\A'}(\AGGF, \X')$ holds in $\RRT$ with $\X'= \X-\Ytop$ and $\X'_f=\X_f$   \\
         }
        
    & \makecell[tll]{
        Recompute $\X'_d$
        \\\emph{None}} 
    \\\bottomrule
\end{tabularx}
\end{normalsize}
\egroup

\begin{myexample}
Consider the example of interactive data analysis session of \Cref{fig:sales-dem-usa} on the tables in \Cref{tab:session1}. Table \factt{T3} is obtained by a filter query $\FILTERQ_{\factt{STORE\_SALES}}(P|\attr{COUNTRY,YEAR})$ and attribute \attr{Amount} in \factt{STORE\_SALES} is aggregable along all dimension attributes except attribute $\attr{Year}$.  By the G-summarizability rule for filter queries for attribute \attr{Amount} in \factt{T3}, we obtain the aggregable property $\aggp_{\attr{Amount}}(\SUM, \X_3)$ where $\X_3$ now contains all attributes of \factt{STORE\_SALES} except \attr{Country} and \attr{Year}. Since 
the table \factt{T4} is obtained by summarizing $\attr{AMOUNT}$ by $\Z=\attr{City, State, Country, Year}$, the aggregate operation over \factt{T3} leading to \factt{T4} is \emph{correct} with respect to the G-summarizability of \attr{Amount}. 

Next, the table \factt{\T5} is obtained by a  merge query adding the attribute \attr{Pop} from table $\dimt{DEM}$ to table \factt{T4}. For attribute \attr{Pop} in \factt{T5}, we have $\aggp_{\attr{Pop}}(\AGGF, \X_5)$, where $\AGGF$ and $\X_5$ are defined according to \Cref{tab:prop_rules_binary_gsum} for right merge (\dimt{DEM} is the "outer" merge table). We have join attributes $\Y= \{\attr{City}, \attr{State}, \attr{Country}, \attr{Year}\}$ with $\Ytop = \{\attr{Country}, \attr{Year}\}$. However, $\pi_\Y(\sigma_{\attr{Country}='USA'\wedge \attr{Year}=2018}(\factt{T4}))\subset \pi_Y(\sigma_{\attr{Country}='USA'\wedge \attr{Year}=2018}(\factt{DEM}))$ (city of 'Palo Alto' is missing in \factt{T4}). 
Thus, we have $\X_5 = \Y$ in the above aggregable property.  
Consequently, the aggregation of \attr{SUM(POP)} along \attr{CITY} is \textbf{incorrect} with respect to the G-summarizability property. 

Table \dimt{DEM'} in \Cref{tab:demprime} is obtained by  aggregating the population in table \dimt{DEM} along attribute \attr{City}.   In table \dimt{DEM}, we have $\aggp_{\attr{Pop}}(\SUM, \X)$ with $\X = \{\attr{City}, \attr{State}, \attr{Country}\}$. So the aggregation along \attr{City} leading to table \dimt{DEM'} is \emph{correct}, and by \Cref{def:propagate-agg-prop}, $\aggp_{\attr{Pop}}(\SUM, \X')$  holds in \dimt{DEM'} with $\X' = \X\cap\Y= \{\attr{State}, \attr{Country}\}$ (function $\SUM$ is distributive). Next, in the merge result of \factt{T4} with \factt{DEM'}, we have $\aggp_{\attr{Pop}}(\AGGF, \X_4)$, where $\AGGF$ and $\X_4$ are defined by the rule for outer merge (\dimt{DEM'} is the outer merge table). We have $\Y= \{\attr{State}, \attr{Country}, \attr{Year}\}$ and $\Ytop = \{\attr{Country}, \attr{Year}\}$. We also have $\pi_Y(\factt{DEM} \ltimes_{\Ytop} \factt{T4}) = \pi_Y(\factt{DEM})$, and since $\aggp_{\attr{Pop}}(\SUM, \X')$ holds in \factt{DEM'}, we get $\X_4 = X' - \Ytop = \{\attr{State}\}$. Finally, since $\Y$ does not literally determine the schema of \factt{T4}, $\AGGF$ must be restricted to one of $\{\COUNTDISTINCT, \MIN, \MAX\}$. 
\end{myexample}

The question that naturally arises is what options should be provided to the end user when G-summarizability is violated (i.e., the  conditions given by aggregable properties are not satisfied). 
A first option is to reject an incorrect aggregate query with respect to G-summarizability and return the grouping set of the aggregable property as an explanation. This is the option we have described in this paper. However, another option could be to accept the aggregate query provided that some metadata is added to the resulting table to enable a non-ambiguous and correct interpretation of the tuples in that table. 
We explain the idea in the next example and leave it for future work.

\begin{myexample}
\label{ex:explainability}
Consider again the interactive session of \Cref{fig:sales-dem-usa} and suppose that the first filter query on  \factt{STORE\_SALES} is: $\attr{state} \neq null$ (expecting that this eliminates all European countries) and \attr{year} = '2018'. 
Next, suppose that the aggregate operation over \factt{T3} sums \attr{Amount} for each partition defined by attributes \attr{Country} and \attr{Year}. It will be difficult for an end user to figure out that an incorrect aggregate value has been computed for country 'USA' if the user ignores that 'Washington DC' has no state. With our current proposition, this aggregate operation will be rejected since the grouping set does not include attribute \attr{State} (we would have $\aggp_{\attr{Amount}}(\SUM, \{\attr{City}, \attr{Country}\})$ in \factt{T3}). However, to disambiguate the result of the aggregate operation, it would be sufficient to "attach" the filter condition $\factt{STORE\_SALES}.\attr{state} \neq null$ as metadata to table \factt{T4} to indicate that the amount for stores in those states has not been accounted for. It is then possible for the end user to query those stores from table  \factt{STORE\_SALES} to visualize them and decide if table \factt{T4} is satisfactory. 

The same principle applies to the result of the  left-merge operation between \factt{T4} and \factt{DEM}. As we have seen in the previous example, we have $\aggp_{\attr{Pop}}(\SUM, \X_5)$ in \factt{T5} and the aggregate operation on \factt{T5} is rejected. However, we could accept the operation and simply mark that measure attribute \attr{SUM(Pop)} now depends on dimension \dimt{SALESORG}. This would indicate that the population is summed for the cities in \dimt{SALESORG}, that is, the cities that have stores. 
\end{myexample}

The idea is therefore to make each analytic table, resulting from an interactive data analysis session, "self-explanatory" with respect to its aggregated attributes.


\section{Related Work}
\label{sec:related-work}
\label{sec:state-of-art-summarizability}

In this section, we focus on previous works that propose conditions on the schema of a fact table, or on the parameters of an aggregate query expressed over that fact table, to determine if the aggregate query returns a correct result with respect to some summarizability definition. 
Previous papers on summarizability use heterogeneous notations and concepts and are sometimes difficult to read because they lack some details or hide some assumptions. To facilitate comparisons with our work, we reformulated each previous proposition using the notations introduced in this paper. 

In our detailed analysis, we establish the following:
\begin{enumerate}
    \item Our data model is more general than the data models considered by previous work.
    \item In the case of a sequence of two aggregate queries, $Q_1$ followed by $Q_2$, our sufficient conditions to determine if $Q_2$ is correct subsume the conditions proposed by previous work.
    \item To our best knowledge, no previous work addressed the case of a sequence made of an arbitrary analytic query followed by an aggregate query, which is addressed by our notion of G-summarizability.
\end{enumerate}

\subsection{Summarizability of a query over a statistical object}
\label{sec:summarize-stats}

The notion of \emph{summarizability} was initially defined by Rafanelli and Shoshani \cite{rafanelli_storm:_1990} for  statistical databases and later refined in their seminal paper \cite{lenz_summarizability_1997}. In their context, \emph{base data}, also referred to as "micro-data", describe all the details about the objects or individuals over which a summarization operation can be applied to produce a so-called \emph{statistical object}, also referred to as "macro-data". 
There are a few constraints. In the base data, an \emph{object of interest} must be identified (e.g., a product, a customer, a store) using some attributes, all other attributes being viewed as "descriptors" of the object. A statistical object is a table defined by a summarized attribute (i.e., an attribute of the base data on which a summarization function is applied) and a set of "category" attributes defining the partitions of the base table on which the summarization function is applied. 
Using the terminology defined in \Cref{sec:model}, base data can be modelled as a non-analytic table and summarization operations are aggregate queries which ignore partitions with null values in their partition identifiers. We shall keep the expression "summarization operation" to distinguish it from our analytic aggregate operation that handles null values as regular values. A statistical object can be modeled as a fact table that results from a summarization operation over the base data where category attributes represent dimensions and the summarized attribute is a measure.  

The fact tables that represent statistical objects in \cite{lenz_summarizability_1997} are however more restricted than the fact tables enabled by our data model. 
Firstly, dimensions are restricted to \emph{strict} hierarchies, that is, each attribute has at most one parent attribute in the hierarchy type, each attribute value of a dimension has at most one parent attribute value in the hierarchy of the dimension, and hierarchy types must have a single bottom and top level attribute.
Secondly, all dimensions in a fact table must be independent (that is, no attribute in some dimension functionally depends on an attribute of another dimension).
Finally, all facts in a fact table have the same dimensions, i.e., the measure attribute does not depend on a subset of the dimensions.

An important distinction, with respect to multidimensional data models, is that there is no notion of managed dimensions in \cite{lenz_summarizability_1997, rafanelli_storm:_1990}. The notion of dimension hierarchy is purely local to a statistical object and depends on the functional dependencies that are supposed to hold in the base data on which the object is built. If these dependencies change, the dimensions hierarchies are adjusted to fit the strictness constraint explained before. 

In \cite{lenz_summarizability_1997, rafanelli_storm:_1990}, summarizability is defined as the property of a summarized attribute $\AGGF(\A)$ of a fact table  (statistical object) $\XXT$ which guarantees that a summarization operation $\AGGG$ over $\AGGF(\A)$produces a  \emph{correct} result. 
Suppose we have a base table $\T_0$ and a fact table $\XXT(S)$, which results from a summarization operation applying aggregation function $\AGGF$ on attribute $\A$ along attributes $\X$: $$\XXT = \AGGQ^*_{\T_0}(\AGGF(A) \mid \X)$$ where $\AGGQ^*$ denotes a summarization operation that does not consider partitions where an attribute of $\X$ has a null value. Let $\Q$ be a summarization query over $\XXT$: $$\Q=\AGGQ^*_{\XXT}(\AGGG(\AGGF(\A)) | \Z)$$ where $\AGGG$ is a function applicable to the summarized attribute $\AGGF(\A)$ of $\XXT$, and $\Z$ is a set of dimension attributes in $\X$.  
Then, the summarization query $\Q$ is said to be \emph{correct}, if the following condition holds: $$\AGGQ^*_{\T_0}(F(\A) | \Z) = \AGGQ^*_{\XXT}(G(F(\A)) | \Z)$$ 

 To guarantee the correctness of summarization query $\Q$, \cite{lenz_summarizability_1997} defines three necessary properties on the summarization query $\Q$ and the dimensions $\D$ of $\T_0$. Let $\X_\D\subseteq \X$ be the set of dimension attributes for dimension $\D$ in $\XXT$, $\X_\D^{bot}\in X_D$ be the bottom level attribute of $\D$ in $\XXT$ and $D^{bot}$ be the bottom level attribute of $\D$ in $\T_0$ (then, $D^{bot} = X_D^{bot}$ or $D^{bot} \typeprec^{*} X_D^{bot}$).
The properties are:

\begin{enumerate}
    \item \emph{Disjointness}. For each dimension $\D$ along which summarization is done in $\Q$, at least one of the following conditions must hold: 
    (a) $\X_\D - Z=\{D^{bot}\}$ consists of the bottom level attribute of $\D$ and the partitions of $\T_0$ using $D^{bot}$ are disjoint with respect to the identifier attributes of the object of interest in $\T_0$; 
    (b) every value in $\T_0$ of a dimension attribute $\A_1$ that is a below an attribute $\A_2$ in $\X_\D - Z$ in $\D$ ($\A_1\typeprec^{*} \A_2$) must map to a single value of its parent attribute (many-to-one mapping);
    \item \emph{Completeness with respect to $F(\A)$}. For each dimension $\D$ of $\XXT$, the domain of each attribute in $\X_\D$ is \emph{complete} in $\T_0$, if both of the following conditions hold: 
    (a)  
    all values of the identifier attributes of the object of interest which are required by $F(\A)$ appear in $\T_0$, 
    and 
    (b) one of the two following conditions holds: 
    if $\X_\D - Z=\{D_{bot}\}$ consists of the bottom level attribute of $\D$ in $\T_0$ then the value of this attribute cannot be \emph{null} in $\T_0$. Otherwise,
     every value of a dimension attribute that is a child of $\X_\D^{bot}$ must map to a parent value in $\X_\D^{bot}$ within $\T_0$. 
    \item \emph{Applicable summary function}. The summary function $\AGGG$ is "applicable" to the summarized attribute $\AGGF(A)$ with respect to all dimensions along which summarization is done in $\Q$. 
\end{enumerate}

We now analyze each one of the previous conditions and relate it to our work.
\paragraph{Disjointness:} 
    In the original formulation of \cite{lenz_summarizability_1997}, the disjointness property requires that the dimension attributes along which summarization is done form disjoint subsets over the "objects of interest" defined in the base table. Two different disjointness conditions are then given, depending on whether the dimension attribute is a bottom attribute of the dimension or not.  
    The goal of the disjointness property is mainly to avoid double counting by overlapping subsets.
    In our work, we address the disjointness property by 
    defining aggregable properties and propagation rules. Each aggregable property describes for a given  attribute $\A$ in fact table $\T_0$ along which attributes it can be aggregated using some function $\AGGF$. The summarizability preserving propagation rules  then produce all aggregable properties of $\AGGF(A)$ in $\XXT$ after the aggregation operation on $\T_0$ is done (see \Cref{def:propagate-agg-prop-try}). 
    Summarizability is defined using the notion of function distributivity and literal functional dependencies.  
    There is no need to choose an object of interest in the base data for defining the scope of summarizability, and any attribute can be summarized.     

\paragraph{Completeness:} In completeness condition (2.a) of  \cite{lenz_summarizability_1997}, it is not clear how the set of "all possible values" for the identifier attributes is determined to assess completeness. Thus, we used the interpretation that the possible values are those listed in some reference directory. Furthermore, condition (2.a) applies to the identifier values required for computing $F(\A)$ which means that the user who is formulating query $\Q$ must decide whether completeness is needed or whether the values listed in $\XXT$ are sufficient to compute a summary attribute. We shall see in \Cref{ex:summarizability} that this condition (2.a) is useless for checking summarizability. 
 The Item (2.b) of the completeness condition 
    is required because summarization operations cannot deal with attributes of $Z$ that have null values. In our work, we do not have such a restriction since our SQL aggregate operations handle null values as regular values.  

\paragraph{Function applicability:} The third condition focuses on testing the compatibility between the type of dimensions and the type of measures used in $\XXT$. The following types of measures can be used by the designer of a fact table: \emph{stock} (i.e., a simple value at a particular point in time), \emph{flow} (i.e., cumulative values over a period of time) and  \emph{value-per-unit} (i.e, determined value for a fixed time). Dimensions can be of type temporal or non-temporal. When a measure attribute $\A$ is aggregated using a given function over some dimension $\D$, the types of $\A$ and $\D$ should be compatible with respect to that function.  
   In our work, applicable functions are captured by the more general notion of aggregable property, which leaves the method to decide which function is applicable to an attribute open. Furthermore, we use propagation rules and default rules to infer the functions that are applicable to an attribute that has been aggregated. Thus, a user is not forced to define the type of the measure attributes in $\XXT$ since aggregable properties for these attributes will be automatically computed using propagation rule (see \Cref{def:propagate-agg-prop-try}).

The following two examples illustrate the conditions of \cite{lenz_summarizability_1997}
and emphasize the differences with our work.

\begin{table}[htb]
    \caption{Table \factt{PRODUCT\_LIST}}
    \label{tab:prod-list2}
    \centering
    \normalsize{
    \begin{tabular}{l|l|l|l|r}
        \toprule
         \underline{\attr{PROD\_SKU}} & \underline{\attr{BRAND}}& \attr{COUNTRY} & \underline{\attr{YEAR}} & \attr{QTY} \\
         \midrule
         cz-tshirt-s & Coco Cola & USA & 2017 & 5 000 \\
         cz-tshirt-s & Coco Cola & USA & 2018 & 7 000 \\         
         cz-tshirt-s & Zora & Spain & 2017 & 5 000 \\
         cz-tshirt-s & Zora & Spain & 2018 & 7 000 \\
         coco-can-33cl & Coco Cola & USA & 2017 & 10 000 \\
         \bottomrule
    \end{tabular}
    }
\end{table}

\begin{myexample}
\label{ex:summarizability2}
Consider the base table \factt{PRODUCT\_LIST} (\attr{PROD\_SKU}, \attr{COUNTRY}, \attr{BRAND}, \attr{YEAR},  \attr{QTY}), whose instance is displayed on \Cref{tab:prod-list2}, and where the "object of interest" is a product identified by \attr{PROD\_SKU}.
To comply with the constraint of strict dimensions defined by \cite{lenz_summarizability_1997}, attribute \attr{PROD\_SKU} must belong to a separate dimension (it determines no other attribute), attributes \attr{BRAND}, \attr{COUNTRY} belong to a dimension \dimt{MKT\_PROD}, and attribute \attr{Year} belongs to a dimension \dimt{TIME}.  

Next, suppose that we define a statistical object 
\factt{PRODUCT\_SUM} (\attr{BRAND}, \attr{YEAR},  \attr{NB\_PROD\_SKU}) built using a summarization query with function $\AGGF=\COUNTDISTINCT$:  
$$\factt{PRODUCT\_SUM}=\AGGQ^*_{PRODUCT\_LIST}(\COUNTDISTINCT(\attr{PROD\_SKU}) | \attr{Brand}, \attr{Year})$$
The result is displayed on \Cref{tab:prod-sum} (we renamed the summarized attribute as \attr{NB\_PROD\_SKU}). The user should associate the summarized attribute \attr{NB\_PROD\_SKU} with a type \emph{flow} since the number of distinct products sold is cumulative over time.

\begin{table}[htb]
    \caption{Table \factt{PRODUCT\_SUM}}
    \label{tab:prod-sum}
    \centering
    \normalsize{
    \begin{tabular}{l|l|r}
        \toprule
         \attr{BRAND} & \attr{YEAR} & \attr{NB\_PROD\_SKU} \\
         \midrule
         Coco Cola & 2017 & 2 \\
         Coco Cola & 2018 & 1 \\         
         Zora & 2017 & 1 \\
         Zora & 2018 & 1 \\
         \bottomrule
    \end{tabular}
    }
\end{table}

The first summarization query $\Q_1$ aggregates \attr{NB\_PROD\_SKU} along \attr{Year} using function $\AGGG=\SUM$ to count the number of distinct products by brand :
$$Q_1 = \AGGQ^*_{PRODUCT\_SUM}(\SUM(\attr{NB\_PROD\_SKU}) | \attr{Brand})$$ 

Since $\attr{Year}$ is a bottom attribute of dimension $\D=\dimt{Time}$ in $\T_0$, Item (1.a) of the  disjointness condition must be tested over $\T_0$. It fails because the product \enquote{cz-tshirt-s} belongs to two different partitions of $\T_0=\factt{PRODUCT\_LIST}$ by \attr{Year}. 
Thus, query $Q_1$ is incorrect. Indeed, the number of distinct products by brand reported by query $Q_1$ (e.g., value $2$ for brand Zora) would be different from the number directly computed from \factt{PRODUCT\_LIST} (e.g., 1 for brand Zora).  

The second summarization query $\Q_2$  aggregates \attr{NB\_PROD\_SKU} along \attr{Brand}: 
$$Q_2 = \AGGQ^*_{PRODUCT\_SUM}(\SUM(\attr{NB\_PROD\_SKU}) | \attr{year})$$ 
Since \attr{Brand} is again a bottom attribute of dimension \dimt{MKT\_PROD} in $\T_0$, Item (1.a) of the disjointness condition must be tested. It fails since  the product \enquote{cz-tshirt-s} maps to different partitions of $\T_0$ by \attr{Brand}. Thus, query $Q_2$ is also incorrect.
Again, the number of distinct products by year reported by query $Q_2$ (e.g., value 2 for year 2018) would be different from the number directly computed from \factt{PRODUCT\_LIST} (e.g., 1 for year 2018).   

By comparison with our work, 
let's assume that the aggregable property $\aggp_{\attr{PROD\_SKU}}(\COUNTDISTINCT \mid \Z)$, where $\Z = \{\attr{BRAND}, \attr{COUNTRY}, \attr{YEAR}\}$, has been validated by the designer of table \factt{PRODUCT\_LIST}. Then, by   \Cref{def:propagate-agg-prop}, we infer that the aggregable property controlling summarizability, $\aggp_{\attr{NB\_PROD\_SKU}}(\SUM \mid \emptyset)$, holds in \factt{PRODUCT\_SUM}. Therefore, we also detect that both $Q_1$ and $Q_2$ are incorrect. However, our detection does not require running any query over the base data, unlike the disjointness condition of \cite{lenz_summarizability_1997}. We only use the knowledge of the attribute graphs of the dimensions and of the aggregable properties defined on fact tables. 
\end{myexample}

\begin{myexample}
\label{ex:summarizability}
Suppose that we have a base object, represented by table \dset{STORE\_SALES} introduced earlier (see \Cref{tab:state-summarizability-sales2}), in which stores are the objects of interest (identified by \attr{STORE\_ID}). A statistical object, modeled by fact table \dset{STORE\_SALES\_YEARLY}  (displayed in \Cref{tab:store_sales_all_years}), is computed using the following summarization query:
$$\dset{STORE\_SALES\_YEARLY}=\AGGQ^*_{STORE\_SALES}(\SUM(AMOUNT) | \attr{City}, \attr{State}, \attr{Country}, \attr{Year})$$

We assume that the summarized attribute \attr{SUM(AMOUNT)} is renamed as \attr{AMOUNT} and has been associated with type \emph{flow}, i.e., it can be summed along any dimension. 
To fulfill the constraints on dimensions, there will be four dimension hierarchies respectively formed of the following attributes: \{\attr{City}\}, \{\attr{State}\}, \{\attr{Country}\}, and \{\attr{Year}\}. 

  \begin{table*}[htb]
  \caption{Results of summarization queries}
      \normalsize{
      \begin{subtable}[b]{.58\linewidth}
    \centering
     \caption{\factt{STORE\_SALES\_YEARLY}}
    \label{tab:store_sales_all_years}

    \begin{tabular}[t]{l|l|l|l|r}
      \toprule
    \attr{City} & \attr{State} & \attr{Country} & \attr{Year} & \attr{Amount} \\
    \midrule
    Dublin & Ohio & USA & 2017 & 3.2 \\
    Dublin & california & USA & 2017 & 5.3 \\
    Dublin & Ohio & USA &  2018 & 8.2 \\
    Dublin & California & USA & 2018 & 6.3 \\
      \bottomrule
    \end{tabular}
\end{subtable}
\hfill
  \begin{subtable}[b]{0.40\linewidth}
  \centering
     \caption{Result of $\Q_3(\factt{STORE\_SALES\_YEARLY})$}
    \label{onto-results-q1}
    \begin{tabular}[t]{l|l|r}
      \toprule
    \attr{Country} & \attr{Year} & \attr{SUM(AMOUNT)} \\
    \midrule
    USA & 2017 & 8.4 \\
    USA & 2018 & 14.5 \\
      \bottomrule
    \end{tabular}
    \end{subtable}
    }
\end{table*}

Consider the following summarization query $\Q_3$ whose result is displayed in \cref{onto-results-q1}: $$Q_3=\AGGQ^*_{STORE\_SALES}(\SUM(AMOUNT) | \attr{Country}, \attr{Year})$$ 

The disjointness condition is obviously satisfied since a store in table \factt{STORE\_SALES} belongs to a single partition by \attr{City} and a single partition by \attr{State}. 
Now, assume that the directory of all possible values for \attr{STORE\_ID} is given by the dimension table \dimt{SALESORG} of \Cref{tab:state-summarizability-sales2}. Then, the  completeness condition (2.a) is violated 
since store $Ca\_02$ is missing in the list of values of \attr{STORE\_ID} in \factt{STORE\_SALES}. However, from a strict summarizability point of view, it is easy to see that the result of $Q_3$ is the same as the result of the same aggregation executed over table \factt{STORE\_SALES}. So completeness constraint (2.a) is useless for checking summarizability.

Completeness constraint (2.b) is also violated since bottom attribute \attr{State},
has a \emph{null} value in \factt{STORE\_SALES} for city \enquote*{Paris}.
Hence, there is no tuple for \enquote*{Paris} in \factt{STORE\_SALES\_YEARLY} and therefore also no tuple for country \enquote*{France} in the result of $Q_3$. The summarizability property is thus clearly violated since the result of $Q_3$ applied to \factt{STORE\_SALES} returns a different result containing the tuple $(France, 2017, 45.1)$. 
In our work, \factt{STORE\_SALES\_YEARLY} would contain a tuple for city \enquote*{Paris} with a null value for \attr{State}, and therefore a tuple for country \enquote*{France} in the result of $Q_3$.

\end{myexample}

\subsection{Summarizability of attributes in fact tables}
\label{subsec:multidim_summarizabilty}
In previous works on multidimensional databases, summarizability is expressed over fact tables in a way similar to our \Cref{def:summarizable-attr} illustrated by \Cref{fig:summ-att-fig1}. In this context, we first review the work of \cite{pedersen_extending_1999, pedersen_foundation_2001} that uses a multidimensional data model close to ours, and generalizes the summarizability conditions of \cite{lenz_summarizability_1997}.   
Note that other solutions propose alternative  analytic data models and methods to modify the representation of dimensions to enforce summarizability. In the following, we do not consider these solutions for which a survey can be found in \cite{mazon_survey_2009}. 
Thus, like in our work, we concentrate on summarizability models  which characterize correct compositions of two aggregation queries over fact tables with fixed dimensions.

Given a summarization query $\Q=\AGGQ^*_{\XXT}(F(\A) | \X)$ over a fact table $\XXT(\S)$, we define \emph{summarizability} as a property of an attribute $\A$ with respect to grouping set $\X$ and function $\AGGF$. More exactly, \cite{pedersen_foundation_2001} considers an attribute $\A$ to be summarizable with respect to  $\X$ and $\AGGF$ if for any subset $Z \subset \X$ of dimension attributes, the following condition holds: 
\begin{align}
\label{eq:summarize_pedersen}
\AGGQ^*_{\XXT}(\AGGF(\A) | \Z) = \AGGQ^*_{\Q}(\AGGF(\AGGF(\A)) | \Z) 
\end{align}
Note that this 
definition is more restrictive than our notion of summarizability in \Cref{def:summarizable-attr}, since it imposes that the same function $\AGGF$ is used in the two queries. 
The following conditions provided by \cite{pedersen_foundation_2001} are 
 sufficient for ensuring \Cref{eq:summarize_pedersen}. Let $\X_\D$ denote the set of dimension attributes for dimension $\D$ in $\X$: 
\begin{enumerate}
\item Function $\AGGF$ is \emph{applicable} to $\A$, 
\item Function $\AGGF$ is \emph{distributive} over the domain of values in $\A$.
\item For every dimension $\D$, all the value mappings between the bottom level attribute of $\D$ in $\XXT$ and any attribute of $\X_\D$ are many to one. 
\item For every dimension $\D$, every value of an attribute in $\XXT$ that is a sub-level of an attribute of $\X_\D$ must be non null.  
\end{enumerate}

We now comment each condition and draw comparisons with  
our work. 
In the first condition, \cite{pedersen_foundation_2001} assumes that we know the functions that are applicable to $\A$. Three types of aggregation functions are distinguished: (1) functions, which  are applicable to data that can be added together (e.g., \SUM, \COUNT, \AVG), (2) functions, which are applicable to data that can be used for average calculations (e.g., \COUNT, \AVG, \MIN, \MAX), and (3) functions which are applicable to data that can only be counted. 
However, unlike our work, the notion of aggregation type does not consider the dimensions along which a summarization can be performed. 
The second condition uses a definition of distributive function that is more restrictive than our \cref{def:distributivefct} because it requires that $\AGGF$ is such that for any two sets, $V_1$ and $V_2$, $F(V_1, \cup V_2) = F(F(V_1) \cup F(V_2))$. Thus, functions like $\COUNT$  are discarded.  
The third condition is similar to the disjointness condition of \cite{lenz_summarizability_1997} which we already compared with our work. 
The fourth condition is equivalent to the second completeness condition (2.b) of \cite{lenz_summarizability_1997} but does not cover the first completeness condition (2.a). As mentioned before, aggregation functions in our work handle null values in dimensions as regular values, and can ignore condition (2.b) to guarantee summarizability.  

\subsection{Multidimensional normal forms}

Several works proposed multidimensional normal forms for analytic tables that provide guarantees for the correctness of summarization queries \cite{lehner_normal_1998, lechtenborger_multidimensional_2003}. These normal forms can be used to design dimension and fact tables over which correct summarization queries can be easily detected and evaluated. In the sequel, we are not interested in the design aspects but we examine the definitions of these normal forms as a way to formulate summarizability conditions. 

We first introduce a few concepts and vocabulary. 
A dimension attribute (also called a dimension level) which can have a \emph{null} value is called \emph{optional} and otherwise called \emph{mandatory}. Dimensions must have a single bottom level type and an implicit top level type, called \emph{ALL}. 
The hierarchy in each dimension is strict using "functional dependencies with nulls" (NFD), that is, each attribute $\A_i$ has at most one parent attribute $\A_j$ and each arc $(A_i,A_j)$ in the attribute graph of the dimension is labelled with a $\mathbf{1}$ (\Cref{sec:attribute-graphs}). 

The novelty of \cite{lehner_normal_1998, lechtenborger_multidimensional_2003} in comparison with \cite{pedersen_foundation_2001} is the following. A dimension is also associated with a (possibly empty) set of \emph{context dependencies}: let $\A_i$ and $\A_j$ be two dimension attributes of a dimension $\D$ such that $\A_i$ is optional, $\A_j \neq ALL$, and $\A_i \typeprec A_j$. If $c \in dom(A_j)$ and $c \neq null$,  then $(A_i, A_j, c)$ is a context dependency for $\D$  stating that 
for every tuple $\XXT$ of the dimension table, $t.\A_j=c \Leftrightarrow t.\A_i \neq null$. Intuitively, the interpretation of a context dependency is that $\A_j$ plays the role of a discriminating attribute in the hierarchy and value $c$ is the discriminating value to indicate when the optional attribute $\A_i$ has a non-null value. Note that the use of an equivalence ($\Leftrightarrow$), in the above formula, is quite strong since it forces the existence of a \emph{single} discriminating value. 

In \cite{lehner_normal_1998, lechtenborger_multidimensional_2003}, a fact table is defined over a set of dimensions at the finest level of detail, that is, all dimension attributes of a dimension are included in the schema of the fact table, and each measure in the fact table is determined (with NFD constraints) by the set of bottom level dimension attributes in each dimension. Thus, fact tables are similar to the "micro data" of \cite{lenz_summarizability_1997}. 
Note that although fact tables in \cite{pedersen_foundation_2001} can represent facts at a coarser granularity, their summarizability conditions impose that the bottom level dimension attributes of each dimension determine (with NFD constraints) all measure attributes. 
In \cite{lechtenborger_multidimensional_2003}, \emph{summarizability constraints} express the dimension hierarchy along which a measure can be aggregated using some aggregation function. However, no formal treatment of summarizability constraints is provided, unlike our use of aggregable properties for analyzing aggregate queries and propagating summarizability constraints to query results. In \cite{lehner_normal_1998}, the same categorization of measure attributes as \cite{lenz_summarizability_1997} is used.

In \cite{lehner_normal_1998}, two different multidimensional normal forms are presented. The first one is called \emph{ Multidimensional Normal Form (MNF)} and provides conditions similar to 
the work of \cite{lenz_summarizability_1997} and \cite{pedersen_foundation_2001}. The second one, called \emph{Generalized Multidimensional Normal Form (GMNF)}, provides more general conditions  which have been slightly extended in \cite{lechtenborger_multidimensional_2003}.

In the following we will describe the GMNF as defined by \cite{lechtenborger_multidimensional_2003}. Let $\XXT$ be a fact table defined over a set of dimensions with a measure (summary) attribute $\A$. Then $\XXT$ is in GMNF if all of the following conditions are satisfied:
\begin{enumerate}
    \item\label{it:lech1} For each dimension $\D$ of $\XXT$: 
    \begin{enumerate}
    \item\label{it:lech1a} for every optional dimension attribute $\A_i$ of $\D$, there exists a context dependency $(\A_i, \A_j, c)$ in $\D$;
    \item\label{it:lech1b} the values of the bottom level dimension attribute in $\XXT$ 
    are complete.
    \end{enumerate}
    \item\label{it:lech2} All dimensions are mutually independent, i.e., there exists no NFD between any two dimension attributes of two distinct dimensions.
    \item\label{it:lech3} The set of (unique) bottom level attributes of all dimensions functionally determines (FD) attribute $\A$.
\end{enumerate}

We comment these conditions. Condition \ref{it:lech1a} and condition \ref{it:lech3} enforce the third and fourth conditions of \cite{pedersen_foundation_2001} presented in \Cref{subsec:multidim_summarizabilty}. Indeed, if all dimension attributes are mandatory, they cannot have null values and all dependencies between dimension attributes become functional dependencies. Condition \ref{it:lech1b} is analogous to the completeness condition of \cite{lenz_summarizability_1997}. Here again, the means to test this requirement are left unspecified and seem to require some external knowledge. Finally,  condition \ref{it:lech2} ensures that dimensions do not share dimension attributes and is not really needed for guaranteeing summarizability.  
Note that by conditions \ref{it:lech2} and \ref{it:lech3} the bottom level dimension attributes in the schema of $\XXT$ form a primary key in $\XXT$ and there is no other primary key for $\A$. 
The novelty with respect to the previous models is brought by \ref{it:lech1a}. It constrains the semantics of every optional dimension attribute $\A_i$ so that there exists at an upper level an attribute $\A_j$ that plays the role of discriminator for $\A_i$. Note that $\A_j$ can itself be an optional attribute, in which case there will again be a context dependency $(\A_j, \A_k, c')$ in $\D$. Eventually, the discriminator attribute will be a mandatory attribute since by definition of context dependency, the upper level attribute cannot be $ALL$. 

We can now reformulate the previous GMNF conditions on dimensions  as summarizability conditions on measure attributes and aggregate queries.
Let $\XXT$ be a fact table in GMNF, and $\Q=\AGGQ^*_{\XXT}(\AGGF(\A) | \X)$ be a summarization query over $\XXT$. Then, $\A$ is \emph{summarizable} with respect to function $\AGGF$ and grouping set $\X$, \ie,  for all $\Z\subseteq \X$: $\AGGQ^*_{\XXT}(\AGGF(\A) | \Z) = \AGGQ^*_{\Q}(\AGGF(\AGGF(\A)) | \Z)$) if the following conditions hold:
\begin{enumerate}
    \item\label{it:lechsumm1} $\AGGF$ is applicable to $\A$ with respect to any subset of $\X$ in the result of $\Q$.
    \item\label{it:lechsumm2} One of the two conditions hold:
    \begin{enumerate}
        \item\label{it:lechsumm2a} $\X$ does not contain any optional dimension attribute, or
        \item\label{it:lechsumm2b} 
        \label{cond-2b-GMNF-summarizability} 
        if $\X$ contains an optional attribute $\A_i$ then, assuming that $(\A_i, \A_j, c)$ is the associated context dependency, a filter condition: $\A_j = c$ must be applied on $\XXT$ before the summarization query is applied
    \end{enumerate}  
\end{enumerate}

Here, the condition~\ref{it:lechsumm1} is determined using either the categorization of $\A$ \cite{lehner_normal_1998} or the summarizability constraints on $\A$ \cite{lechtenborger_multidimensional_2003}, the latter being more precise as we stated earlier. The condition~\ref{it:lechsumm2} states that summarizability holds provided that a (filtered) subset of the fact table is considered, and this subset is given by the context dependencies of the dimensions over which the fact table is defined.

\begin{myexample}
\label{GMNF-query-example}
Consider a product dimension, \dimt{PROD\_NEW}, whose attribute graph is depicted in \Cref{fig:prod_new}. A new optional attribute \attr{Video\_Res} provides the video resolution for products with screens where $\attr{Video\_Res} \typeprec \attr{Category}$ and $\attr{Prod\_Sku} \typeprec \attr{Video\_Res}$. Consider the fact table \factt{PROD\_NEW\_SALES} over \dimt{PROD\_NEW} whose instance is displayed in \Cref{fig-GMNF-example}. 

\begin{figure}[htbp]
    \centering
    \includegraphics[width=0.63\linewidth]{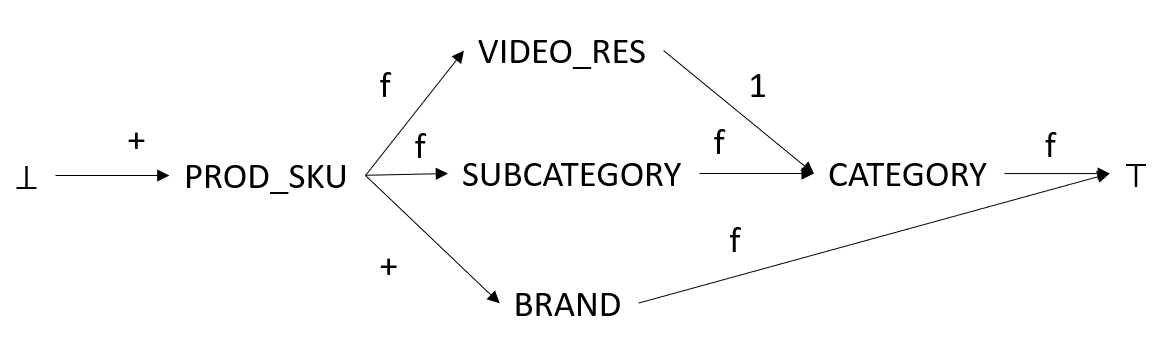}
    \caption{Attribute graph of \dimt{PROD\_NEW}}
    \label{fig:prod_new}
\end{figure}

Assume that the designer of the dimension table \dimt{PROD\_NEW} defined a context dependency :  $(\attr{Video\_Res},\attr{Category},\text{\enquote*{video}})$. Then, table \factt{PROD\_NEW\_SALES} is in GMNF if we assume that all products are listed in the table. It is easy to see that conditions \ref{it:lech2} and \ref{it:lech3} are satisfied on the attribute graph. Condition \ref{it:lech1} is also satisfied because the only optional attribute \attr{Video\_Res} has an upper level discriminating attribute with value \enquote*{video}.

    \begin{table}[t]
    \caption{\factt{PROD\_NEW\_SALES}}
    \label{fig-GMNF-example}
    \normalsize{
    \begin{tabular}[t]{l|l|l|l|l|r}
      \toprule
    \underline{\attr{Prod\_Sku}} & \attr{Subcategory} & \attr{Category} & \attr{Video\_Res} & \attr{Brand} & \attr{Amount} \\
    \midrule
    p\_01 & Video projector & video & 1920x1080 & Epson & 42 \\
    p\_02 & TV & video & 3840x2160 & Philips & 58 \\
    p\_05 & TV & video & 3840x2160 & Samsung & 90 \\
    p\_03 & Radio & audio & - & Philips & 45 \\
    p\_04 & CD-player & audio & - & Samsung & 5 \\
    \bottomrule
    \end{tabular}
    }
    \end{table}

Consider a summarization query $\Q = \AGGQ_{\factt{PROD\_NEW\_SALES}}^*(SUM(\attr{AMOUNT}) \mid X)$. Then $\SUM$ is still applicable to the resulting attribute \SUM(\attr{AMOUNT}). If $\X = \{\attr{Category}, \attr{Brand}\}$ then \attr{AMOUNT} is summarizable with respect to $\SUM$ and $\X$ because $\X$ only contains mandatory attributes.
If $\X = \{\attr{Video\_Res}, \attr{Brand}\}$ then, by summarizability condition \ref{cond-2b-GMNF-summarizability}, \factt{PROD\_NEW\_SALES} must be first filtered with a filter: \attr{Category} = 'Video' before applying $\Q$. Afterwards, \attr{AMOUNT} is summarizable  with respect to $\SUM$ and $\X$. \bernd{added:} Otherwise, because partitions with identifiers containing null values are ignored, a second query taking the sum along \attr{Video\_Res} would generate an incorrect value for Philips ($58$ instead of $103$) and Samsung ($90$ instead of $95$). 
 
\end{myexample}

We now compare GMNF with our work in the same context. Going back to \Cref{GMNF-query-example}, attribute \attr{AMOUNT} is \emph{literally} determined  by the minimal subset of dimension attributes \{\attr{Prod\_Sku}\}, which also determines all other dimension attributes of \factt{PROD\_NEW\_SALES}. Since  $\SUM$ is applicable to \attr{AMOUNT}, by \Cref{def:aggprop} on \Cpageref{def:aggprop}, $\aggp_{\mattr{AMOUNT}} (\SUM, Z)$ holds in \factt{PROD\_NEW\_SALES}, where $Z$ is the set of all dimension attributes of \factt{PROD\_NEW\_SALES}. Consider the query $Q_1$ of \Cref{GMNF-query-example}. By 
\Cnameref{prop:summarizabile-att}%
, attribute \attr{AMOUNT} is summarizable with respect to $\SUM$ and $\X = \{\attr{Category}, \attr{Brand}\}$ since $\X \subset Z$ and $\SUM$ is distributive.  
Now if $\X = \{\attr{Video\_Res}, \attr{Brand}\}$, since $\X \subset Z$,  attribute  \attr{AMOUNT} is also summarizable  with respect to $\SUM$ and $\X$, without requiring any pre-filtering of \factt{PROD\_NEW\_SALES}. The reason is our usage of SQL aggregation operations that considers null values as regular values. 
Indeed, it is easy to see that the summarizability condition is satisfied by looking at the result of $\Q$, displayed in \Cref{fig-query-result-Q2}.
To conclude the comparison, observe that fact table \factt{STORE\_SALES} of \Cref{tab:state-summarizability-sales2} 
is not in GMNF since condition 1 is violated. The optional  \attr{State} has a null value for different countries and it is not possible to create a \emph{single} context dependency for attribute \attr{State} using either attribute \attr{Country} or \attr{Continent}. Thus, in our work, by accepting partition identifiers with null values, we handle cases of summarizability that are rejected by the conditions based on GMNF.

\begin{table}[htb]
\centering
\caption{Query result of $Q_2$}
\label{fig-query-result-Q2}
\normalsize{
\begin{tabular}[t]{l|l|r}
  \toprule
\underline{\attr{Video\_Res}} &  \underline{\attr{Brand}} & \attr{Amount} \\
\midrule
1920x1080 & Epson & 42 \\
3840x2160 & Philips & 58 \\
3840x2160 & Samsung & 90 \\
- & Philips & 45 \\
- & Samsung & 5\\
\bottomrule
\end{tabular}
}
\end{table}

\subsection{Reasoning over constraints on dimensions}
In \cite{hurtado_capturing_2005}, the summarizability constraints on dimensions generalize the idea of context dependencies introduced in \cite{lehner_normal_1998,lechtenborger_multidimensional_2003}. 
The multidimensional data model 
is restricted as follows. All dimension hierarchies have one top level attribute called \attr{ALL} and possibly multiple bottom level attributes. As in \cite{pedersen_foundation_2001, lehner_normal_1998, lechtenborger_multidimensional_2003}, a dimension attribute can have multiple parent dimension attributes in the hierarchy (such dimensions are called "heterogeneous"), and there can be both, mandatory and optional dimension attributes. Every child-parent attribute mapping should be functional (i.e., every value only maps to one parent value). This is equivalent to the existence of an NFD dependency from any attribute $\A_i$ to attribute $\A_j$ where $\A_i \typeprec A_j$. As in \cite{lehner_normal_1998, lechtenborger_multidimensional_2003}, fact tables are defined over dimensions at the finest level of detail, that is, the schema of the fact table includes the bottom level attributes of the dimensions. Measure attributes are determined by \emph{all} the dimensions and can only be aggregated using distributive functions (defined as in \Cnameref{def:distributivefct}). Dimensions are also supposed to be mutually independent in a fact table. 
Summarizability is defined as a property of dimensions and any fact table built over summarizable dimensions has summarizable measures. 
Let $\D$ be a dimension, $\X$ be a subset of dimension attributes in $\D$, and $\B$ a dimension attribute in $\D$ such that $\A_i \typeprec \B$ for some  attribute $\A_i\in\X$.
Attribute $\B$ is \emph{summarizable from $\X$ in $\D$} if and only if for every fact table $\XXT$ defined over $\D$, every measure attribute $\M$ of $\XXT$, every set $\X' \subset  \X$,
and every distributive aggregate function $\AGGF$ using $\AGGG$ that is applicable to $M$, we have: 
\begin{equation}
\label{hurtado-eq}
    \AGGQ^{*}_{\XXT}(F(M) | \B) = \AGGQ^{*}_{\YYT}(G(F(M)) | \B) \mathrm{\;where\;}     \YYT = \AGGQ^{*}_{\XXT}(\AGGF(\M) | \X' \cup \B)
\end{equation}
 

The above definition of summarizability relates to summarizability conditions as follows. In \Cnameref{def:summarizable-attr}, we consider 
a fact table $\YYT = \AGGQ^*_{\XXT}(\AGGF(\M) \mid \X)$ resulting from a summarization query over a fact table $\XXT$, where $\AGGF$ is a distributive function using \AGGG. Then we consider that \emph{$\M$ is summarizable with respect to $\X$ and $\AGGF$} when $\AGGQ^{*}_{\XXT}(\AGGF(\M) | \Z) = \AGGQ^{*}_{\YYT}(\AGGG(\AGGF(\M)) | \Z)$ for any $Z \subset X$. In this case, query $\Q=\AGGQ^{*}_{\YYT}(\AGGG(\AGGF(M)|\Z)$ is considered to be correct by \cite{hurtado_capturing_2005}. In the above definition, Condition \ref{hurtado-eq} must hold for any grouping set $\X' \subset \X$ in the query defining $\YYT$, but $\Z$ is restricted to $\B$. Therefore, the condition to determine if query $\Q$ is correct, is to enforce that \emph{every attribute $\B$ of any dimension $\D$ in $\Z$ is summarizable from  $\X_\D$, where $\X_\D$ is the set of dimension attributes of $\D$ in $\X$} 
(\Cref{hurtado-eq} must hold for all attributes $\B$ of any dimension $\D$ and any subset of attributes $\X'\subset \X_D$). 
 
\cite{hurtado_capturing_2005} show that the summarizability Condition \ref{hurtado-eq} can also be expressed independently of the fact tables which refer to a given dimension:  attribute $\B$ is summarizable from $\X$ in dimension $\D$ if and only if for every bottom level attribute $\A_{\bot}$ of $\D$, the following equality holds, where $\pi$ denotes the relational duplicate elimination projection and $\bowtie$ denotes the null-eliminating join:
 \begin{equation}
 \label{hurtado-dim-eq}
\pi_{A_{\bot}, B}(D) = \bigcup_{A_i \in X}(\pi_{A_{\bot}, A_i}(D) \bowtie \pi_{A_i, B}(D))     
 \end{equation}

\begin{myexample}
Consider the product dimension \dimt{PROD\_NEW} in \Cref{tab:prod-schema}. Using \Cref{hurtado-dim-eq}, we can show that attribute \attr{Category} is summarizable from $\X=\{\attr{Subcategory}\}$ because 
$\pi_{\attr{Prod\_Sku}, \attr{Category}}(\dimt{PROD\_NEW})$ is equal to the join $\pi_{\attr{Prod\_Sku}, \attr{Subcategory}}(\dimt{PROD\_NEW})  \bowtie\allowbreak \pi_{\attr{Subcategory}, \attr{Category}}(\dimt{PROD\_NEW})$.
However, attribute \attr{Category} is not summarizable from $\X=\{\attr{Video\_Res}\}$ because the natural join between 
$\pi_{\attr{Prod\_Sku}, \attr{Video\_Res}}(\dimt{PROD\_NEW})$ and  $\pi_{\attr{Video\_Res}, \attr{Category}}(\dimt{PROD\_NEW})$  eliminates products 'p\_03' and 'p\_04'. 

    \begin{table}[htb]
    \caption{\dimt{PROD\_NEW}}
    \label{tab:prod-schema}
    \normalsize{
    \begin{tabular}[t]{l|l|l|l|l}
      \toprule
    \underline{\attr{Prod\_Sku}} & \attr{Subcategory} & \attr{Category} & \attr{Video\_Res} & \attr{Brand} \\
    \midrule
    p\_01 & Video projector & Video & 1920x1080 & Epson \\
    p\_02 & TV & Video & 3840x2160 & Philips \\
    p\_05 & TV & Video & 3840x2160 & Samsung \\
    p\_03 & Radio & Audio & - & Philips \\
    p\_04 & CD-player & Audio & - & Samsung \\
    \bottomrule
    \end{tabular}
    }
    \end{table}

\end{myexample}

 To efficiently check summarizability, \cite{hurtado_capturing_2005} proposes to 
 transform the summarizability problem into the problem of verifying the satisfaction of a set of dimension constraints by some dimension $\D$. Let $\A_i$ be a dimension attribute of a dimension $\D$ 
 and $\langle \A_i, \A_{i+1}, ..., \A_{i+n} \rangle$ denote a path in $\D$ such that $\A_k\typeprec \A_{k+1}$, $i\leq k < i+n$. 
 Then, the following  dimension constraints can be defined on $\A_k$ and $\D$: 
\begin{enumerate}
    \item $D \models \langle \A_i, \A_{i+1}, ..., \A_j \rangle$ means that for every attribute value $v$ of $\A_i$, 
    there exists a tuple $t$ in $\D$ such that $t.\A_i = v$ and $t.\A_{i+1}$, ..., $t.\A_{j}$ are non-null (all values of $\A_i$ roll-up to a value $\A_j$ through a value of $\A_{i+1}$ ...).
    We shall say that $\A_i$ \emph{rolls up to} $\A_j$.  
    \item $D \models \langle \A_i, ..., \A_j = k \rangle$ means that for every attribute value $v$ of $\A_i$, 
    there exists a tuple $t$ in $\D$ such that $t.\A_i = v$ and $t.\A_j$ is not null if and only if $t.\A_j = k$.
\end{enumerate}

Constraints can then be composed using the usual Boolean logical connectives. Now, assume that a set of constraints are specified on the schema of $\D$. 
To determine if an attribute $\B$ is summarizable from $\X = \{A_1, ..., A_n\}$ in $\D$, one must determine if, for each bottom level attribute $\A_{\bot}$ of $\D$, the following constraint is satisfied : 
\begin{align}
D \models \langle A_{\bot}, ..., B \rangle \implies  
(\langle A_{\bot}, ..., A_1, ..., B \rangle \oplus  
... \oplus 
(\langle A_{\bot}, ..., A_n, ..., B \rangle)
\end{align}
where  $\oplus$ denotes an exclusive disjunction (XOR). 
Intuitively, if all values of $\A_{\bot}$ roll-up to a value of $\B$ then  all these values either roll-up through values of $\A_1$ or (exclusive) through values of $\A_2$ ... or through values of $\A_n$. 
We use the following examples to illustrate the use of constraints to determine summarizability.

\begin{myexample}
Suppose that in dimension \dimt{PROD\_NEW}, \attr{VIDEO\_RES} has now \attr{Subcategory} as parent in the hierarchy type (\Cref{fig:prod_new_hierarchy_bis}). 
Then the two constraints (a) and (b) shown in \Cref{fig:prod_new_const_bis} are expressed  on \dimt{PROD}. Note that the disjunction constraint (a) is more expressive than the context dependency of \cite{lehner_normal_1998, lechtenborger_multidimensional_2003} because it is not restricted to a single value.
\begin{figure*}[htb]
\centering
\begin{minipage}[b][4cm][b]{.35\textwidth}
    \includegraphics[width=0.65\textwidth]{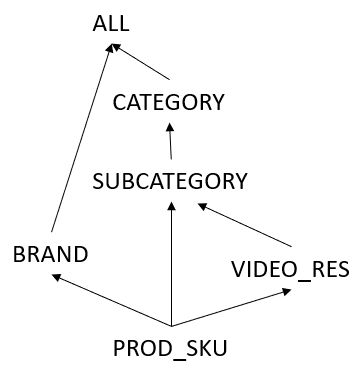}
    \subcaption{Hierarchy type of \dimt{PROD\_NEW}}
    \label{fig:prod_new_hierarchy_bis}
\end{minipage}
\begin{minipage}[b][4cm][b]{.55\textwidth}
\normalsize{
    \begin{tabular}{l|l}
    \toprule
    \multicolumn{2}{c}{Rules}\\
    \midrule
    (a) &  \makecell[tl]{$\langle \attr{Prod\_Sku}, \attr{Video\_res}, \attr{Subcategory} =$ 'TV' $\rangle$ $\oplus$ \\
    $\langle \attr{Prod\_Sku}, \attr{Video\_res}, \attr{Subcategory} =$ 'Video projector' $\rangle$}\\
    (b) & \makecell[tl]{$\langle \A_i, \A_j \rangle$, for all other edges ($\A_i, \A_j$)}\\
    \bottomrule
    \multicolumn{2}{c}{Semantics}\\
    \midrule
        (a) & A value of \attr{Prod\_Sku} rolls up to \attr{Video\_res} and \attr{Subcategory} \\ 
         & only for the 'Video Projector' and 'TV' values of \\
         &  \attr{Subcategory} \\
        (b) & All other attributes directly roll up to their parent attribute \\
    \bottomrule
\end{tabular}
   \subcaption{Constraints on \dimt{PROD\_NEW}}
   \label{fig:prod_new_const_bis}
}
\end{minipage}

\caption{The dimension schema of \dimt{PROD\_NEW}}
\end{figure*} 

The attribute \attr{CATEGORY} is summarizable from $\X = \{\attr{SUBCATEGORY}\}$ because the following constraint is satisfied (all products roll up to a category through some subcategory):
\begin{align}
\dimt{PROD}\models \langle \attr{PROD\_SKU}, \ldots, \attr{CATEGORY}\rangle \Rightarrow \langle \attr{PROD\_SKU}, \attr{SUBCATEGORY}, \attr{CATEGORY}\rangle
\end{align}

Using the constraints in (b), we can compose $\langle \attr{PROD\_SKU}, \attr{SUBCATEGORY}\rangle$ and $\langle \attr{SUBCATEGORY}, \attr{CATEGORY}\rangle$ to yield the final constraint. Thus, if the table \factt{PROD\_NEW\_SALES} is first aggregated with a grouping set $\{\attr{SUBCATEGORY}\, \attr{CATEGORY}\}$, then a query that further aggregates this result with a grouping set $\{\attr{CATEGORY}\}$ is correct. 

However, attribute \attr{SUBCATEGORY} is not summarizable from $\X = \{\attr{VIDEO\_RES}\}$ because the following constraint cannot be satisfied:
%
\begin{align}
\dimt{PROD}\models \langle \attr{PROD\_SKU}, \attr{SUBCATEGORY}\rangle \Rightarrow \langle \attr{PROD\_SKU}, \attr{VIDEO\_RES}, \attr{SUBCATEGORY}\rangle
\end{align}

Thus, if table \factt{PROD\_NEW\_SALES} is first aggregated with a grouping set $\{\attr{SUBCATEGORY}, \attr{VIDEO\_RES}, \attr{CATEGORY}\}$, then a query $\Q$ that further aggregates this result with a grouping set $\{\attr{SUBCATEGORY}, \attr{CATEGORY}\}$ is incorrect.
If the tuples of \factt{PROD\_NEW\_SALES} with \attr{SUBCATEGORY} attribute values 'TV' or 'Video projector' are filtered out then the summarizability constraint can be satisfied and previous query $\Q$ will be correct.

\end{myexample}

It is clear that the data model and constraints proposed by \cite{hurtado_capturing_2005} subsume the data model with context dependencies of \cite{lehner_normal_1998, lechtenborger_multidimensional_2003}. 
We already showed that our summarizability conditions are more expressive than context dependencies by considering $null$ values as regular values for aggregation. The same arguments apply to dimension constraints.
In addition, \cite{hurtado_capturing_2005} has the following limitations with respect to our work. First, 
any non-null value of a mandatory attribute must map to a single parent value in the hierarchy. This discards the use of dimension tables like \dimt{SALESORG} in \Cref{tab:state-salesorg2}. Second, measure attributes cannot depend on a subset of the dimensions of a fact table. This discards the use of fact tables resulting from interactive user queries like the result of the left-merge of \factt{T2} with \factt{DEM'}in \Cref{tab:T3-bis}, presented in the Introduction section. Finally, unlike other works, the notion of applicability of an aggregation function to an attribute is not covered. 

\section{Conclusions}

In this article, we introduce a new framework for controlling the correctness of aggregation operations during sessions of interactive analytic queries. 
Our framework adopts an \emph{attribute-centric view}, whereby \emph{aggregable properties} of attributes are used to describe and  control the interaction between measures, dimensions and aggregation functions. 
As a first advantage, aggregable properties enable the designers of analytic tables to describe the wide variety of semantic properties of measure attributes with respect to their dimensions defined in previous work \cite{kimball_data_2013, horner_analysis_2004, niemi_detecting_2014, stevens_theory_1946, pedersen_foundation_2001, lenz_summarizability_1997}. Another advantage of aggregable properties is their ability to guarantee that aggregate queries over some attributes can only be expressed if these attributes are summarizable. 
We provide two definitions of summarizable attributes. Our first definition covers the case when an aggregate query is defined over the result of another aggregate query; it subsumes the definitions of previous work on summarizability \cite{lenz_summarizability_1997, pedersen_foundation_2001, lehner_normal_1998, lechtenborger_multidimensional_2003, hurtado_capturing_2005}. Our second definition introduces the new notion of \emph{G-summarizability} which applies in the case of an aggregate query defined over the result of an arbitrary analytic queries. The two definitions are complementary. Our main technical results are the definition of \emph{propagation} rules that automatically compute the aggregable properties of attributes in the result of an analytic query knowing the aggregable properties of the attributes in the operand tables of the query. We progressively refine our propagation rules to handle the semantic properties of measures, and the summarizability and G-summarizability properties of attributes. 

There is a number of perspectives for future work. 
First, aggregable properties could be extended to handle other correctness issues of aggregate queries.  Currently, we rely on literal functional dependencies (LFD) for analyzing the summarizability properties of query results. 
Simpson's paradox~\cite{wagner1982simpson} is an example of incorrect causal interpretation of aggregated attribute values where a statistical observation like ratio or bias on the measures from several partitions might disappear or be inverted on the aggregated measures over these partitions.
Aggregable properties could be extended to guard against this kind of statistical errors by exploiting existing causal dependencies between table attributes \cite{pearl2016causality}  (representing features) in addition to LFD.

Another direction of research is the "explainability" of aggregated values in an analytic table resulting from an interactive data analysis session. Aggregable properties provide explanations for the decision to forbid an incorrect  aggregate query on a table. They also help the user to backtrack in her session to find an intermediate result over which a desired aggregation can be expressed. However, as shown in \Cref{ex:explainability}, a priori non-summarizable aggregate queries could be accepted provided that adequate annotations are added to the result table so that aggregated values can be properly interpreted. Generating such minimal   annotations and propagating them is still open.

Finally, although a large part of our data model has already been prototyped in SAP HANA  ~\cite{liu_rutian_semantic_2020}, a significant effort is needed to integrate our framework into self-service data preparation tools and analytic database systems.  
 
Another possible direction is the  implementation of our framework as an independent software component, \eg Python library, which could be part of data preparation pipelines for Deep Learning applications (Python notebooks).

\bibliographystyle{ACM-Reference-Format}
\bibliography{references}

\end{document}